\documentstyle[12pt]{article}
\begin{document}
\titlepage

\title{\begin{flushright}
Preprint PNPI-2449,  2001
\end{flushright}
\bigskip\bigskip\bigskip
Manifest calculation and the finiteness
of the superstring Feynman diagrams}
\author{G.S. Danilov \thanks{E-mail address:
danilov@thd.pnpi.spb.ru}\\ Petersburg Nuclear Physics
Institute,\\ Gatchina, 188300, St.-Petersburg, Russia}

\maketitle
\begin{abstract}
The multi-loop amplitudes for the closed, oriented superstring
are represented by finite dimensional integrals of explicit
functions calculated through the super-Schottky group parameters
and interaction vertex coordinates on the supermanifold.
The integration region is proposed to be consistent
with the group  of the local symmetries of the amplitude
and with the unitarity equations. It is shown that, besides the
$SL(2)$ group, super-Schottky group and modular one, the total
group of the local symmetries includes an isomorphism between
sets of the forming group transformations,
the period matrix to be the same.  The singular integration
configurations are studied.  The calculation of the integrals over
the above configurations is developed preserving all the local
symmetries of the amplitude, the amplitudes being free from
divergences.  The nullification of the 0-, 1-, 2- and 3-point
amplitudes of massless states is verified.  Vanishing the
amplitudes for a longitudinal gauge boson is argued.

\end{abstract}
\newpage

\section{Introduction}
Superstrings \cite{gsw} are currently considered for the unified
interaction particle  theory. Nevertheless, despite great
efforts \cite{ver,as,momor,vec,pst,ntw,mand} during years, the
multi-loop superstring interaction amplitudes were not
calculated explicitly so as to be suitable for applications.

It has been attempted \cite{ver} to  construct the integration
measures (partition functions) from modular forms on the Riemann
surface that requires a complicated module description.  Further
still, the world-sheet supersymmetry is lost as far as in this
case the amplitudes depend on the choice of a basis of the
gravitino zero modes \cite{as,momor}. The 0-, 1-, 2- and 3-point
massless state amplitudes of higher genera are not seen to be
nullified contrary to the requirement \cite{martpl} following
from the space-time supersymmetry. In the supercovariant scheme
\cite{bshw} the calculation is complicated because of Grassmann
moduli, especially, in the Ramond case. As the result, for an
extended period only the Neveu-Schwarz spin structures where
known explicitly \cite{vec,pst,dan90}.

All the contributions to the amplitude, the Ramond sector
included, have been explicitly calculated in
\cite{danphr,dannph,dan96}.  The superstring amplitudes have
been given by  integrals of a sum over super-spin structures
\cite{danphr,dannph}
defined by super-Schottky groups on the complex $(1|1)$
supermanifold \cite{bshw}.  Every term of the sum is
the integration measure times the vacuum value of the
vertex product calculated through the superfield vacuum
correlators.  The super-spin structures \cite{danphr,dannph} are
superconformal extensions of ordinary spin ones \cite{swit}.
Being an $SL(2)$ transformation, the super-Schottky group
element "$s$" is, generically, determined \cite{vec,pst,volkov}
by its multiplier $k_s$ along with limiting points
$U_s=(u_s|\mu_s)$ and $V_s=(v_s|\nu_s)$ where $\mu_s$ and
$\nu_s$ each are the Grassmann partner of $u_s$ or,
respectively, of $v_s$.  The integrations are performed over the
above parameters and over coordinates of the interaction
vertices on the complex $(1|1)$ supermanifold, any $(3|2)$
variables being fixed due to the $SL(2)$
symmetry.

The integration region over the Schottky parameters was not,
however, fully known. Moreover, the calculation of the integrals
remained to be solved because of degenerated configurations
where the leading approximation integrand is proportional to
either a vacuum function, or to the 1-point one. Unlike the
genus-1 case, the higher genus functions under discussion do not
vanish locally in the space of super-Schottky group parameters.
In this case the amplitude seemingly diverges like the boson
string amplitude \cite{bk}. In the superstring, due to Grassmann
integrations, the result is, however, finite or divergent being
dependent on the choice of the integration variables (see a
discussion of this point in the beginning of Section 7 of the
present paper).  In this case one must be guided by the
requirement to preserve the local symmetries of the amplitude.
In particular, if divergences appear, the $SL(2)$ symmetry is
broken due to the cut-off parameter. As was observed
\cite{dan99}, the divergences have an evident tendency for the
cancellation, if the calculation preserves the explicit $SL(2)$
symmetry.  Nevertheless, in \cite{dan99} the total cancellation
of the divergences was not verified. Moreover, an explicit
regularization procedure \cite{dan99} for the integrals is not
necessary as far as the result is expected to be finite.

At present paper the integration region over the Schottky
parameters is given to be restricted by the group of the local
symmetries of the amplitude and by the unitarity.
The above group of the local symmetries is considered.
The singular configurations are
studied. The calculation of the integrals over the singular
configurations is proposed. In this case we, step-by-step,
integrate over variables of every handle in such a way that for
every step, the integral is convergent. So the multi-loop
superstring amplitudes are obtained to be finite.  We argue that
the calculation preserves all the local symmetries of the
amplitude.  The finiteness of the superstring amplitudes has
been expected \cite{gsw,mand,berk} from years, but till now it
was clearly seen only for the one-loop case.
Unlike one-loop amplitudes, the finiteness of the
multi-loop ones is provided not by local sum rules but by
the vanishing of certain convergent integrals of a sum over
super-spin structures.  We verify the required nullification of
the vacuum amplitude, vacuum-dilaton transition constant and of
the 2- and 3- point amplitudes for the massless boson states. We
show the vanishing of the amplitudes for the emission of a
longitudinal boson due to the gauge invariance.

The present paper provides rich opportunities for applications
of the superstring perturbation series. In particular, the
obtained expressions can be used for the calculation of the
infinity genus amplitudes and for the summation of the
perturbation series under various asymptotics conditions. Being
calculated in the infra-red energy limit, they give explicit and
rather compact amplitudes in any perturbative order for the
10-dimensional $IIA$ super-gravity corresponding to the
superstring considered (to be given in another place). So, among
other things, the superstring perturbation series is the
wonderfully effective instrument for the explicit calculation in
the quantum field theory \cite{kapl}.  In this paper the closed,
oriented superstring theory is considered. An extension of the
results to the open and/or non-oriented superstring will be
given elsewhere. The perturbation series can also be constructed
in heterotic and compactified superstring theories. Moreover,
the series can be built for the Dirichlet boundary conditions.
The perturbation series for relevant compactified (super)string
can be used for calculations in four-dimensional quantum field
theories \cite{kapl}.

As it is known, the group of the local symmetries of the
amplitude includes the $SL(2)$ group, changes of any
given interaction vertex coordinate by super-Schottky group
transformations and the modular group. We show that, in
addition, the total
genus-$n$ group ${\cal R}_n$ contains also a group $\{\tilde
G\}$ of isomorphic replacements
$\Gamma_s\to G_s\Gamma_sG_s^{-1}$ of the given
$\{\Gamma_s\}$ set of forming super-Schottky group
transformations by the $\{G_s\Gamma_sG_s^{-1}\}$ set.  Here
$\{G_s\}$ is a relevant set of the super-Schottky group
transformations (not every  $\{G_s\}$ set
originates the isomorphism, and, therefore, is relevant!)
Evidently, the considered isomorphism does not touch the period
matrix.

For the Riemann surface theory, there are known\cite{siegal}
the constraints making the period matrix to be  inside the
fundamental domain of the modular group. In the present paper we
extend the constraints \cite{siegal} to the superstring case
where the period matrix depends on the super- spin structure
\cite{pst,danphr}. The period matrix being given through
the super-Schottky group parameters, the above constraints bound
a region for the Schottky variables. The discussed constraints
are, however, invariant under
the above $\{\tilde G\}$
group transformations preserving the period matrix.  Moreover, we
demonstrate  module configurations destroying
the unitarity equations (see Section 4 of the present paper).
In this case one of the limiting points
of the forming group transformation $\Gamma_s$ penetrates deep
enough into the Schottky circle for the forming transformation
$\Gamma_r$ with $r\neq s$ while another point lies outside both
Schottky circles assigned to $\Gamma_r$. The considered
configurations need to be excluded from the integration region.
Evidently, they can not be reduced by the $\{\tilde G\}$
transformations to those configurations, which are consistent
with the unitarity equations. Hence the fundamental domains for
the direct product of $SL(2)$ group times $\{\tilde G\}$
do not all are equivalent to each other, but they
are divided into the classes of the equivalent domains.

The unitarity equations are already saturated due to the
region where all the Schottky circles of the
forming group transformations are separated from each other, but
it is not a fundamental region for the local symmetry group.  We
argue that the integration region can be restricted by
configurations having no group limiting  points inside the common
interior of any pair of Schottky circles of the forming
group transformations. Really the discussed
region can be varied within a certain range, as it is usually
for a  group fundamental region.

The modular symmetry constraints \cite{siegal}, along with the
$\{\tilde G\}$ symmetry, bounds the absolute value of the
multiplier of any Schottky group transformation by a magnitude
smaller than unity. In this case, for not very high genera, we
show that the Schottky circles of the same forming
transformation are separated from each other. For the sake of
simplicity we assume the same to be in the general case though
now we have not a full proof for this.

When the $k$ multiplier for the boson loop is nullified,
the divergence appear for the single spin structure. The
divergence is, however, canceled for the sum
over relevant spin structures, just as in the Neveu-Schwarz
sector \cite{vec}.
Divergences might be also due to singularities in the
$\{u_s,v_s\}$ limiting points of the forming group
transformations. As has been noted above, generally, the
discussed singularities for the genus-$n>1$ amplitudes are not
canceled locally  even for the total sum over the spin
structures. Nevertheless, we give the integration procedure
calculating the amplitude to be finite and consistent with its
local symmetries.

When all the limiting group points lay
within a finite domain, the singularities in
$\{u_s,v_s\}$ do not appear in the integration region until
certain of $(v_s-u_s)$ differences go to zero. Generally, the
integrand is, however, singular when all the limiting
points of $n_1\leq n$ transformations
go to the same point $z_0$. For the amplitudes
having more than three legs, the basic configurations dangerous
for the divergences are the configurations with no more than
one interaction vertex coordinate going to $z_0$.

When limiting points $u_1$ and $v_1$
of the sole transformation go to
each other, the integrand is singular in
 $\tilde u_1=v_1-u_1$ (or
in $w_1=\tilde v_1-\nu_1\mu_1$) at $\tilde v_1\to0$ (or at
$w_1\to0$). By a direct calculation
using manifest expressions, one can, however, verify
that the singularity is canceled.
The total cancellation of the
singularity is achieved after the summation over the spin
structures of the considered handle and, if the interaction
vertex coordinate goes to $z_0$, once certain integrations to be
performed, $u_1$ or $v_1$  being fixed (see the Section 6).
Therefore, the contribution to the amplitude of the $n_1=1$
configuration is finite provided that the integral over
$\tilde v_1$ (or $w_1$) is calculated after integrations
over other variables of the considered configuration
to be taken.

For the $n_1=2$
we consider the integral
over variables of the discussed configuration keeping
$\tilde v_2$ (or $w_2$) to be fixed. So the integral is
a function of $\tilde v_2$ (or
$w_2$).
The integration over
$\tilde v_1$ (or $w_1$) is taken after the integration
over  the remaining variables of the first handle to be
performed. In this case the integral of the sum over the spin
structures is convergent due to the
cancellation of the singularity at $\tilde v_1\to0$ or $w_1\to0$
for the $n_1=1$ integral with fixed $\tilde v_1$ of the sum over
the spin structure of the handle.
Indeed, at $\tilde
v_1=0$ (or $w_1=0$)  the integrand of the $n_1=2$ integral is a
sum of terms bi-linear in the $n_1=1$ integrals (see the proof
in the beginning of Section 7).  Moreover, we show (Section 7)
that the $n_1=2$ integral is convergent being taken of the sum
over the spin structures either of two handles.
In the calculation of the amplitude we have
deal only with these convergent integrals and with the integral
of the total sum over spin structures of both handles, which is
convergent, too.  As we show, the last integral being
a function of $\tilde v_2$ (or $w_2$), is finite
at $\tilde v_2=0$ (or $w_2=0$). Thus the contribution to the
amplitude from the discussed configuration is finite, as far as
this contribution to the amplitude is just obtained by
the additional integration over $\tilde v_2=0$ (or $w_2=0$) of
the discussed integral with $\tilde v_2$ (or $w_2$)
to be fixed.

The required finiteness of the integral with fixed $\tilde
v_2\to0$ (or $w_2\to0$) at $\tilde v_2\to0$ ($w_2\to0$) is
manifested through the change of the integration variables
by relevant transformations from $SL(2)$ and $\{\tilde G\}$
groups, as well as by a super-Schottky group transformation
of the interaction vertex coordinate (if it goes to $z_0$).
The discussed finiteness of the
integral with fixed $\tilde v_2\to0$ (or $w_2\to0$) at $\tilde
v_2\to0$ ($w_2\to0$) is again shown to be due to the
cancellation of the singularity for the $n_1=1$ integrals.
Apart from $SL(2)$ transformations, the remaining
transformations depend, however, on the spin structure because
of fermion-boson mixing, which is present for non-zero
$\{\mu_s,\nu_s\}$ Grassmann parameters. At the same time, we can
not perform the transformations of the integration variables for
the integral of the single spin structure since this integral is
divergent.  Nevertheless, the transformations dependent on the
spin structure of the first handle can be made for the
convergent integral of the sum over the spin structures of the
second handle.  And the transformations dependent on the spin
structure of the second handle can be made for the convergent
integral of the sum over the spin structures of the first one.
Being a group product of the above
transformations,
an arbitrary $\{\tilde G\}$ or super-Schottky group
transformation can be performed only for the integral of the
total sum over the spin structures of the configuration. Hence
the cancellation of the singularity at $\tilde v_2\to0$
($w_2\to0$) is verified only for the integral of the total sum
over the spin structures.  Correspondingly, the finiteness of
the contribution to the amplitude from the discussed $n_1=2$
configuration is achieved only for the integrals of the total
sum over the spin structures of the configuration. In this case
one calculates the integral over the variable of the 1-st
handle keeping $\tilde v_1$ (or $w_1$) to be fixed. Then one
calculate the integral over $\tilde v_1$ (or $w_1$). After this
one calculates the integral over the variable of the 2-nd handle
keeping $\tilde v_2$ (or $w_2$) to be fixed, and after this the
integral over $\tilde v_2$ (or $w_2$) is calculated.  The
calculation preserves all the local symmetries of the amplitude
including the modular symmetry.
Being, generally, dependent
on the spin structure \cite{dannph}, modular
transformations can be performed only for the integral of the
total sum over the spin structures, like the above discussed
$\{\tilde G\}$
and super-Schottky group transformations.

In the general case $n_1>2$ one integrates,
step-by-step, over the limiting points of every forming
transformation $s$ (except the points fixed due to the $SL(2)$
symmetry) once the summation over its spin structures to be
performed. The integral is first calculated with
fixed $\tilde v_s= v_s-u_s$ or, on equal terms, with fixed
$w_s=\tilde v_s-\nu_s\mu_s$. For every step, the integral being
a function of $\tilde v_s$ (or $w_s$), is shown to be
non-singular at $\tilde v_s=0$ (or $w_s=0$). Hence it can
be further integrated over $\tilde v_s$ (or $w_s$), the result
being finite. When some $(u_s,v_s)$ pairs go to the infinity
the integral is finite again
since the infinite
point can be reduced to the finite one by the relevant $L(2)$
transformation.  As far as the calculation preserves all the
local symmetries, the amplitude is independent of the
fundamental region of the ${\cal R}_n$ group, which is employed
as the integration region.  In particular, the amplitude is
independent of those the $(3|2)$ variables, which are fixed.

The 0-, 1-, 2- and 3-point massless state
amplitude is given by the integrals with the $(3|2)$
fixed parameters $(u_1|\mu_1)$
and $(v_1|\nu_1)$ assigned to any given handle along with one of
two local limiting points ($u_2$ or $v_2$) of any one of the
remaining handles. We show this integral is convergent, and
it is zero at $u_1\to v_1$. Thus, being invariant under the
$SL(2)$ and $\{\tilde G\}$ group and under the super-Schottky
group (for 1-, 2- and 3-point amplitudes), the integral vanishes
identically for any $v_1$ and $u_1$, as it is required.

There are different super-extensions of ordinary spin
structures, but not all they are  suitable for
the superstring, especially, because the space of
half-forms does not necessarily have a basis when there are odd
moduli \cite{hodkin}. The super-Schottky groups for
all superspin structures have been constructed in
\cite{danphr,dannph,dan93}. In the Neveu-Schwarz case they
have been given before \cite{vec,pst,martnp}.
Due to the  fermion-boson agitation, the genus-$n>1$ super-spin
structures are different from
ordinary spin structures \cite{swit} where boson fields are
single-valued on Riemann surfaces while fermion fields being
twisted about $(A,B)$-cycles, may only receive the sign.
Really the super-Schottky group description of the supermanifold
is non-split in the sense of \cite{crrab}. The transition to a
split description is singular \cite{dan97}, the
superstring being non-invariant under the transition discussed.
As the result, the world-sheet supersymmetry is lost in
\cite{ver} where a split module description was implied.
Contrary to \cite{ver}, our calculation preserves
the world-sheet supersymmetry, as well as other local
symmetries of the amplitude.

In the calculation we use partial functions and superfield
vacuum correlators obtained \cite{danphr} by
equations, which were derived from
the requirement that
multi-loop superstring amplitudes are independent of a choice
of both the {\it vierbein} and the gravitino field.
As it is usually, the superfield vacuum correlators is
calculated through the holomorphic Green functions.  For the
Ramond type handle, the round about the Schottky circle being
given by a non-split transformation
\cite{danphr,dannph,dan93},  the holomorphic Green
functions in the Ramond sector can not be represented by the
Poincar{\'e} series \cite{ales}.  Correspondingly, the
integration measure is not a product over known expressions
\cite{vec,dan90} in terms of super-Schottky group
multipliers.  Nevertheless, the discussed genus-$n$ functions
can be given \cite{danphr} by a series over integrals of
relevant genus-1 function products.  For the Neveu-Schwarz
sector the above series can be reduced to the Poincar{\'e}
series. Now we continue the study of the holomorphic Green
functions and integration measures. In particular, we present
them (Appendices B and C of the paper ) in the form, which is
more convenient for application than the expressions in
\cite{danphr}.  We also derive them
through the functions of lower genera $n_i\geq1$. We employ
these expressions for the calculation of the
integrals over singular configurations.

The paper is organized as it follows. Sections 2 contains a
brief review of the super-Schottky group parameterization
\cite{danphr,dannph} using in the paper. The expression for
the multi-loop superstring amplitude is given. In Sections 3 and
4 the constraints on the
integration region
are discussed. The
integration region over the
Schottky parameters is proposed.
In Section 5 the integration measures and the
superfield vacuum correlators are derived through the lower
genus functions.
The integration measures and the
superfield vacuum correlators are calculated for the degenerated
configurations dangerous for divergences. In Section 6 the
strategy calculating the integrals is discussed. Vanishing the
amplitude of the emission of a longitudinally polarized boson is
argued.  Configurations dangerous for divergences are collected.
The cancellation of divergences for easy configurations is
demonstrated. In Section 7 the finiteness of the amplitudes is
shown. The vanishing of the 0-, 1-, 2- and 3-point functions is
verified. The preservation of the local symmetries of the
amplitudes is argued.

\section{Expression for the multi-loop superstring amplitude}

As it was noted, we employ
super-Schottky groups variables. The super-Schottky group
determines the super-spin structure on the complex
$(1|1)$ supermanifold mapped by the $t=(z|\vartheta)$
coordinate. The genus-$n$ super-spin structure presents a
superconformal extension of the relevant genus-$n$ spin
one given by the set of transformations
$\Gamma^{(0)}_{a,s}(l_{1s})$ and $\Gamma^{(0)}_{b,s}(l_{2s})$
(where $s=1,\dots n$), which correspond to the round of
$A_s$-cycle and, respectively, of the $B_s$-cycle on the Riemann
surface. They depend on the theta function characteristics
$l_{1s}$ and $l_{2s}$ assigned to the given handle $s$.  A
discrimination is made only between those (super-)spin
structures, for which field vacuum correlators are distinct. So
$l_{1s}$ and $l_{2s}$ can be restricted by 0 and 1/2. In this
case $l_{1s}=0$ is assigned to the Neveu-Schwarz handle while
$l_{1s}=1/2$ is reserved for the Ramond one. In doing so
\begin{equation}
\Gamma^{(0)}_{b,s}(l_{2s})=\left\{z\to \frac{a_sz+b_s}
{c_sz+d_s}\,,\quad\vartheta\to -\frac{(-1)^{2l_{2s}}}{c_sz+d_s}
\vartheta\right\},\,\, \Gamma^{(0)}_{a,s}(l_{1s})=\left\{z\to
z\,,\quad\vartheta\to (-1)^{2l_{1s}}\vartheta \right\}
\label{zgama}
\end{equation}
where $a_sd_s-b_sc_s=1$.
Furthermore,
\begin{equation}
a_s=\frac{u_s-k_sv_s}{\sqrt
{k_s}(u_s-v_s)}\,,\quad d_s=\frac{k_su_s-v_s}
{\sqrt{k_s}(u_s-v_s)}\quad{\rm and}\quad
c_s=\frac{1-k_s}{\sqrt{k_s}(u_s-v_s)}
\label{uvk}
\end{equation}
where $k_s$ is a complex multiplier and $|k_s|\leq1$. Further,
$u_s$ is the attractive limiting (unmoved) point of
(\ref{zgama}) while $v_s$ is the repulsive limiting one.  The
set of transformations (\ref{zgama}) along with their group
products form the Schottky group. The
$\Gamma^{(0)}_{b,s}(l_{2s})$ transformation in (\ref{zgama})
turns the boundary of the Schottky  circle $C_{v_s}$ into the
boundary of $C_{u_s}$ where
\begin{equation}
C_{v_s}=\{z:|c_sz+d_s|=1\}\quad{\rm
and}\quad C_{u_s}=\{z:|-c_sz+a_s|=1\}
\label{circ}
\end{equation}
for $s=1,\dots n$.
Using (\ref{uvk}), one can see that $v_s$ lies inside $C_{v_s}$
and outside $C_{u_s}$. Correspondingly, $u_s$ is inside
$C_{u_s}$ and outside $C_{v_s}$. Since
$\Gamma^{(0)}_{a,s}(l_{1s})$ in (\ref{zgama}) corresponds to the
round of the Schottky circle, in the Ramond case a square root
cut appears on $z$-plane between $u_s$ and $v_s$. The
fundamental region of a group is that one, which does not
contain points congruent\footnote{The congruent points, or
curves, or domains are those related by a group transformation
other that the identical transformation \cite{ford}.} under
the group, and such that the neighborhood of any point
on the boundary contains points congruent to the points of the
region \cite{ford}. In particular, a fundamental region
of the Schottky group on the complex $z$ plane is the
exterior of all the Schottky circles associated with the given
group. The group invariant integral of the conformal $(1,1)$
tensor being  calculated over the fundamental region of the
group, does not depend on the choice of the
fundamental region above \cite{ford}.

In the superstring theory the forming transformations
(\ref{zgama}) are replaced by $SL(2)$ transformations
$\Gamma_{a,s}(l_{1s})$ and $\Gamma_{b,s}(l_{2s})$ with
\cite{danphr,dannph,dan93}
\begin{equation}
\Gamma_{a,s}(l_{1s})=\tilde\Gamma^{-1}_s
\Gamma^{(0)}_{a,s}(l_{1s})\tilde\Gamma_s\,,
\qquad \Gamma_{b,s}(l_{2s})=
\widetilde\Gamma^{-1}_s\Gamma^{(0)}_{b,s}
(l_{2s})\widetilde\Gamma_s
\label{gamab}
\end{equation}
where $\Gamma^{(0)}_{b,s}(l_{2s})$ and
$\Gamma^{(0)}_{a,s}(l_{1s})$ are given by (\ref{zgama}) while
$\widetilde\Gamma_s$ depends, among other things, on two
Grassmann parameters $(\mu_s,\nu_s)$ as it follows
\begin{equation}
\tilde\Gamma_s=\left\{
z=z^{(s)} +\vartheta^{(s)}\varepsilon_s(z^{(s)})\,,\quad
\vartheta=\vartheta^{(s)}\left(1+
\frac{\varepsilon_s\varepsilon'_s}2\right)+
\varepsilon_s(z^{(s)})
\right\}\,,
\label{tgam}
\end{equation}
\begin{equation}
\varepsilon_s(z)=\frac{\mu_s(z-v_s)-\nu_s(z-u_s)}
{u_s-v_s}\,,\qquad
\varepsilon'_s=\partial_z\varepsilon_s(z)
\label{teps}
\end{equation}
Thus $(u_s|\mu_s)$ and $(v_s|\nu_s)$ are limiting points of
transformations (\ref{gamab}). The set of the transformations
(\ref{gamab}) for $s=1,\dots,n$ together with their group
products forms the genus-$n$ super-Schottky group. If
$l_{1s}=1/2$, then both transformations (\ref{gamab}) are
non-split, as it was already discussed in the Introduction. The
$\Gamma_{a,s}(l_{1s})$ transformation relates superconformal
$p$-tensor $T_p(t)$ with its value $T_p^{(s)}(t)$ obtained from
$T_p(t)$ by $2\pi$-twist about $C_{v_s}$-circle (\ref{circ}).
So, $T_p(t)$ is changed under the $\Gamma_{a,s}(l_{1s})=
\{t\rightarrow t_s^a\}$ and $\Gamma_{b,s}=\{t\rightarrow
t_s^b\}$ transformations as follows
\begin{equation}
T_p(t_s^a)=T_p^{(s)}(t)Q_{\Gamma_{a,s}(l_{1s})}^p(t),\qquad
T_p(t_s^b)=T_p(t)Q_{\Gamma_{b,s}(l_{2s})}^p(t).
\label{stens}
\end{equation}
Here $Q_G(t)$ is the factor, which the spinor left derivative
$D(t)$ receives under the $SL(2)$ transformation $G(t)=
\{t\rightarrow t_G=(z_G(t)|\vartheta_G(t))\}$:
\begin{equation}
Q_G^{-1}(t)=D(t)\vartheta_G(t)\,;\qquad
D(t_G)= Q_G(t)D(t)\,,\qquad
D(t)=\vartheta\partial_z+\partial_\vartheta
\label{supder}
\end{equation}
It follows from (\ref{supder}) that for the group product
$G=G_1G_2$,
\begin{equation}
Q_{G_1G_2}(t)=Q_{G_1}(G_2(t))Q_{G_2}(t)
\label{suprel}
\end{equation}
Furthermore, $\Gamma_{b,s}(l_{2s})$ turns the boundary of the
$\hat C_{u_s}$ "circle" to the boundary of $\hat C_{v_s}$ where
\begin{equation}
\hat C_{v_s}=\{t:|c_sz^{(s)}+d_s|^2=1\}\quad{\rm and}\quad
\hat C_{u_s}=\{t:|-c_sz^{(s)}+a_s|^2=1\}
\label{hcirc}
\end{equation}
and $z^{(s)}$ is defined by (\ref{tgam}). Moreover, the same is
true for the "circles"
\begin{equation}
\hat C_{v_s}'=\{t:|Q_{\Gamma_{b,s}}(t)|^2=1\}\quad{\rm and}\quad
\hat C_{u_s}'=\{t:|Q_{\Gamma_{b,s}^{-1}}(t)|^2=1\}
\label{hcirc1}
\end{equation}
where the super-derivative factors (\ref{supder}) correspond to
$\Gamma_{b,s}(l_{2s})$ and, respectively, its inverse
transformation. "Circles" (\ref{hcirc}) and (\ref{hcirc1})
differ from (\ref{circ}) only in terms proportional to the
Grassmann quantities. Being constructed for every group
transformation, both the "circles" can be used to define the
boundary of the fundamental region.  For applications it is
useful solely to keep in mind that the fundamental region
can be given by the step function factor through relevant
functions $\ell_G(t)$ and $\ell_G(G(t))$ as it follows
\begin{equation}
B_{L,L'}^{(n)}(t,\bar t;\{q,\bar q\})=
\prod_G\theta(|\ell_G(t)|^2-1) \theta(1-|\ell_G(G(t))|^2)
\label{zbound}
\end{equation}
where $\theta(x)$ is step function defined to be $\theta(x)=1$ at
$x>0$ and $\theta(x)=0$ at $x<0$. Further, $L=\{l_{1s},l_{2s}\}$
is the super-spin structure for the right movers while $L'$ is
the same for the left ones. The product is taken over all
group products $G$ of the $\Gamma_{b,s}(l_{1s})$ transformations
except $G=I$.  Evidently, one can exclude from $\{G\}$ a
transformation inverse to the given one of the set. It is
implied that all the group limiting points lay exterior to the
region, and the region does not contain points related with each
other by the group transformation. The region (\ref{zbound}) is
relevant for the integration region for the group invariant
integral of the $(1/2,1/2)$ super-tensor.  Generically, the
argument of step function (\ref{zbound}) depends on Grassmann
parameters. The expansion in a series over the above Grassmann
ones originates $\delta$-functions and their derivatives, which
give rise to the boundary terms in the integral. The integral is
independent of the boundary sharp that can be directly verified
for infinitesimal variations of the boundary.  As it is usually
\cite{ford}, one can replace any part of the fundamental region
by a congruent part and still have a fundamental region.  The
integral discussed is independent of the fundamental region,
which it is taken over.

The superstring amplitude (\ref{ampl}) is calculated by the
integration over a fundamental region
of the total group ${\cal R}$ of local symmetries, for details
see Sections 3 and 4.  In this case the $\{N_0\}$ set of
$(3|2)$ variables among the
group limiting points and interaction vertex coordinates are
fixed due to the $SL(2)$ symmetry.  Simultaneously, the
integrand is multiplied by a factor $H(\{N_0\})$. For the sake
of simplicity, we assume to be fixed two any variables
$z_1^{(0)}$ and $z_2^{(0)}$ along with their Grassmann partners
$\vartheta_1^{(0)}$ and $\vartheta_2^{(0)}$, and one more
variable $z_3^{(0)}$, as well.  In this case \cite{danphr}
\begin{equation}
H(\{N_0\})=(z_1^{(0)}-z_3^{(0)})(z_2^{(0)}-z_3^{(0)})
\left[1-\frac{\vartheta_1^{(0)}\vartheta_3}
{2(z_1^{(0)}-z_3^{(0)})}-\frac{\vartheta_2^{(0)}\vartheta_3}
{2(z_2^{(0)}-z_3^{(0)})}\right]\,.
\label{factorg}
\end{equation}
So, generically, (\ref{factorg}) depends on the integration
variable $\vartheta_3$, which is the Grassmann partner of
$z_3^{(0)}$. The $n$-loop, $m$-point amplitude
$A_m^{(n)}(\{p_j,\zeta^{(j)}\})$ for the interaction states
carrying 10-dimensional momenta $\{p_j\}$ and the polarization
tensors $\zeta^{(j)}$, is given by\footnote{Through the paper
the overline denotes the complex conjugation.}
\begin{eqnarray}
A_m^{(n)}(\{p_j\},\zeta^{(j)}\})=\frac{g^{2n+m-2}}{2^nn!}\int
|H(\{N_0\})|^2\sum_{L,L'}
Z_{L,L'}^{(n)}(\{q,\overline q\})
<\prod_{r=1}^mV(t_r,\overline t_r;p_r;\zeta^{(r)})>
\nonumber\\
\times
\hat
B_{L,L'}^{(n)}(\{q,\overline q\})\tilde
B_{L,L'}^{(n)}(\{q,\overline
q\})
\prod_{j=1}^mB_{L,L'}^{(n)}(t_j,\bar
t_j;\{q,\bar q\})
(dqd\overline qdtd\overline t)'
\label{ampl}
\end{eqnarray}
where $\{q\}=\{k_s,u_s,v_s,\mu_s,\nu_s\}$ is the set of the
super-Schottky group parameters, $g$ is the coupling constant,
$H(\{N_0\})$ is defined by (\ref{factorg}) and $L$ ($L'$) labels
the super-spin structures of right (left) movers. Further,
$Z_{L,L'}^{(n)}(\{q,\overline q\})$ is the partition function,
and $<...>$ denotes the vacuum expectation of the product of the
interaction vertices $V(t_r,\bar t_r;p_r;\zeta^{(r)})$.  The
integration is performed over those variables, which do not
belong to the $\{N_0\}$ set. The step function factor $\hat
B_{L,L'}^{(n)}(\{q,\overline q\})$ keeps the period matrix to be
interior to the fundamental region of the modular group. As far
as the period matrix is given through the super-Schottky group
parameters $\{q\}$, this factor restricts the integration region
over $\{q\}$.
The discussed factor depends on the spin structures as far as
the period matrix depends on the spin structure by terms
proportional to Grassmann parameters. Further, $\tilde
B_{L,L'}^{(n)}(\{q,\overline q\})$ more bounds the
integration region due to the $\{\tilde G\}$ symmetry and the
unitarity as it has been mentioned in the Introduction. Both
factors are discussed in Sections 3 and 4. The step multiplier
$B_{L,L'}^{(n)} (t_j;\{q,\overline q\})$ is given by
(\ref{zbound}) at $t=t_j$.  Generically, $\ell_G(t_j)$ can
depend on $j$. The discussed multiplier is assigned every $t_j$
including $t_j$ of
the $\{N_0\}$ set, as well.
Indeed, due to the invariance under the super-Schottky
group changes of any one vertex coordinate, $t_j$
can be fixed insides region (\ref{zbound}). Really
the discussed step factor must be assigned to fixed
$t_j$ of
the $\{N_0\}$ set for the
unitarity equations to be true, see Section 4.

The $1/2^n$ factor in (\ref{ampl}) is due to the symmetry of the
integrand under the interchange between $(u_s|\mu_s)$ and
$(v_s|\nu_s)$, and $1/n!$ is due to the symmetry under the
interchange of the handles. Both symmetries are particular
cases of the modular symmetry \cite{siegal} (see the next
Section). The above factors are consistent with the
unitarity equations (as an example, see Appendix A of the
paper).  For any boson variable $x$ we define $dxd\overline
x=d(Re\,x)d(Im\,x)/(4\pi)$.  For any Grassmann variable $\eta$
we define $\int d\eta\eta=1$.  The super-spin structure being
odd\footnote{The (super)-spin structure is even (odd), if
$4l_1l_2=4\sum_{s=1}^ nl_{1s}l_{2s}$ is even (odd).}, the
integrand has some features due to the spinor zero modes, which
present in this case. We mainly discuss the even super-spin
structures as far as the odd super-spin one can be obtained by a
factorization of relevant even super-spin structure
\cite{dan96}.

Our classification over the super-spin structures implies that
\begin{equation}
|\arg
k_s|\leq\pi\,,\quad|\arg(u_s-u_r)|\leq\pi\,,
\,\,|\arg(u_s-v_r)|\leq\pi\,,\,\,|\arg(v_s-v_r)|\leq\pi\,,
\quad(s\neq r).
\label{argum}
\end{equation}
To be accurate, (\ref{argum}) are given  for the "bodies" of the
corresponding quantities, as far as they may have "soul" parts
proportional to Grassmann parameters. Using the Green functions
\cite{danphr,dannph}, one can see that $|\arg k_s|\leq\pi$
discriminates between the Green functions for
$(l_{1s}=0,l_{2s}=0)$ and for $(l_{1s}=0,l_{2s}=1/2)$.  The rest
constraints discriminate between $(l_{1s}=l_{1r}=1/2,
l_{2s}=l_{2r}=0)$ and $(l_{1s}=l_{1r}=1/2, l_{2s}=l_{2r}=1/2)$.
The adding of $\pm2\pi$ to any one of the quantities in
(\ref{argum}) presents a modular transformation \cite{dannph},
see Section 3 below. We use (\ref{argum}) instead of known
constrains \cite{siegal} for the real part of the
period matrix.

We consider the massless boson interaction amplitudes. Thus, for
the normalization used, the known expression \cite{fried} of
the interaction vertex through
the string superfields $X^N(t,t')$ for $N=0,\dots 9$  is as
follows
\begin{equation}
V(t,\overline
t;p;\zeta)=4\zeta_{MN} [D(t)X^M(t,\overline
t)][\overline{D(t)}X^N(t,\overline t)]
\exp[ip_RX^R(t,\overline
t)]
\label{vert}
\end{equation}
where $p=\{p^M\}$ is 10-momentum of the interacting boson while
$\zeta_{MN}$ is its polarization tensor,
$p^M\zeta_{MN}=p^N\zeta_{MN}=0$, and $p^2=0$. The spinor
derivative $D(t)$ is defined in (\ref{supder}). The summation
over twice repeated indexes is implied.  We use the "mostly
plus" metric. The dilaton $\zeta_{MN}$ tensor is equal to the
transverse Kronecker symbol $\delta_{MN}^\perp$. The vacuum
expectation in (\ref{ampl}) is calculated in term of the
genus-$n$ scalar superfield vacuum correlator given through the
holomorphic Green function $R_L^{(n)}(t,t';\{q\})$ and
super-holomorphic functions $J_r^{(n)}(t;\{q\};L)$ having periods
(here $r=1,\dots n$). In this case
\begin{eqnarray}
J_r^{(n)}(t_s^b;\{q\};L) = J_r^{(n)}(t;\{q\};L)+2\pi i
\omega_{sr}(\{q\},L)\,,
\nonumber\\
J_r^{(n)}(t_s^a;\{q\};L)=J_r^{(n)(s)}(t;\{q\};L)+
2\pi i\delta_{rs}
\label{trjs}
\end{eqnarray}
where $t_s^a$ and $t_s^b$ are the same as in (\ref{stens}) and
$\omega_{sr}(\{q\},L)$  is the corresponding element of the
period matrix.  The Green function is changed under  the
transformations (\ref{stens}) as
\begin{eqnarray}
R_L^{(n)}(t_r^b,t';\{q\})=R_L^{(n)}(t,t';\{q\})+
J_r^{(n)}(t';\{q\};L)\,,
\nonumber\\
R_L^{(n)}(t_r^a,t';\{q\})=R_L^{(n)(r)}(t,t';\{q\})\,.
\label{rtrans}
\end{eqnarray}
In this case the Green function is normalized as it
follows
\begin{equation}
R_L^{(n)}(t,t';\{q\})=\ln(z-z'-\vartheta\vartheta')+
\tilde R_L^{(n)}(t,t';\{q\})
\label{lim}
\end{equation}
where $\tilde R_L^{(n)}(t,t';\{q\})$ has no a singularity at
$z=z'$.  The scalar superfield vacuum correlator $\hat
X_{L,L'}(t,\overline t;t',\overline t';\{q\})$ is given
by\footnote{The string tension is taken to be $1/\pi$}
\begin{equation}
4\hat X_{L,L'}(t,\overline t;t',\overline t';\{q\})=
R_L^{(n)}(t,t';\{q\})+\overline{R_{L'}^{(n)}(t,t';\{q\})}
+I_{LL'}^{(n)}(t,\overline
t;t',\overline t';\{q\})\,,
\label{corr}
\end{equation}
\begin{eqnarray}
I_{LL'}^{(n)}(t,\overline
t;t',\overline t';\{q,\bar q\})= [J_s^{(n)}(t;\{q\};L) +
\overline{J_s^{(n)}(t;\{q\};L')}]
[\Omega_{L,L'}^{(n)}(\{q,\overline q \})]_{sr}^{-1}
\nonumber\\ \times
[J_r^{(n)}(t';\{q\};L)+\overline{J_r^{(n)}(t';\{q\};L')}]
\label{illin}
\end{eqnarray}
where $\Omega_{L,L'}^{(n)}(\{q,\overline q\})$ being calculated
in terms of the $\omega^{(n)}(\{q\},L)$ period matrix, is
\begin{equation}
\Omega_{L,L'}^{(n)}(\{q,\overline q\})=
2\pi
i[\overline{\omega^{(n)}(\{q\},L')}-
\omega^{(n)}(\{q\},L)].
\label{grom}
\end{equation}
As it is usually, the Green function at the same point $z=z'$ is
defined to be the $\tilde R_L^{(n)}(t,t';\{q\})$ at $z=z'$. The
dilaton emission amplitudes include the vacuum pairing
$\tilde I_{L,L'}^{(n)}(t,\bar t;\{q\})$ of the superfields in
front of the exponential in (\ref{vert}). In line with aforesaid
\begin{equation}
\tilde
I_{L,L'}^{(n)}(t,\bar t;\{q,\bar q\})=
2D(t)D(\bar t')I_{L,L'}^{(n)}
(t,\overline t;t',\overline t';\{q\})|_{t=t'}
\label{ngcor}
\end{equation}
where the definitions are given in (\ref{corr}) and in
(\ref{illin}). The dilaton-vacuum transition constant is
determined by the integral of (\ref{ngcor}) over the
supermanifold. Integrating by parts the right side of
(\ref{ngcor}) with the following using of the relations
(\ref{trjs}), one obtains the discussed constant to be
$n$ times the vacuum amplitude \cite{gsw}.

Due to a separation in right and left movers,
the  integration measure in (\ref{ampl}) is
represented as
\begin{equation}
Z_{L,L'}^{(n)}(\{q,\overline
q\})=(8\pi)^{5n}[\det\Omega_{L,L'}^{(n)}(\{q,\overline q
\})]^{-5} Z_L^{(n)}(\{q\}) \overline {Z_{L'}^{(n)}(\{q\})}
\label{hol}
\end{equation}
where $Z_L^{(n)}(\{q\})$ is a holomorphic function of the
$q$ moduli and the $\Omega_{L,L'}^{(n)}(\{q,\overline q \})$
matrix is given by (\ref{grom}). The holomorphic partition
function in (\ref{hol}) is given by
\begin{equation}
Z_L^{(n)}(\{q\})=\hat Z_L^{(n)}(\{q\})
\prod_{s=1}^n(u_s-v_s-\mu_s\nu_s)^{-1}
\label{zinv}
\end{equation}
where $\hat Z_L^{(n)}(\{q\})$ is invariant under the
$SL(2)$ transformations as far as
\begin{equation}
du_sdv_sd\mu_s d\nu_s/(u_s-v_s-\mu_s\nu_s)
\label{sliv}
\end{equation}
is $SL(2)$ invariant.  Explicit
$\hat Z_L^{(n)}(\{q\})$ and other functions of interest
are given in Section 5. In two following Sections we
consider the integration region.

\section{Modular symmetry constraints}

The modular transformation of the supermanifold, generically,
presents a globally defined, holomorphic superconformal
change $t\to\hat t$ of the coordinate along with holomorphic
changes $q\to\hat q$ of the super-Schottky group parameters and
by a change $L\to\hat L$ of the super-spin structure. Like the
modular transformation of the Riemann surface\cite{siegal}, it
determines the going to a new basis of non-contractable cycles.
So the period matrix $\omega(\{q\},L)$ is changed, as it is
usually, by
\begin{equation}
\omega(\{q\},L)=[A\omega(\{\hat
q\},\hat L)+B] [C\omega(\{\hat q\},\hat L)+D]^{-1}
\label{modtr}
\end{equation}
where integer $A$, $B$, $C$ and $D$ matrices obey \cite{siegal}
the relations
\begin{equation}
C^TA=A^TC\,,\qquad
D^TB=B^TD\,,\qquad D^TA-B^TC=I\,.
\label{mrel}
\end{equation}
The right-top "$T$" symbol labels the transposing. In this case
$t(\hat t;\{\hat q\};\hat L)$ and $q(\{\hat q\};\hat L)$ both
depend on the superspin structure by terms proportional to
Grassmann parameters \cite{dannph}.  Then the period matrix also
depends on the super-spin structure, as has been noted in the
Introduction.  For zero Grassmann parameters the period matrix
is reduced to $\omega^{(0)}(\{q\})$ whose matrix elements
$\omega_{sp}^{(0)}(\{q\})$ are  given through the Schottky
parameters $\{q\}$ by \cite{vec,danphr}
\begin{equation}
2\pi i\omega_{sp}^{(0)}(\{q\})= \delta_{sp}\ln k_p+
{\sum_\Gamma}^{''}\ln\frac{[u_s-g_\Gamma(u_p)]
[v_s-g_\Gamma(v_p)]}{[u_s-g_\Gamma(v_p)][v_s-g_\Gamma(u_p)]}
\label{omega0}
\end{equation}
where $\delta_{sp}$ is the Kronecker symbol. The summation in
(\ref{omega0}) is performed over all those the transformations
$g_\Gamma$ of the Schottky group, whose leftmosts are not group
powers of $g_s$, or the rightmosts are not group powers of
$g_r$. Besides, $g_\Gamma\neq I$, if $s=p$. So the addition of
$\pm2\pi$ to $\arg k_s$ adds $\pm1$ to
$\omega_{ss}^{(0)}(\{q\})$. For $r\neq s$, due to the term with
$\Gamma=I$ in (\ref{omega0}), the addition of $\pm2\pi$ to the
argument of the difference between  limiting points in
(\ref{argum}) adds $\pm1$ to $\omega_{sr}^{(0)}(\{q\})$. Hence
the discussed changes of arguments (\ref{argum}) just correspond
to transformations of the period matrix by (\ref{modtr}) with
$C=0$ and $A=D=I$. Evidently, it is true for non-zero Grassmann
parameters too, and we do not enlarge on this matter. We note
only that for non-zero Grassmann parameters,
the addition of $\pm2\pi$ to the
argument of the difference between  limiting points in
(\ref{argum})
is accompanied \cite{dannph}
by a certain change of $t$ and of $\{q\}$. Relations
(\ref{argum}) replace constraints
$|Re\,\omega_{sp}^{(0)}(\{q\})|\leq1/2$ for the period matrix
\cite{siegal}. In the both cases the sum in (\ref{ampl}) includes
all the distinct spin structures without a double counting, but
constraints (\ref{argum}) are much more convenient for
applications than
constraints $|Re\,\omega_{sp}^{(0)}(\{q\})|\leq1/2$.

Further constraints appear due to transformations (\ref{modtr})
with
\begin{equation}
B=C=0\,,\quad A=F\,,\quad D^{-1}=
F^T\,,\quad \det  F=\pm1\,,
\label{parttr}
\end{equation}
$F$
being an integer matrix. For the diagonal $F$ matrix,
$F_{ss}=-1$ corresponds to the replacement of the corresponding
group transformation by its inverse that
interchanges between $(u_s|\mu_s)$ and $(v_s|\nu_s)$.
Indeed,
from (\ref{rtrans}), under the above replacement,
$J_s^{(n)}(t;\{q\};L)$ receives the sign. Hence from
(\ref{trjs}), the period matrix elements
$\omega_{rs}(\{q\},L)$ with $r\neq s$ also receive the sign
that just correspomds to
modular transformation discussed
(for zero Grassmann parameters this follows
directly from (\ref{omega0})).
The $F$ matrix having a sole non-zero non-diagonal
matrix element $F_{s_1s_2}=F_{s_1s_2}=1$ and
$F_{s_1s_1}=F_{s_2s_2} =0$, corresponds to the
interchange $s_1\rightleftharpoons s_2$ between the handles.
Thus the Schottky parameters is bounded by constraints
\cite{siegal}, which discriminate between repulsive and
attractive limiting points, as well as between the handles. In
their stead we prefer to introduce  the $1/(2^nn!)$ factor in
(\ref{ampl}).  Remaining $F$ matrices correspond to transitions
to new basic cycles, which are certain sums over the
former basic ones.  For non-zero Grassmann parameters,
unlike the case of zero Grassmann ones, both $t$ and
$\{q\}$ are changed. Indeed, in this case
$\Gamma_{as}(l_{1s}=1/2)$ does not commute with transformations
(\ref{gamab}) assigned to another handle.  Hence the
group transformations for rounds about resulted $(A,B)$-cycles
being a sum of the former ones, do not commute with each
other, if $t$ and $\{q\}$ do not changed.  The required changes
of $t$ and $\{q\}$ could be calculated by the method developed
in \cite{dannph}, but it is not a subject of the present paper.

For zero Grassmann parameters, among period matrices related by
(\ref{parttr}), one takes \cite{siegal} the matrix having the
smallest imaginary part $y_{jj}(\{q,\bar q\})$. In this case
\begin{equation}
[Fy(\{q,\bar q\}) F^T]_{jj}\geq y_{jj}(\{q,\bar q\})\,.
\label{imagin}
\end{equation}
As it is usually, $y_{jj}(\{q,\bar q\})$ is non-negative.
Starting with $j=1$, one, step-by-step, constructs a
constraints for $j=2,\dots,n$. Calculating the integral,
one sums with the $1/n!$ factor over permutations of the
handles.  Further constraints are due to transformations with
$C\neq0$. In this case, among the $\omega^{(0)}(\{q\})$ period
matrices related by (\ref{modtr}), one takes the matrix, which
gives the greatest magnitude of $\det y(\{q,\bar q\})$. Due to
(\ref{modtr}) and (\ref{mrel}), the corresponding constraints
are given \cite{siegal} by
\begin{equation}
|\det
[C\omega^{(0)}(\{q\})+D]|^2\geq1
\label{bound}
\end{equation}
where $C$ and $D$  are the matrices in (\ref{modtr}). Important
particular constraints are obtained from (\ref{imagin}) when,
for given $r$ and $j$, one takes $F_{jr}=F_{rj}=\pm1$ and either
$F_{jj}=0$, or $F_{rr}=0$. All other non-diagonal elements of
$F$ are assumed to be zeros. Then
\begin{equation}
min[y_{jj}(\{q,\bar
q\}),\,y_{rr}(\{q,\bar
q\})]\pm2y_{rs}(\{q,\bar
q\})\geq 0\,.
\label{imnd}
\end{equation}
Also, using special $C$ and $D$ in (\ref{bound}), one derives
important constraints for the principal minors
$[\det(F\omega^{(0)}(\{q\})F^T+\tilde B)]_{s_1\dots s_k}$ of the
$\det(F\omega^{(0)}(\{q\})F^T+\tilde B)$ as it follows
\begin{equation}
[\det(F\omega^{(0)}(\{q\})F^T+\tilde
B)]_{s_1\dots s_k}[\det(
\overline{F\omega^{(0)}(\{q\})F^T+
\tilde B)}]_{s_1\dots s_k}\geq1\
\label{minors}
\end{equation}
where $F$ is an integer matrix, $\det F=\pm1$. There is no
summation over the $s_1\dots s_k$ indices, and the integer
matrix $\tilde B$ is chosen from the condition that
$Re\,|F\omega^{(0)}(\{q\})F^T+\tilde B|\leq1/2$.

For non-zero Grassmann parameters, $\omega^{(0)}(\{q\})$ is
replaced by $\omega(\{q\},L)$ while
$\overline{\omega^{(0)}(\{q\})}$ is replaced by
$\overline{\omega(\{q\},L')}$. All the other extensions contain
$\overline{\omega(\{q\},L)}$ and/or $\omega(\{q\},L')$. Hence
they are not consistent with the modular symmetry. Indeed,
(\ref{modtr}) relates $\overline{\omega(\{q\},L)}$ with
$\overline{\omega(\{q_L\},\hat L)}$, which is different from
$\overline{\omega(\{q_{L'}\}, \hat L)}$
due to terms proportional to the Grassmann parameters
\cite{dannph}. So the imaginary part of the period matrix is
replaced by
\begin{equation}
y(\{q,\bar
q\},L,L')=\frac{1}{2i}[\omega(\{q\},L)-
\overline{\omega(\{q\},L')}]\,.
\label{yll}
\end{equation}
In this case the corresponding step factor in (\ref{ampl}) is
given by
\begin{eqnarray}
\hat B_{L,L'}^{(n)}(\{q,\overline q\})=\left[\prod_{C,D}
\theta(\det\{[C\omega(\{q\},L+D]
[C(\overline{\omega(\{q\},L')}+\tilde B)+D]\}-1)\right]
\nonumber\\
\times
\left[\sum\prod_{ F}\theta([ Fy(\{q,\bar q\},L,L')
F^T]_{jj}
-y_{jj}(\{q,\bar q\},L,L'))\right]\prod_{j=1}^n
\theta(y_{jj}(\{q,\bar q\},L,L'))
\label{sbound}
\end{eqnarray}
where the sum is performed over permutations of the handles
while the products are calculated over all the matrices $C$, $D$
and $F$ in (\ref{modtr}) and (\ref{parttr}). The step function
$\theta(x)$ is the same as in (\ref{zbound}).

\section{Integration region}

The period matrix is preserved under isomorphic changes
\begin{equation}
\Gamma_{a,s}(l_{1s})\to
G_s\Gamma_{a,s}(l_{1s})G_s^{-1}\,,\quad \Gamma_{b,s}(l_{2s})\to
G_s\Gamma_{b,s}(l_{2s})G_s^{-1}
\label{morph}
\end{equation}
of the set of the forming group transformations (\ref{gamab})
where $G_s$ is a relevant transformation \footnote{Not every
$\{\tilde G_s\}$ set is relevant to determine the isomorphism.
As an example, the set formed by $G_1G_2G_1G_2^{-1}G_1^{-1}$ and
$G_2$ is not isomorphic to the $(G_1,G_2)$ set since $G_1$ can
not be represented by a group product constructed by using
$G_1G_2G_1G_2^{-1}G_1^{-1}$ and $G_2$.} from the super-Schottky
group.  Generically, $G_s$ depends on $s$. The discussed
isomorphism only replaces the limiting points $U_s=(u_s|\mu_s)$
and $V_s=(v_s|\nu_s)$ of the transformation (\ref{gamab}) by
$G_sU_s$ and  by $G_sV_s$. As far as this is globally defined
holomorphic transformation, the amplitude integral (\ref{ampl})
is invariant under the $\{\tilde G\}$ group of the isomorphic
changes discussed. In particular, the holomorphic Green function
and the scalar functions are not changed since (\ref{morph})
does not touch (\ref{trjs}) and (\ref{rtrans}).  Moreover, from
the equations \cite{danphr} for the partition functions, one can
derive that the invariant partition function $\hat
Z_L^{(n)}(\{q\})$ in (\ref{zinv}) is also unchanged. Only the
multiplier behind $\hat Z_L^{(n)}(\{q\})$ receives a factor,
which is just canceled by the Jacobian of the transformation, as
it also follows from the below  constructing of the considered
group.

In doing so, for given $s_1$ and $s_2\neq s_1$, one changes the
transformations (\ref{gamab}) for $s=s_1$ by (\ref{morph}) with
$G_{s_1}$ to be any one from transformations (\ref{gamab}) for
$s=s_2$. The resulted set $S(s_1|s_2)$ is evidently isomorphic to
the former set $S_0$ of transformations (\ref{gamab}) for
$s=1,\dots,n$. Moreover, $U_{s_1}$ and $V_{s_1}$ are changed by
the $SL(2)$ transformation independent of $U_{s_1}$ and
$V_{s_1}$. Thus factor (\ref{sliv}) is unchanged. In this way
one obtains the set of different $S(s_1|s_2)$ sets. Applying
this procedure to every set, one builds further sets. As an
example, one obtains the $S(s_1,s_2|s_2,s_1)$ set where the
transformations for $s=s_2$ from the $S(s_1|s_2)$ set  are
changed by (\ref{morph}) with $G_{s_2}$ to be the $s=s_1$
transformation from the same $S(s_1|s_2)$ set. In this case the
$s_2$ transformation of the $S(s_1,s_2|s_2,s_1)$ set depends,
among other things, on the starting transformation (\ref{gamab})
for the same $s=s_2$.  By construction, all the sets are
isomorphic to each other, the amplitude being invariant under
the transformations discussed.  So every set can be used as the
set of forming group transformations.  In this way one
constructs the desired infinite (super)-group $\{\tilde G\}$
generating forming group transformation sets by action of
$\{\tilde G\}$ on the given set (\ref{gamab}).

Since (\ref{zbound}) and (\ref{sbound}) are preserved under the
$\tilde G$ transformations, a further constraint of the
integration region is necessary to exclude domains related by
$\tilde G$ each to other. In particular, due to the $\tilde G$
symmetry, either one of two limiting points of every forming
transformation (\ref{gamab}) can be restricted to be the
exterior of Schottky circles assigned to all the other forming
transformations. For certain configurations, both limiting point
of every given  forming transformation appear exterior to the
Schottky circles above. These configurations can not be
obtained by $\tilde G$ transformations of configurations where
this is not so. Moreover, the integral over the last
configurations destroys the unitarity equations, as demonstrated
for the case when all the Schottky multipliers go to zero.

In this case, from
(\ref{uvk}), the boundaries of Schottky circles (\ref{circ}) are
given by  the conditions
\begin{equation}
|z-v_s|^2=|k_s||u_s-v_s|^2\,,\qquad
|z-u_s|^2=|k_s||u_s-v_s|^2\,.
\label{circlim}
\end{equation}
Period matrix elements (\ref{omega0}) are reduced to
\begin{equation} 2\pi
i\omega_{jr}^{(0)}\to
\ln\frac{(u_j-u_r)(v_j-v_r)}{(u_j-v_r)(v_j-u_r)}\,,
\qquad \omega_{jj}\to\ln k_j
\label{kzero}
\end{equation}
Moreover, from (\ref{imnd}), one derives that
\begin{equation}
2\left|\ln\bigg|\frac{(u_j-u_r)(v_j-v_r)}{(u_j-v_r)(v_j-u_r)}
\bigg|\right|\leq min\bigg(-\ln|k_j|,\,-\ln|k_r|\bigg)
\label{ndgbound}
\end{equation}
Below we assume $j=1$ and $r=2$. Constraint (\ref{ndgbound})
allows, for instance, the discussed configuration
\begin{equation}
|u_1-u_2|\sim |k_2|^{1-\delta}\,,\quad
|v_1-u_2|\sim|k_2|^{1/2-\delta}\,,\quad|k_2|>|k_1|\,,
|v_1-v_2|\sim|u_1-v_2|\sim1
\label{antun}
\end{equation}
where $k_2\to0$ and $\delta<<1$. In this case $v_2$ is exterior
to $C_{u_1}$ and to $C_{v_1}$ while $u_2$ lies inside $C_{u_1}$.
Other constraints (\ref{sbound}) also do not prevent the above
configuration.  In parallel with (\ref{antun}), the boundary of
(\ref{ndgbound}) is reached for only two limited points, say
$u_1$ and $u_2$, to go to each other. In this case $|u_1-u_2|
\sim\sqrt{|k_2|}$, all the other distances being of order of
the unity. Both regions originate discontinuities. To demonstrate
this, we discuss the discontinuities in the energy invariant
$s=-(p_1+p_2)^2=-(p_1+p_2)^2$  for the two-loop scattering
amplitude. We ignore specifics due to the Grassmann moduli that
is not of very importance for the matter discussed.
We fix the complex coordinates $z_1$, $z_2$ and
$z_3$ of the vertices to be exterior to the Schottky circles.
The integration is performed over the Schottky parameters
and over $z_4=z$, certain group limiting points being assumed to
go to $z_3$. In addition, $z$ is assumed nearby one of
Schottky circles.  If the circle is not nearby $z_3$, then $z$
is changed by a relevant Schottky group transformation moving
$z$ to the interior of a circle to be nearby $z_3$.
First, we consider the configuration
\begin{equation}
|u_1-z_3|\sim|u_2-z_3|\sim
min(|k_1|^{1/2-\delta}\,,|k_2|^{1/2-\delta})\,,\quad
|v_1-z_3|\sim|v_2-z_3|\sim|v_1-v_2|\sim1
\label{crclm}
\end{equation}
where $\delta$ being positive, goes to zero. In this case the
Schottky circles  have no a common interior. We define variables
$y_1\sim y\sim1$ to be given by $(u_1-z_3)= y_1(u_2-u_1)$ and
$(z-u_1)=y(u_2-u_1)$.  The integration over the "small"
variables being taken, the discontinuity is represented by an
integral over four complex $y_1$, $y$, $(v_1-z_3)$ and
$(v_2-z_3)$. The integral just corresponds to the
unitarity equation. Indeed, this unitarity equation is by-linear
in $2\to3$ tree amplitudes, every amplitude being the integral
over two complex variables (an example is given in Appendix A).
More discontinuities would appear, if $z_3$ is allowed to
penetrate into the Schottky circles.  For instance,
due to configuration where $z_3$ being
interior to $C_{v_1}$, $g_1(z_3)$, is, simultaneously, to be
exterior to all the Schottky circles. And $C_{u_1}$ along with
$C_{u_2}$ both go to $g_1(z_3)$. In this case the unitarity
equations would be broken. Hence constraints (\ref{zbound}) for
the fixed vertex coordinates hold the unitarity equations.

For the configuration (\ref{antun}), there are only $(u_1-z_3)$
and $(z-z_3)$ to be of the same order magnitudes. So only two
variables of order the unity can be constructed. They are
$(v_1-z_3)$ and $(\tilde z-z_3)$ where $z-z_3=(\tilde
z-z_3)(u_1-z_3)$. Once the integration over the small variables
being performed,
the discontinuity is given by the integral over $v_1\sim1$ and
$\tilde z\sim1$, which is not a product of two $2\to3$ tree
amplitudes.  Moreover (see Appendix A), for the tachyon-tachyon
scattering amplitude from the boson string theory the
configuration (\ref{antun}) originates the false threshold at
$s=6m_{th}^2$ where $m_{th}^2=-8$ is the square of the tachyon
mass. A like discontinuity appears, if $u_2$ and $v_2$ in
(\ref{antun}) both are
the $\{\tilde G\}$ image of the
limiting points of a forming transformation (instead of to be
the limiting points of the forming one). Along with
(\ref{antun}), the discussed configuration must be removed from
the integration region.

By above, the
unitarity equations are saturated due to the region when the
Schottky circles of the forming transformations do not overlap
each other. But this region being no a fundamental region for
the symmetry group, can not be the  total
integration region.
The relevant region for the integration one
seems the ${\cal G}_n$ region where all the limiting group
points of $\{\tilde G\}$ lay inside the Schottky circles of the
forming transformation (\ref{gamab}) and outside the common
interior of any pair of the circles above. Indeed, we
demonstrate that for the integrals of the group covariant
expression over the boundary of the region, the boundary can be
replaced by pieces, which are congruent to each other under the
$\{\tilde G\}$ group. Hence the group invariant integral over
the region discussed is the same under those infinitesimal
variations of the circles, which are related by the
corresponding group transformation (\ref{gamab}).  In addition,
the interior of the region does not contain the points related
by the $\{\tilde G\}$ transformations. So the region possesses
properties of the fundamental region desired. We demonstrate a
range, which the boundary of the region can be varied within.

For this purpose we consider the
boundary "1" with $u_2=u_2^{(1)}$ laying on $C_{u_1}$, and the
boundary "2" including the $u_2=u_2^{(2)}$ point on $C_{v_1}$.
The $u_2^{(2)}$ point is obtained by the
$\Gamma_{b,1}^{-1}(l_{21})$ transformation (\ref{gamab}) of
$u_2^{(1)}$. On the boundary "1" the remaining variables  cover
a domain ${\cal C}[u]$ and, respectively, over ${\cal C}[v]$. We
add ${\cal C}[u]$  by the ${\cal C}^{(1)}[v]$ domain, which is
the image of ${\cal C}[v]$ under that $\tilde G=\tilde G_2$
transformation. The transformation only changes both limiting
points $U_2=(u_2|\mu_2)$ and $V_2=(v_2|\nu_2)$ by
$\Gamma_{b,1}(l_{21})$.  Respectively, we add ${\cal C}[v]$ by
the ${\cal C}^{(-1)}[u]$ domain to be the image of ${\cal C}[u]$
under the inverse transformation $\tilde G_2^{-1}$.
Due to the $\tilde G$ symmetry, the integral over ${\cal
C}[v]+{\cal C}[u]$ differs only by the factor $1/2$ from the
integral over ${\cal C}'[v]+ {\cal C}'[u]$ where ${\cal
C}'[v]={\cal C}[v]+{\cal C}^{(-1)}[u]$ and ${\cal C}'[u]={\cal
C}[u]+{\cal C}^{(1)}[v]$.  So we can replace ${\cal C}[u]$ by
${\cal C}'[u]$ and ${\cal C}[v]$ by ${\cal C}'[v]$ with dividing
the integral by 2. By construction, the boundaries ${\cal
C}'[u]$ and ${\cal C}'[v]$ are just transformed to each other
under the above transformation $\tilde G_2$, as it required for
the fundamental region boundaries.  The kindred consideration
can be performed for those boundaries, which include the group
limiting points obtained by a $\tilde G$ change of the forming
set.  So the boundary of the region can be represented by
pieces, which are congruent to each other under the $\{\tilde
G\}$ group. Thus the boundary integral is nullified,
the integral over ${\cal G}_n$
the region discussed is the same under those infinitesimal
variations of the circles, which are related by
group transformations.

To clarify a
range, which the boundary of the region can be varied within,
we consider
the genus-2 configuration
(\ref{crclm})
with $|k_2|<|k_1|$. Then the ${\cal
C}[u]$ region restricts $v_2$ to lay between $C_{u_1}$ and some
closed curve ${\ell}_u$ rounding $C_{u_1}$. The above curve is
determined by (\ref{ndgbound}).  Respectively, ${\cal C}[v]$
restricts $v_2$ to lay between $C_{v_1}$ and a closed curve
${\ell}_v$ around $C_{u_1}$, the  curve being determined by
(\ref{ndgbound}).  When $u_2^{(1)}$ penetrates into $C_{u_1}$,
the $u_2^{(2)}$ point goes away from $C_{v_1}$.  The penetration
of $u_2^{(1)}$ into $C_{u_1}$ is allowed until $u_2^{(2)}$ meets
${\ell}_v$. At this moment, an integration region over $v_2$ can
not be deformed continuously that gives a natural restriction
for the penetration $u_2^{(1)}$ into $C_{u_1}$. One can see that
configuration (\ref{antun}) is not reached.
The considered example holds that
the ${\cal G}_n$ region is relevant for the integration one.

The limiting points of any super-Schottky group transformation
$G$ are obtained by the action $G^n$ at $n\to\pm\infty$ on an
arbitrary point including the limiting points of the
transformations (\ref{gamab}) to be among them. So, in the
${\cal G}_n$ region all the group limiting points lay outside
the overlapping of Schottky circles assigned to the forming
transformations.  Using (\ref{suprel}), one shows that, if the
leftmost of the transformation is a positive (negative) power of
$\Gamma_{b,s}(l_{2s})$, the attractive limiting point lies
inside the $\hat C_{u_s}$ circle\footnote{As an example,
$u_{g_1g_2}$ being the attractive point of the $g_1g_2$
transformation, $|Q_{g_1^{-1}}(u_{g_1g_2})|=
|Q_{g_1^{-1}}(g_1g_2(u_{g_1g_2}))|=|Q_{g_1}^{-1}(g_2u_{g_1g_2})|
\leq1$ due to the constraints above, $Q_G(z)=c_Gz+d_G$.
Thus $u_{g_1g_2}$ lies inside $C_{v_1}$.}
(the $\hat C_{v_s}$ one). The attractive (repulsive) limiting
point of $\Gamma_{b,s}(l_{2s})$ lies exterior to the $\hat
C_{u_G}$ circle\footnote{As an example, $|Q{g_1g_2}(u_{g_3})|=
|Q_{g_1}(g_2(u_{g_3}))Q_{g_2}(u_{g_3})|\geq1$.} of any group
transformation $G$ having the leftmost no to be a positive
(negative) power of $\Gamma_{b,s}(l_{2s})$.  So, we take $\tilde
B^{(n)}(\{q,\overline q\})$ in (\ref{ampl}) as
\begin{eqnarray}
\tilde
B^{(n)}(\{q,\overline q\})= \prod_{s=1}^n
\Biggl(\prod_{G\in\{G_s\}}
\theta(|\tilde\ell_s
(U_G)|^2-1)
\theta(1-|\tilde\ell_s
(\Gamma_s[U_G])|^2)\Biggl)
\nonumber\\
\times
\Biggl(\prod_{G\in\{G_s'\}}
\theta(1-|\tilde\ell_s
(U_G)|^2)\Biggl)
\prod_{G\in\{G_s''\}}
\theta(|\tilde\ell_s
(\Gamma_s[U_G])|^2-1)
\label{bunit}
\end{eqnarray}
where  $\tilde\ell_s(t)$ is circle (\ref{hcirc}) or
(\ref{hcirc1}), or $\tilde\ell_s(t)$ is obtained by a continuous
deformation of the circle within the region allowed by
(\ref{sbound}).  In this case $\Gamma_s[U_G]$ denotes the
$\Gamma_{b,s}(l_{2s})$ transformation (\ref{gamab})
of $U_G$. Furthermore, $\{G_s\}$ is formed by those group
products of $\Gamma_{b,r}(l_{2r})$, whose leftmost is not a
power of $\Gamma_{b,s}(l_{2s})$. Leftmost of any transformation
from the $\{G_s'\}$ set is a positive power of
$\Gamma_{b,s}(l_{2s})$.  Every $\{G_s''\}$ transformation  has
leftmost to be a negative power of $\Gamma_{b,s}(l_{2s})$. The
sets include the given transformation along with its inverse
one.  Hence in (\ref{bunit}) only the attractive limiting points
$U_G$ present. They are defined by the condition that $G^n[t]\to
U_G$ at $n\to\infty$.  In place of (\ref{bunit}), one can use
any region obtained through a replacement of a part of
(\ref{bunit}) by the part congruent to it under the $\{\tilde
G\}$ group transformation.

In the considered region the radius of the Schottky circle of
any group transformation $G$ is finite, if its limiting points
lay at a finite distance from limiting points of the forming
group transformation. Then (\ref{minors}) forbids the multiplier
$|k_G|\to1$. Indeed, choosing relevant $F$ in (\ref{minors}),
one obtains that $\omega_{GG}^{(0)}(\{q\})\geq1$ where
$\omega_{GG}^{(0)}$ is given by (\ref{omega0}) for $j=r=G$.
Furthermore, $(u_G-v_G)\to0$ at $k_G\to1$, as far as the radius
of the circles assigned to $G$ is finite.  Thus the sum in
(\ref{omega0}) for the matrix element discussed is nullified
forcing $k_G$ to be small. The limiting points going to infinity
can be moved to the same finite point by a relevant $L(2)$
transformation that does not change the sum (\ref{omega0}).  So
the sum vanishes, the multiplier being small again.  For any
integer $\tilde n$, the multiplier $k_G\to\exp[2\pi/\tilde n]$
is too excluded since the multiplier of $G^{\tilde n}$ goes to
the unity that is forbidden by above.  As far as $\tilde n$ can
be arbitrary large, $|k_G|\to1$ is excluded at all.  Along with
the $k\to0$, the $k\to 1$ region contributes to the unitarity
equations. So it is naturally that this region is a copy of the
region where $k\to0$.

When all the multipliers are small, one see from (\ref{omega0})
that
\begin{equation}
2\pi i\omega_{jj}^{(0)}(\{q\})\approx \ln k_j
+2\sum\limits_{s\neq j}\frac{k_s(u_s-v_s)^2(u_j-v_j)^2}
{(u_j-u_s)(v_j-u_s)(u_j-v_s)(v_j-v_s)}+\dots
\label{omkzer}
\end{equation}
Only the leading term being considered, constraints
(\ref{minors}) require $|k_j|\leq1/230$. In this case Schottky
circles (\ref{circ}) of the same transformation to be separated
from each other.  The leading correction  is most when the
boundary of (\ref{circlim}) is achieved.  For $n$ limiting
points to be closed to each other, the correction is roughly
$\sim n|k|d_{uv}/\hat d \sim n\sqrt{|k|}$. Here $\hat d$ is a
characteristic distance between closed limiting points of
distinct forming transformations while $d_{uv}$ is a
characteristic size for the $|u_s-v_s|$ distances. For the
genera being not much high, the correction is small. In this case
the Schottky circles $C_{u_s}$ and $C_{v_s}$ have no the common
interior.  When the genus is increased, $\hat d$ grows that
might reduce the correction. In the general case we have not
the estimation for the correction term, but it seems natural to
expect that constraints (\ref{sbound}) and (\ref{bunit}) always
forbid for $C_{u_s}$ and $C_{v_s}$ to be overlapped.

Constraints (\ref{sbound}) and (\ref{bunit}) along with
(\ref{zbound}) fully determine the integration region
in (\ref{ampl}). One can also use any region obtained by a
replacement of an arbitrary part of the given region by a part
congruent under the group ${\cal R}$ of the local symmetries of
the amplitude. Subtle details due to divergences
in the particular spin structure are discussed in Section 7.
The $\{N_0\}$ set being changed, the integral is reduced to the
initial form due to due to the $SL(2)$ symmetry, the symmetry
under the $\{\tilde G\}$ group and the symmetry under the
super-Schottky group transformations of any given interaction
vertex coordinate. So the amplitude is independent of the choice
of the $\{N_0\}$ set.  The local symmetries are employed in the
following calculation of the integrals over the degenerated
configurations, but details of the integration region will
not be important for this purpose.

\section{Green functions and the integration measures}

In
\cite{danphr,dannph}
the Green function $R_L^{(n)}(t,t';\{q\})$, the period matrix
and the scalar functions have been obtained
in terms of genus-1 functions. Now
we represent them through the genus-$n_i$ functions
where $\sum_i n_i=n$ and $n_i\geq1$. In this case $n$ handles
are divided into groups of $n_i$ ones, every group being given
by the set $\{q\}_i$ of the super-Schottky group parameters and
by its super-spin structure $L_i$ assumed to be even.  So
$\{q\}=\{\{q\}_i\}$ and $L=\{L_i\}$. Using the obtained
formulas, we derive convenient expressions of the above
quantities for degenerated configurations  mentioned in the
Introduction.  In the following Sections these
expressions will be applied to the calculation of integrals over
the degenerated configurations of interest.

Along with $R_L^{(n)}(t,t';\{q\})$ of Section 2, we consider
$K_L^{(n)}(t,t';\{q\})$ defined to be
\begin{equation}
K_L^{(n)}(t,t';\{q\})=D(t')R_L^{(n)}(t,t';\{q\})
\label{kr}
\end{equation}
where the spinor derivative is defined by (\ref{supder}).
Furthermore, we build (see also \cite{danphr}) a
matrix operator $\hat K=\{\hat K_{sr}\}$ where $\hat K_{sr}$ is
an integral operator vanishing at $s=r$. For $s\neq r$, the
kernel of $\hat K_{sr}$ is $\tilde
K_{L_s}^{(n_s)}(t,t';\{q\}_s)dt'$.  Here $\tilde
K_{L_s}^{(n_s)}(t,t';\{q\}_s)$ is related by (\ref{kr}) with
$\tilde R_{L_s}^{(n_s)}(t,t';\{q\}_s)$, which is the
non-singular part (\ref{lim}) of the Green function. So
\begin{equation}
K_{L_s}^{(n_s)}(t,t';\{q\}_s)=
\frac{\vartheta-\vartheta'}
{z-z'}+\tilde K_{L_s}^{(n_s)}(t,t';\{q\}_s)\,.
\label{polk}
\end{equation}
Like \cite{danphr}, we define kernels together with the
differential $dt'=dz'd\vartheta'/2\pi i$. The discussed operator
performs the integration with $\tilde
K_{L_s}^{(n_s)}(t,t';\{q\}_s)$ over $t'$ along $C_r$-contour,
which surrounds the limiting points associated with the
considered group $r$ of the handles and the cuts between
limiting points for the Ramond handles.  The desired relation
for the Green function is
\begin{eqnarray}
R_L^{(n)}(t,t';\{q\})=\ln(z-z'-\vartheta\vartheta')+
\sum_{s}\tilde
R_{L_s}^{(n_s)}(t,t';\{q\}_s)
\nonumber\\
+\sum_{r,s}\int_{C_s}[(1-\hat K)^{-1}\hat
K]_{rs}(t,t_1)dt_1\tilde R_{L_r}^{(n_r)}(t_1,t';\{q\}_r)
\label{rprt}
\end{eqnarray}
where $[(1-\hat K)^{-1}\hat K]_{rs}(t,t_1)dt_1$ is the kernel of
the operator
\begin{equation}
(1-\hat K)^{-1}\hat K=
\hat K+\hat K\hat K+\dots
\label{opr}
\end{equation}
In the Neveu-Schwarz sector where Green function (\ref{kr}) has
the poles solely  \cite{vec,pst,dan90}, the integrals in
(\ref{rprt}) are calculated without difficulties. In this case,
using for Green functions (\ref{polk}) Poincar{\'e} series
\cite{vec,pst,dan90}, one reproduces the series
for $R_L^{(n)}(t,t';\{q\})$. In fact, to prove (\ref{rprt}),
one needs only check that (\ref{rprt}) satisfies to
(\ref{rtrans}). In particular, to verify (\ref{rtrans}) under
the transformations assigned to the $r$-th group of the handles,
we represent (\ref{rprt}) as
\begin{eqnarray}
R_L^{(n)}(t,t';\{q\})=
R_{L_r}^{(n_r)}(t,t';\{q\}_r)+
\sum_{s\neq r}\int_{C_s}
K_{L_r}^{(n_r)}(t,t_1;\{q\}_r)
\tilde R_{L_s}^{(n_s)}(t_1,t';\{q\}_s)dt_1
\nonumber\\
+\sum_{p\neq r}\sum_{s}
\int_{C_p}K_{L_r}^{(n_r)}(t,t_1;\{q\}_r)dt_1\int_{C_s}
[(1-\hat K)^{-1}\hat
K]_{ps}(t_1,t_2)dt_2 \tilde R_{L_s}^{(n_s)}(t,t';\{q\}_s)
\label{rpart}
\end{eqnarray}
where $R_{L_r}^{(n_r)}(t,t';\{q\}_r)$ and
$K_{L_r}^{(n_r)}(t,t_1;\{q\}_r)$ are total Green functions
including the singular term in (\ref{lim}) and (\ref{polk}).
Indeed, once, using (\ref{opr}), one calculates the contribution
to (\ref{rpart}) of the pole term in (\ref{polk}),
eq.(\ref{rprt}) appears. Relations (\ref{rtrans}) for the
transformations of the $r$-th group discussed  are evidently
satisfied for (\ref{rpart}), the scalar function
$J_{j_r}^{(n)}(t;\{q\};L)$ associated with the $j_r$ handle of
the $r$-th supermanifold being
\begin{eqnarray}
J_{j_r}^{(n)}(t;\{q\};L)=J_{j_r}^{(n_r)}(t;\{q\}_r;L_r)+
\sum_{s\neq r}\int_{C_s}D(t_1)
J_{j_r}^{(n_r)}(t_1;\{q\}_r;L_r)dt_1
\tilde R_{L_s}^{(n_s)}(t_1,t;\{q\}_s)
\nonumber\\
+\sum_{p\neq r}\sum_{s}
\int_{C_p}D(t_1)
J_{j_r}^{(n_r)}(t_1;\{q\}_r;L_r)dt_1\int_{C_s}
[(1-\hat K)^{-1}\hat
K]_{ps}(t_1,t_2)dt_2 \tilde R_{L_s}^{(n_s)}(t_2,t;\{q\}_s)\,.
\label{tjr}
\end{eqnarray}
Hence (\ref{rpart}) and (\ref{rprt})  both are the
correct expressions for $R_L^{(n)}(t,t'\{q\})$. The period
matrix is calculated from (\ref{tjr}).
For this purpose one considers the
difference $J_{j_r}^{(n)}(t;\{q\};L)-J_{j_r}^{(n)}(t_0;\{q\};L)$
where $t_0$ is a fixed parameter. The above difference is
given by
\begin{eqnarray}
J_{j_r}^{(n)}(t;\{q\};L)-J_{j_r}^{(n)}(t_0;\{q\};L)=
J_{j_r}^{(n_r)}(t;t_0\{q\}_r;L_r)
+\int_{C_s'}D(t_1)J_{j_r}^{(n_r)}
(t_1;\{q\}_r;L_r)
dt_1
\nonumber\\
\times
R_{L_s}^{(n_s)}(t_1,t;t_0;\{q\}_s)
\nonumber\\
+\sum_{p\neq r}
\int_{C_p}D(t_1)
J_{j_r}^{(n_r)}(t_1;\{q\}_r;L_r)dt_1\int_{C_s'}
[(1-\hat K)^{-1}\hat
K]_{ps}(t_1,t_2)dt_2R_{L_s}^{(n_s)}(t_2,t;t_0;\{q\}_s)
\label{jar}
\end{eqnarray}
where both $z$ and $z_0$ lay inside the $C_s'$ contour while
\begin{eqnarray}
R_{L_s}^{(n_s)}(t_1,t;t_0;\{q\}_s)=
R_{L_s}^{(n_s)}(t_1,t;\{q\}_s)-
R_{L_s}^{(n_s)}(t_1,t_0;\{q\}_s)\,,
\nonumber\\
J_{j_r}^{(n_r)}(t;t_0\{q\}_r;L_r)=
J_{j_r}^{(n_r)}(t;\{q\}_r;L_r)
-J_{j_r}^{(n_r)}(t_0;\{q\}_r;L_r)
\label{jar1}
\end{eqnarray}
where $R_{L_s}^{(n_s)}(t_1,t;\{q\}_s)$ is the total Green
function (\ref{lim}) including the singular term. To prove
(\ref{jar}), one, using (\ref{opr}), calculates the
contribution from the $\ln[(z_2-z-\vartheta_2\vartheta)/
(z_2-z_0-\vartheta_2\vartheta_0)]$ term due to the singularity
of the Green function.  The corresponding integral is
transformed to the one along the cut between
$z_2=z-\vartheta_2\vartheta$ and
$z_2=z_0-\vartheta_2\vartheta_0$. Then it is found to be
\begin{equation}
\int
D(t_2)f(t_2)[\theta(z_2-z-\vartheta_2\vartheta)-
(z_2-z_0-\vartheta_2\vartheta_0)]dz_2d\vartheta_2=
f(t)-f(t_0)
\label{calom}
\end{equation}
where $f(t)$ denotes either the Green function, or
$J_{j_r}^{(n_r)}(t_1;\{q\}_r;L_r)$. As the results, one obtains
(\ref{tjr}). From (\ref{jar}),
the $\omega_{j_rj_s}^{(n)}(\{q\};L)$ element of the period
matrix is found to be
\begin{eqnarray}
2\pi
i\omega_{j_rj_s}^{(n)}(\{q\};L)=\delta_{j_rj_s}\ln k_{j_r}
+(1-\delta_{j_rj_s})\int_{C_s}D(t)
J_{j_r}^{(n_r)}(t;\{q\}_r;L_r)dt
J_{j_s}^{(n_s)}(t;\{q\}_s;L_s)
\nonumber\\
+\sum_{p}\int_{C_p}D(t)
J_{j_r}^{(n_r)}(t;\{q\}_r;L_r)dt\int_{C_s}[(1-\hat K)^{-1}\hat
K]_{pr}(t,t')dt'
J_{j_s}^{(n_s)}(t';\{q\}_s;L_s)\,.
\label{omjr}
\end{eqnarray}
For the odd super-spin structure, due to the spinor zero
mode, there is no the Green function satisfying (\ref{rtrans})
and, at the same time, having no non-physical poles.  In this
case  further terms need to be added in (\ref{rprt}) providing
true properties of $R_L^{(n)}(t,t';\{q\})$. In particular,
these terms appear in the genus-$n$ Green function given in
terms of the genus-1 functions when $L$ includes genus-1 odd
spin structures, see Appendix B of the present paper.

The integration measure is given by (\ref{hol}) and
(\ref{zinv}) with $\hat
Z_L^{(n)}(\{q\})$ to be \cite{danphr}
\begin{equation}
\hat
Z_L^{(n)}(\{q\})= \tilde Z^{(n)}(\{q\},L)\prod_{s=1}^n
\frac{ Z^{(1)}(k_s;l_{1s},l_{2s})}{k_s^{(3-2l_{1s})/2}}
\label{zhol}
\end{equation}
where the $(l_{1s},l_{2s})$ theta characteristics are either 0,
or 1/2, while \cite{danphr}
\begin{equation}
Z^{(1)}(k;l_{1},l_{2})=(-1)^{2l_{1s}+2l_{2s}-1}16^{2l_{1s}}
\prod_{p=1}^\infty
\frac{[1+(-1)^{2l_2}k^pk^{(2l_1-1)/2}]^8}{[1-k^p]^8}\,.
\label{z1h}
\end{equation}
The expression of $\tilde Z_L^{(n)}(\{q\})$ through the
genus-$n_i$ functions is derived using Appendix C along with the
Green functions given in the Appendix B.  When all the $L_i$
super-spin structures are even, the desired expression is given
by (see Appendix C for more details)
\begin{equation}
\ln\hat Z_L^{(n)}(\{q\}))=\sum_{i}\ln
\hat Z_{L_i}^{(n_i)}(\{q\}_i))-5trace\ln(I-\hat K) +
trace\ln(I-\hat G)
\label{ipf}
\end{equation}
where $\hat K$ is the same as in (\ref{rprt}). The  $\hat G$
operator is constructed similar to $\hat K$, the non-singular
part $\tilde G_{L_s}^{(n_s)}(t,t';\{q\}_s)$ of the
$G_{L}^{(n)}(t,t';\{q\})$ ghost superfield Green function
\cite{danphr} being employed instead of $\tilde
K_{L_s}^{(n_s)}(t,t';\{q\}_s)$. In this case
\begin{equation}
G_{L}^{(n)}(t,t';\{q\})= \frac{\vartheta-\vartheta'}{z-z'}
+\tilde G_{L_s}^{(n_s)}(t,t';\{q\}_s)
\label{greg}
\end{equation}
where the last term on the right side has no singularity at
$z=z'$.  The Green function is a superconformal 3/2-tensor under
transformations (\ref{stens}) of $t'$, but it is not a
superconformal (-1) tensor under the transformations of $t$.
Indeed, in the last case the Green function receives additional
terms being the sum of a polynomials in $t$ multiplied by a
relevant superconformal 3/2-tensor in $t'$ (see eq.(63) in
\cite{danphr} and Appendix C of the present paper).

The obtained expressions can be applied to the degenerated
configurations where all the limiting points of $n_i$ forming
group transformations go to the same point. As the basic case,
we consider $n_1<n$ forming group ones, the limiting points
going to $z_0$. We say that they form the degenerated $n_1$
configuration, its super-spin structure being $L_1$. The
remaining $n_2=n-n_1$ handles form the $n_2$ configuration, its
super-spin structure being $L_2$.  We  assume no more than one
interaction vertex to be near $z_0$. As was noted in the
Introduction, the considered configuration is the main one,
which might originate divergences in the group limiting points
of the amplitude integral (\ref{ampl}).

By Section 4, the Schottky multipliers are not closely to the
unity in their absolute values.  Moreover, the $\tilde G$
symmetry bounds $z_0$ to lay exterior to Schottky circles
assigned to the $n_2$ configuration.  When $z_0$ lies at a
finite distances from the group limiting points of the $n_2$
configuration, we imply that $\rho_1<<\rho$.  In this case
$\rho_1$ is the maximal size of the $n_1$ configuration while
$\rho$ is the minimal distance between $z_0$ and any point of
essence assigned to the $n_2$ one. So $\rho\leq\rho_2$ where
$\rho_2$ is the maximal size of the $n_2$ configuration.  If
$z_0\to\infty$, then $\rho_1\leq\rho$ and $\rho>>\rho_2$.  This
case is, however, reduced to the finite $z_0$ case by a relevant
$L(2)$ transformation. Hence, for brevity, we discuss finite
$z_0$. The super-spin structures $L_1$ and $L_2$ are taken to
be even. We obtain for this configuration the leading
approximated integration measure, vacuum correlator, scalar
functions and period matrix, as well as the leading corrections
for the above quantities. To clarify the method, we
present a more detailed calculation of the leading corrections
for the holomorphic partition function (\ref{ipf}). In this case
the sum on the right side of (\ref{ipf}) gives the
factorized partition
function while two rest terms are
corrections. In particular, the correction due to the second
term is given by
\begin{equation}
J_{cor}^n=-trace\ln(I-\hat K)=
\int_{C_I}\Biggl(\int_{ C_{II}}\tilde
K_{L_1}^{(n_1)}(t_1,t_2)dt_2 \tilde
K_{L_2}^{(n_2)}(t_2,t_1)\Biggr) dt_1+\dots
\label{kcor}
\end{equation}
where $dt=dzd\vartheta/2\pi$. In this case $\tilde
K_{L_1}^{(n_1)}(t,t')$ and $\tilde K_{L_2}^{(n_2)}(t,t')$ are
defined by (\ref{kr}) on the genus-$n_1$ supermanifold and,
respectively, on the genus-$n_2$ one. The $C_I$ contour bounds
the domain occupied by the degenerated handles while the
$C_{II}$ contour bounds the domain of the rest handles.
To calculate the first term on the right side of (\ref{kcor}),
the $C_I$ contour is gone on a distance $\sim\rho$ from the
degenerated handles. Thus $\tilde K_{L_1}^{(n_1)}(t,t')$ can be
approximated by its asymptotics, $z$ and $z'$ both being far
from $z_0$.  Due to (\ref{kr}), the above asymptotics is related
with the asymptotics  of $\tilde R_L^{(n_1)}(t,t')$ in
(\ref{lim}), which are given by
\begin{equation}
\tilde R_L^{(n)}(t,t';\{q\})\approx
\frac{\tilde a_L^{(n)}(\{q\})+
\tilde\alpha_L^{(n)}(\{q\})(\vartheta+\vartheta')}{(z-z_0)
(z'-z_0)}
-\frac{\tilde b_L^{(n)}(\{q\})\vartheta\vartheta'}
{(z-z_0)(z'-z_0)}
\Biggl[\frac{1}
{z-z_0}-\frac{1}
{z'-z_0}\Biggl]\,.
\label{asr}
\end{equation}
We assign a smallness $\sim\sqrt{\rho_1}$ to each of the
integrated Grassmann parameters of the degenerated $n_1$
configuration.  Along with the estimation  $\sim1/\sqrt{\rho_1}$
for its differential, it correctly determines the magnitude of
the integral over the variables discussed. Hence
\begin{equation}
\tilde a_L^{(n)}(\{q\})\sim\rho_1^2\,,\qquad
\tilde b_L^{(n)}(\{q\})\sim\rho_1^2\,,\qquad
\tilde\alpha_L^{(n)}(\{q\})\sim\rho_1\sqrt{\rho_1}
\label{coef}
\end{equation}
The asymptotics of $K_{L_1}^{(n_1)}(t,t')$ is given by
(\ref{kr}) and  (\ref{asr}) at $n=n_1$ and $L=L_1$. The
integrals over $z_1$ and $z_2$ are calculated by the Cauchy
theorem upon going the contours to surround the poles on the
right side of (\ref{asr}).  Thus
\begin{equation}
J_{cor}^n \approx [\tilde a_{L_1}^{(n_1)}(\{q\}_1)
\partial_z\partial_{z'}-
2\tilde b_{L_1}^{(n_1)}(\{q\}_1)
\partial_ z\partial_{\vartheta}
\partial_{\vartheta'}
-2\tilde\alpha_{L_1}^{(n_1)}(\{q\}_1)
\partial_z\partial_{\vartheta'}]
\tilde
R_{L_2}^{(n_2)}(t,t';\{q\}_2)
\label{rescor}
\end{equation}
where $z=z'=z_0$ and $\vartheta=\vartheta'=0$. The indices "1"
and "2" are assigned to the corresponding configuration. Due to
(\ref{coef}), the terms in (\ref{rescor}) are $\sim
\rho_1^2\rho_2^2/\rho^4$ and  $\sim
\rho_1\sqrt{\rho_1}\rho_2\sqrt{\rho_2}/\rho^3$.  The last term
in (\ref{ipf}) is calculated in the same fashion using Appendix
C. Being no more than $\sim
\rho_1^3\sqrt{\rho_1}\rho_2^3\sqrt{\rho_2}/\rho^7$, it
can be neglected. Thus
\begin{equation}
Z_L^{(n)}(\{q\}))\approx Z_{L_1}^{(n_1)}(\{q\}_1))
Z_{L_2}^{(n_2)}(\{q\}_2))[1+5J_{cor}^n]
\label{ipfas}
\end{equation}
where $J_{cor}^n$ is given by (\ref{rescor}).
Other corrections are calculated in the kindred manner. In
particular, from (\ref{omjr}), the corrections for the
$\omega_{j_2j_2'}^{(n)}(\{q\};L)$ period matrix elements are
also proportional to the coefficients in (\ref{asr}). Here
$j_2$ labels the cycles assigned to the $n_2$
configuration. Really we shall see in Section 6 that corrections
proportional to the coefficients in (\ref{asr}) are too small to
originate divergences. More large corrections are determined by
the asymptotics at $z\to\infty$ of the scalar function
\begin{equation}
J_r^{(n)}(t;\{q\};L)\approx
\frac{\hat S_r^{(n)}(\{q\};L)+
\hat\Sigma_r^{(n)}(\{q\};L)\vartheta}{z-z_0}
\equiv \frac{\hat J_r^{(n)}(\{q\};L;\vartheta)}{z-z_0}\,.
\label{asj}
\end{equation}
In this case $\hat S_r^{(n_1)}(\{q\}_1;L_1)
\sim\rho_1$
and $\hat\Sigma_r^{(n_1)}(\{q\}_1;L_1)\sim\sqrt\rho_1$.
Below $j_1$ is reserved for the cycles assigned to the
degenerated $n_1$ configurations.  Corrections for
$\omega_{j_1j_1'}^{(n)}(\{q\};L)$ are quadratic in the
coefficients of (\ref{asj}) since the sole $\sim\rho_1$ term
$\sim\hat\Sigma_{j_1}^{(n_1)}(\{q\}_1;L_1)
\hat\Sigma_{j_1'}^{(n_1)}(\{q\}_1;L_1)$ vanishes. Indeed, the
above term is multiplied by $\partial_\vartheta
\partial_{\vartheta'}\tilde
R_{L_2}^{(n_2)}(t,t';\{q\}_2)$ at $z=z'$, which is nullified due
to the bose symmetry of the Green function.  So
\begin{eqnarray}
\omega_{j_1j_1'}^{(n)}(\{q\};L)\approx
\omega_{j_1j_1'}^{(n_1)}(\{q_1\};L_1)\,,\qquad
\omega_{j_2j_2'}^{(n)}(\{q\};L)\approx
\omega_{j_2j_2'}^{(n_2)}(\{q_2\};L_2)\,,
\nonumber\\
2\pi
i\omega_{j_1j_2}^{(n)}(\{q\};L)\approx-\int
\hat J_{j_1}^{(n_1)}(\{q\}_1;L_1;\vartheta_1)
d\vartheta_1
D(t_1)J_{j_2}^{(n_2)}(t_1;\{q\}_2;L_2)
\label{omas}
\end{eqnarray}
where the integral is calculated at $z_1=z_0$, for further
definitions see (\ref{asj}). The Green functions and the scalar
functions for $|z-z_0|>>\rho_1$ and $|z'-z_0|>>\rho_1$ are found
to be
\begin{eqnarray}
J_{j_2}^{(n)}(t;\{q\};L)\approx
J_{j_2}^{(n_2)}(t;\{q\}_2;L_2)\,,\quad
R_L^{(n)}(t,t';\{q\})\approx R_{L_2}^{(n_2)}(t,t';\{q\}_2)\,,
\nonumber\\
J_{j_1}^{(n)}(t;\{q\};L)\approx-\int
\hat J_{j_1}^{(n_1)}(\{q\}_1;L_1;\vartheta_1)
d\vartheta_1 D(t_1)
R_{L_2}^{(n_2)}(t_1,t;\{q\}_2)
\label{scal}
\end{eqnarray}
where $z_1=z_0$. The indices "1" and "2" mark to the
corresponding configurations.  When $|z-z_0|\sim\rho_1$ and
$|z'-z_0|>>\rho_1$ the correction for the Green function is
calculated through the asymptotics of $\tilde
R_L^{(n)}(t,t';\{q\})$ at $z'\to\infty$. The above asymptotics
is given by
\begin{equation}
\tilde R_L^{(n)}(t,t';\{q\})\approx
\frac{\tilde S^{(n)}(t;\{q\};L)+
\tilde\Sigma^{(n)}(t;\{q\},L)\vartheta'}{(z'-z_0)}
\equiv
\frac{\tilde E_L^{(n)}(t;\{q\};\vartheta')}{(z'-z_0)}\,.
\label{asymr}
\end{equation}
In more details, the coefficients are represented as
\begin{eqnarray}
\tilde S^{(n)}(t;\{q\},L)=
\vartheta\tilde\Sigma_1^{(n)}(z;\{q\};L)+
\tilde S_1^{(n)}(t;\{q\},L)\,,
\nonumber\\
\tilde\Sigma^{(n)}(t;\{q\};L)=\vartheta\tilde S_2^{(n)}
(z;\{q\};L)+
\tilde\Sigma_2^{(n)}(z;\{q\};L);
\label{cfnt}
\end{eqnarray}
where the "tilde" terms  determine the asymptotics of the
"tilde" Green function in (\ref{lim}).  By dimensional reasons,
$\tilde S_2^{(n_1)}(z;\{q\}_1;L_1)\sim1$,
$\tilde\Sigma_1^{(n_1)}(z;\{q\}_1;L_1)\sim
\tilde\Sigma_2^{(n_1)}(z;\{q\}_1;L_1)
\sim\sqrt{\rho_1}$ and $\tilde
S_1^{(n_1)}(t;\{q\}_1;L_1)\sim\rho_1$. Further, at $z\to\infty$
the functions discussed are related with the
coefficients in (\ref{asr}) by $(z-z_0)\tilde
S_1^{(n)}(t;\{q\},L)\to\tilde a_L^{(n)}(\{q\})$,
$(z-z_0)^2\tilde S_2^{(n)}(t;\{q\},L)\to\tilde
b_L^{(n)}(\{q\})$,
$(z-z_0)\tilde\Sigma_1^{(n)}(t;\{q\},L)\to\tilde
\alpha_L^{(n)}(\{q\})$ and
$\tilde\Sigma_2^{(n)}(t;\{q\},L)\to
\tilde\Sigma_1^{(n)}(t;\{q\},L)$. In the case discussed,
\begin{equation}
R_L^{(n)}(t,t';\{q\})=
R_{L_2}^{(n_2)}(t_0,t';\{q\}_2)
-\int E_{L_1}^{(n_1)}(t;\{q\}_1;\tilde\vartheta)
d\tilde\vartheta
D(\tilde t)R_{L_2}^{(n_2)}(\tilde t,t';\{q\}_2)
+\dots
\label{sgras}
\end{equation}
where $\tilde z=z_0$, $t_0=(z_0|0)$ and, in addition,
\begin{eqnarray}
E_{L_1}^{(n_1)}(t;\{q\}_1;\tilde\vartheta)=
S^{(n_1)}(t;\{q\}_1;L_1)+\Sigma^{(n_1)}(t;\{q\}_1;L_1)
\tilde\vartheta\,,\quad
S^{(n_1)}(t;\{q\}_1;L_1)=
\nonumber\\
-(z-z_0)+\tilde
S^{(n_1)}(t;\{q\}_1;L_1)\,,\quad
\Sigma^{(n_1)}(t;\{q\}_1;L_1)=\vartheta
+\tilde\Sigma^{(n_1)}(t;\{q\}_1;L_1)
\label{etild}
\end{eqnarray}
with definitions given in (\ref{asymr}) and in (\ref{cfnt}).
The sum $[\ln(z'-z_0)+
E_{L_1}^{(n_1)}(t;\{q\}_1;\tilde\vartheta)]$ gives the
asymptotics at $\tilde z\to\infty$ of the Green function
$R_{L_1}^{(n_1)}(t,\tilde t;\{q\}_1)$ related with
$\tilde R_{L_1}^{(n_1)}(t,\tilde t;\{q\}_1)$ by
(\ref{lim}).
The scalar functions at $|z-z_0|\sim\rho_1$ are given by
\begin{eqnarray}
J_{j_1}^{(n)}(t;\{q\};L)=
J_{j_1}^{(n_1)}(t;\{q\}_1;L_1)
- \int
\hat J_{j_1}^{(n_1)}(\{q\}_1;L_1;\tilde\vartheta)
d\tilde\vartheta D(\tilde t)
\tilde R_{L_2}^{(n_2)}(\tilde t,t;\{q\}_2)+\dots\,,
\nonumber\\
J_{j_2}^{(n)}(t;\{q\};L)=
J_{j_2}^{(n_2)}(t_0;\{q\}_2;L_2)-
\int
E_{L_1}^{(n_1)}(t;\{q\}_1;\tilde\vartheta)
d\tilde\vartheta D(\tilde t)
J_{j_2}^{(n_2)}(\tilde t;\{q\}_2;L_2)\dots
\label{nscal}
\end{eqnarray}
with the definitions given in (\ref{sgras}). The correction
$\sim \hat\Sigma_{j_1}^{(n_1)}(\{q\}_1;L_1)
\tilde\Sigma_{j_1}^{(n_1)}(\{q\}_1;L_1)\sim\rho_1$
in the first line is absent  because, being proportional to
$\partial_\vartheta\partial_{\vartheta'}\tilde
R_{L_2}^{(n_2)}(t,t';\{q\}_2)$ at $z=z'$, it is nullified.
The vacuum correlator (\ref{corr}) is given
through ${\cal X}_{L_1,L_1'}^{(n_1)}(t,\overline
t;\vartheta_1;\{q,\bar q\}_1)$ defined to be
\begin{eqnarray}
{\cal X}_{L_1,L_1'}^{(n_1)}(t,\overline
t;\vartheta_1;\{q,\bar q\}_1)=
E_{L_1}^{(n_1)}(t;\{q\}_1;\vartheta_1)+
[J_s^{(n_1)}(t;\{q\}_1;L_1)+
\overline{J_s^{(n_1)}(t;\{q\}_1;L_1')}]
\nonumber\\
\times
[\Omega_{L_1,L_1'}^{(n_1)}(\{q,\bar q \}_1)]_{sr}^{-1}
\hat
J_r^{(n_1)}(\{q\}_1;L_1;\vartheta_1)
\equiv X_{L_1,L_1'}^{(n_1)}(t,\overline t;\{q,\bar
q\}_1)+\Xi_{L_1,L_1'}^{(n_1)}(t,\overline t;\{q,\bar
q\}_1)\vartheta_1\,;
\nonumber\\
X_{L_1,L_1'}^{(n_1)}(t,\overline t;\{q,\bar
q\}_1)=-(z-z_0)+
\tilde X_{L_1,L_1'}^{(n_1)}(t,\overline t;\{q,\bar
q\}_1)\,,
\nonumber\\
\Xi_{L_1,L_1'}^{(n_1)}(t,\overline t;\{q,\bar
q\}_1)=\vartheta+
\tilde\Xi_{L_1,L_1'}^{(n_1)}(t,\overline t;\{q,\bar
q\}_1)\,.
\label{corhx}
\end{eqnarray}
In this case the "tilted" quantities are calculated through the
"tilted" ones in (\ref{etild}). From (\ref{corr}), (\ref{sgras})
and (\ref{nscal}) the correlator is given by
\begin{eqnarray}
\hat X_{L,L'}(t,\overline t;t',\overline
t';\{q\})\approx \hat X_{L_2,L_2'}(t_0,\overline t_0;t',\overline
t';\{q\}_2) -\int[{\cal
X}_{L_1,L_1'}^{(n_1)}(t,\overline
t;\vartheta_1;\{q,\bar q\}_1)d\vartheta_1
\nonumber\\
\times
D(t_1)
\hat X_{L_2,L_2'}(t_1,\overline{t_1};t',\overline
t';\{q\}_2)
\nonumber\\
-\int\overline{
[{\cal
X}_{L_1',L_1}^{(n_1)}(t,\overline
t;\vartheta_1;\{q,\bar q\}_1)}d\bar\vartheta_1
\overline{D( t_1)} \hat
X_{L_2,L_2'}(t_1,\overline{t_1};t',\overline t';\{q\}_2)
+\dots
\label{corl}
\end{eqnarray}
where $t_0=(z_0|0)$, $t_1=(z_0|\vartheta_1)$ and the
"dots" encode the terms  at $t=t_0$ due to  the leading
corrections for the $(j_2,j_2')$ matrix elements of
(\ref{grom}) and for the scalar functions carrying the $j_1$
index. The above terms are easy found using (\ref{grom}),
(\ref{omas}) and (\ref{nscal}). From (\ref{rescor}) and
(\ref{omas}), the integration measure is mainly factorized.  For
$n_1=1$, the genus-1 factor
\begin{equation}
Z_{L_1,L_1'}^{(1)}(\{q,\overline
q\}_1)=Z_{tore}(k,\bar k;L_1,L_1')|u-v-\mu\nu|^{-2}
\label{torus}
\end{equation}
differs from the integration measure $Z_{tore}(k,\bar
k;L_1,L_1')$ on the torus \cite{gsw} by the $|u-v-\mu\nu|^{-2}$
multiplier, which is due to the integration over Killing genus-1
modes.  Furthermore, only  even $L_1$ super-spin structures
might originate the divergences.  Otherwise the integrand is not
singular, see Appendix C.  Moreover, the contribution to
(\ref{ampl}) of odd super-spin structures is obtained by the
factorization of the relevant even super-spin ones, see the end
of Appendix C. So, it is sufficient to check the cancellation of
divergences for the even super-spin structures. In doing so we
use eqs. (\ref{rescor}) and (\ref{ipfas}) for the holomorphic
partition function, expression (\ref{omas}) for the period
matrix, eqs.  (\ref{scal}) and (\ref{nscal}) for the scalar
functions, eq. (\ref{sgras}) for the Green function and
eq.(\ref{corl}) for the vacuum correlator.

\section{Integrals of the superstring theory}

The amplitude (\ref{ampl}) includes Grassmann integrations along
with the ordinary ones. In this case the result is finite or
divergent depending on the used integration variables, as this
is seen for an easy integral
\begin{equation}
I_{(ex)}=\int\frac{dxdyd\alpha
d\beta d\bar\alpha d\bar\beta}
{|z-\alpha\beta|^p}\theta(1-|z|^2)
\label{examp}
\end{equation}
where $z=x+iy$ while $\alpha$ and $\beta$ are complex Grassmann
variables. The complex parameter $p$ characterizes the strength
of the singularity. Integrals of this kind really appear in
(\ref{ampl}). In particular, from (\ref{zinv}),
the integration measure contains singularity (\ref{examp})
with $p=2$. The kindred expression  is originated by
the singularity (\ref{lim}) at $z=z'$ of the Green function. In
this case $p=s_{jl}/4+2$ where $s_{jl}=-(p_j+p_l)^2$ is the
square center mass energy in the given reaction channel and the
add 2 is due to the vacuum contractions of the fields in front
of the exponential in (\ref{vert}).  For the sake of simplicity,
we bound the integration region in (\ref{examp}) by
$|z|^2\leq1$.  Once the Grassmann integrations being performed,
one obtains the integral
\begin{equation}
I_{(ex)}=p^2\int\frac{dxdy} {4|z|^{p+2}}\theta(1-|z|^2)\,,
\label{exam}
\end{equation}
which is divergent at $z=0$, if $Re\,p>0$. On the other side, in
(\ref{examp}) one can turn to the variable $\tilde
z=z-\alpha\beta$.  Then the Grassmann variables will present
only in the step function
$\theta(|\tilde z+\alpha\beta|^2)$. After
the integration over the Grassmann variables, the integral
is turned to the boundary integral at $|z|^2=1$.
Thus for any $p$ the result of the integration is finite being
\begin{equation}
I_{(ex)}=-\int\frac{d\tilde
xd\tilde yd\alpha d\beta d\bar\alpha d\bar\beta} {|\tilde
z|^p}\alpha\beta\bar\alpha\bar\beta \left[\delta(|\tilde
z|^2-1)+|\tilde z|^2\frac{d\delta(|\tilde z|^2-1)}{d|\tilde
z|^2}\right]=-\frac{\pi p}{2}\,.
\label{ex}
\end{equation}
So (\ref{examp}) depends on the integration
variables, at least for $Re\,p>0$. For $Re\,p<0$ when
(\ref{exam}) is convergent, (\ref{exam}) and (\ref{ex}) both
give the same result. One could, however, change
the integration variable $z$ by
$z+\sum_{i=1}^N\delta_i\delta_i^{(1)}$ where $\delta_i$ and
$\delta_i^{(1)}$ are arbitrary Grassmann numbers. When $p$ is
not an negative even number, the resulted integrand has the
singularity $\sim|z|^{-(p+2+2N)}$. So, for $2N>-p$, the
integral is divergent. For a negative even  $p$ the integral can
be reduced to the singular integral by a change $z=\tilde
z+\alpha\beta$ with the following $|\tilde z|=|\hat z|^{p_1}$
and $\arg \tilde z= \arg\hat z$ where $p_1>0$ is no integer or
half-integer,  the strength of the resulted singularity being
$(p_1p-1)$.  This integral can be transformed to the divergent
one. The result is, however, the same, if the integration
variable change remains the integral to be convergent.

Calculating the amplitude, we are guided by preserving its local
symmetries. Divergences break the conformal symmetry due to a
cutoff parameter, the amplitude depends on $\{N_0\}$ set
(\ref{factorg}) that falls the theory.  In the proposed
calculation all the  divergences are  cancelled and the local
symmetries are preserved.

Due to the singularity (\ref{lim}) at $z=z'$ of the Green
function, the integrand in (\ref{ampl}) is singular when $m_1>1$
interaction vertices go the same point on the complex $z$ plane.
When the vertices are accompanied by degenerated Schottky
circles, singularities present also for
the integration measure.  If $1<m_1<(m-1)$ for the $m>3$
point amplitude, then the strength of the singularity depends on
the energy 10-invariant of the given reaction channel. The
integral is calculated \cite{gsw} for those energies below the
reaction threshold, where it is convergent.  The result is
analytically continued to energies above the threshold.  In this
case the amplitude receives singularities required by the
unitarity equations (as an example, see Appendix A).  Due to the
energy-momentum conservation, there is no a domain for the
energy 10-invariants where the integrals over all the singular
regions are convergent simultaneously. Hence the amplitude is
obtained by the summing of the pieces obtained by the analytical
continuation from the distinct regions of the 10-invariants. For
instance, the calculation of the scattering amplitude
includes an analytical continuation
from the relevant energy region of the integrals over nodal
regions. Each of the integrals
gives rise to the cut in $s$, $t$ and
$u$-Mandelstam invariant. Every  region  contains
a pair of the corresponding vertices going to each other
(see the discussion of the unitarity equations in Section 4 and
in Appendix A). The analytical continuation procedure
discussed is evidently consistent with the local symmetries of
the amplitude.

Moreover, the amplitudes for the emission of a longitudinal
gauge boson are nullified as it is required. Indeed, from
(\ref{vert}), in this case the integrated function is the
super-derivative in $t_j$ of a local function of the $t_j$
coordinate assigned to the longitudinal boson discussed.
When $z_j$ and certain
other vertex coordinates both go to the same point, the
integral is calculated for those energy invariants, which it
is convergent for.  Hence the integral is
reduced to the integral over boundary of a singular region
discussed.  The boundary
integral is analytically continued to considered energies.
Having no singularity discussed, the boundary integral is
independent from which energy region it was be continued. Hence
the boundary integrals being collected together, are canceled.

By aforesaid, the divergences of the $m$-point amplitude may
appear only when a number $m_1$ of the vertices at the same
point is $m_1=0$, $m_1=1$ and $m_1\geq(m-1)$. The $m_1\geq(m-1)$
case is, however, out of the integration region, if $\{N_0\}$
set in (\ref{ampl}) is formed by $(3|2)$ ones from the vertex
coordinates. So first we consider configurations of degenerated
Schottky circles, no more than one vertex being nearby. From
(\ref{hol}), (\ref{zinv}) and (\ref{zhol}),  the integration
measure is singular in Schottky multipliers $k_s$ and in group
limiting points, as well.

Due to (\ref{sbound}) and (\ref{bunit}), the singularity in
$k_s$ appears only at $k_s\to0$. In this case, from (\ref{zhol})
and (\ref{z1h}), the holomorphic integration measure is $\sim
k_s^{-(3-2l_{1s})/2}$. Half-integer powers of $k_s$ at
$l_{1s}=0$ are cancelled after the summation over $2l_{2s}=0$ and
$2l_{2s}=1$ since the sum is unchanged when $\sqrt
k_s\to-\sqrt k_s$, see Section 3 (for the vacuum amplitude this
directly follows from (\ref{z1h})).  From (\ref{grom}) and
(\ref{omjr}), the non-holomorphic factor in (\ref{hol}) is
$\sim1/(\ln|k_s|)^5$ at $k_s\to0$. So  the integrand
(\ref{ampl}) at $k_s\to0$ is $\sim 1/[|k_s|^2(\ln|k_s|)^5]$, the
integral (\ref{ampl}) over small $|k_s|$ being finite as it has
been observed in \cite{vec} for the Neveu-Schwarz sector.

Singularities in the group limiting points are due to the
configurations discussed in Section 5 where $n_1$ degenerated
handles (carrying the even super-spin structure $L_1$) go to the
same point $z_0$.  As far as $z_0\to\infty$ is reduced to a
finite $z_0$  by a relevant $L(2)$ transformation, we mainly
consider finite $z_0$. By aforesaid, we assume no more than one
vertex nearby $z_0$.  When no to be the vertex, we say this is
the vacuum  configuration.  If the $z_j$ coordinate of the
corresponding interaction vertex $j$ goes to $z_0$, we say
it is the $j$-th configuration. With our convention for the
$\{N_0\}$ set (see Section 2) all the Grassmann module
parameters of the $n_1$ configuration are the integration
variables while, for the $j$-th configuration, the $\vartheta_j$
Grassmann partner of $z_j$ may be fixed. So we discuss the
integral over the group limiting points of the $n_1$
configuration with given $z_0$ and $\vartheta_j$ (for the $j$-th
one). We bound from top the size of the $n_1$
configuration by the $\Lambda<<\rho$ cut-off where (see Section
5) $\rho$ is the characteristic distance from $z_0$ to points
assigned to the $n_2$ configuration. Due to (\ref{bunit}), and
(\ref{zbound}), the integrand is non-singular until no one of
the differences $\tilde v_s=(v_s-u_s)$ is equal to zero.  Hence
we take the above differences as the integration variables. On
equal terms one can also take the super-difference $w_s=\tilde
v_s-\nu_s\mu_s$ instead of $\tilde v_s$. To be detailed,
the remaining variables are chosen to be $u_s$, $\mu_s$ and
$\nu_s$.  When $\{\tilde v_s\}$ (or $\{w_s\}$) are fixed, the
integrals over the above variables are non-singular. By
dimensional reasons, the  result of the integration may,
however, be singular at $\tilde v_s\to0$ (or at $w_s\to0$) that
originates the divergences for the integral over $\{\tilde
v_s\}$ (or over $\{w_s\}$). All $\tilde v_s$ (or $w_s$) being of
the same order $\sim\rho_1\to0$, the singularity is due to the
integration over the region in a size $\sim\rho_1$.  The easiest
way to estimate the integrals is, as has been noted already, to
assign the $\sim\sqrt{\rho_1}$ smallness to each one of the
integrated Grassmann variable of the degenerated genus-$n_1$
supermanifold. Simultaneously, the differential of the
variable is $\sim1/\sqrt{\rho_1}$. So
leading terms in (\ref{cont}) and (\ref{jint}) might originate
the divergence $\sim(\rho/\rho_1)^2$ where $\rho>>\rho_1$
characterizes distances from $z_0$ to points associated with the
$n_2$ non-degenerated configuration. The corrections terms might
originate the divergences $\sim(\rho/\rho_1)$ and
$\sim\ln(\rho/\rho_1)$. We propose, however, a calculation,
which avoids the divergences. Before we need to discuss in
more details the singular configurations of interest, which are
the vacuum configuration and the $j$-th one.

For the vacuum configuration the integrand (\ref{ampl})  can be
represented as
\begin{equation}
{\cal F}_0^{n_1,n_2}(\{q,\bar q\};\{p_r,\zeta^{(r)}\})=
\sum_{P,\overline P'}O_0^{(n_1)}(P,\overline
P';\{q,\bar q\}_1)Y_{P,\overline
P'}^{(n_2)}
\label{cont}
\end{equation}
where $O^{(n_1)}(P,\overline P';\{q\}_1)$ is calculated for the
degenerated configuration while $Y_{P,\overline P'}^{(n_2)}$
depends only on $z_0$, parameters of the $n_2$ configuration and
on characteristics of the interaction states.  As above,
$\{q\}_1$ is the set of the module variables assigned to the
degenerated configuration. The sum over corrections $(P)$
for holomorphic functions and over corrections $(P')$ for the
anti-holomorphic ones, includes $P=1$ (and
$P'=1$) corresponding to the leading term. Eq.(\ref{cont})
follows directly from the calculation of the
holomorphic functions in the
previous Section. Among other things, the sum in
(\ref{cont}) includes terms due to the corrections for the
boundary (\ref{sbound}). For the vacuum configuration
(\ref{cont}) the leading term $O_0^{(n_1)}(1,\bar 1 ;\{q,\bar
q\}_1)$  is
\begin{equation}
O_0^{(n_1)}(1,\bar 1
;\{q,\bar q\}_1)= \sum_{L_1,L_1'}\tilde
Z_{L_1,L_1'}^{(n_1)}(\{q,\overline q\}_1)
\label{lidvac}
\end{equation}
where zero point function $\tilde Z_{L,L'}^{(n)}(\{q,\overline
q\})$ including step functions (\ref{sbound}) and (\ref{bunit})
is given by
\begin{equation}
\tilde
Z_{L,L'}^{(n)}(\{q,\overline q\})=\frac{g^{2n}}{2^nn!}
Z_{L,L'}^{(n)}(\{q,\overline q\})
\hat B_{L,L'}^{(n)}(\{q,\overline q\})\tilde
B_{L,L'}^{(n)}(\{q,\overline q\})\,,
\label{tilpf}
\end{equation}
integration measure $Z_{L,L'}^{(n)}(\{q,\overline q\})$ being
given by (\ref{hol}) and (\ref{zinv}).
For the $j$-th configuration containing the dilaton
emission vertex, the leading term
$O^{(n_1)}(DJ,\overline{DJ};1,\bar1;\{q,\bar q\}_1)$ is due to
pairing (\ref{ngcor}) of fields (\ref{vert}) in front of the
exponential calculated for the $n_1$ configuration. In this case
\begin{equation}
O_j^{(n_1)}(DJ,\overline{DJ};1,\bar1;\{q,\bar q\}_1)=
\sum_{L_1,L_1'}\tilde
Z_{L_1,L_1'}^{(n_1)}(\{q,\overline q\}_1)
\tilde I_{L_1,L_1'}^{(n_1)}(t_j,\bar
t_j;\{q,\bar q\}_1)
B_{L_1,L_1'}^{(n_1)}(t_j,\bar t_j;\{q,\bar q\}_1)
\label{lidj}
\end{equation}
where $t_j=(z_j|\vartheta_j)$ is the vertex coordinate going to
$z_0$, for other definitions see also (\ref{ngcor}) and
(\ref{tilpf}). We define the integrand together with the step
factor (\ref{zbound}).
The integrand of (\ref{ampl}) for the $j$-th configuration
(including the corrections) is given by
\begin{eqnarray}
{\cal F}_{(j)}^{n_1,n_2}(t_j,\bar t_j;\{q,\bar
q\};\{p_r,\zeta^{(r)}\})=
\sum_{P,\bar P'}O_j^{(n_1)}(DJ,\overline{DJ};P,\bar
P';\{q,\bar q\}_1)\tilde
Y_{P,\bar P'}^{(n_2)}(j)
\nonumber\\
+\sum_{P,\bar P'}
\tilde O_j^{(n_1)}(
P,\bar P';\{q,\bar q\}_1)
\tilde Y_{P,\bar P'}^{(n_2)}(j)
\label{jint}
\end{eqnarray}
where $t_0=(z_0|0)$. As above, $P$ lists holomorphic
corrections, $\bar P'$ lists anti-holomorphic ones, and $\tilde
Y_{P,\bar P'}^{(n_2)}(j)$ depends only  on the parameters of the
$n_2$ configuration, the interaction particle characteristics
and on $z_0$. Every term for the first sum on the right side
is proportional to $\tilde
I_{L_1,L_1'}^{(n_1)}(t_j,\bar t_j;\{q,\bar q\}_1)$. So the first
sum presents only when the configuration includes the dilaton
emission vertex. The remaining terms  are included in the second
sum. In the second sum the term $P=1$ (and $P'=1$) is absent.
When no one of $\{q\}_1$
belongs to the $\{N_0\}$ set, one can take $z_0=z_j$. Otherwise
we identify $z_0$ with  a group limiting point. In more
details, the $(P,\bar P')$ term in (\ref{cont}) is represented
as
\begin{equation}
O_0^{(n_1)}(P,\overline P';\{q,\bar
q\}_1)= \sum_{L_1,L_1'} P(\{q\}_1;L_1)\hat A_{P,P'}
(\{q,\bar q\}_1;L_1,L')
\overline{P'(\{q\}_1;L_1')}
\label{vaccor}
\end{equation}
where $\hat A_{P,P'}(\{q,\bar q\}_1;L_1,L')$ offers the $SL(2)$
and $\{\tilde G\}$ symmetries while $P(\{q\}_1;L_1)$ is the
function of $\{q\}_1$ in (\ref{asr}) or in (\ref{asj}), or
products constructed using the functions above.
As it has been noted,
$P=1$ or $P'=1$ is assigned to the
leading term for the corresponding movers. In a like fashion,
the term for the first sum on the right side of (\ref{jint})
is given by
\begin{eqnarray}
O_j^{(n_1)}(DJ,\overline{DJ};P,\bar
P';\{q,\bar q\}_1)=
\sum_{L_1,L_1'}\sum_{r,s}
B_{L_1,L_1'}^{(n_1)}(t_j,\bar t_j;\{q,\bar q\}_1)
D(t_j)J_r^{(n_1)}(t_j;\{q\}_1;L_1)
\nonumber\\
\times
\tilde A_{P,P'}^{(r,s)}(\{q,\bar q\}_1;L_1,L')
P(t_j;\{q\}_1;L_1)
\overline{D(t_j)J_s^{(n_1)}(t_j;\{q\}_1;L_1')
P'(t_j;\{q\}_1;L_1')}
\label{corj1}
\end{eqnarray}
where $\tilde A_{P,P'}^{(r,s)}(\{q,\bar q\}_1;L_1,L')$ has the
$SL(2)$ and $\{\tilde G\}$ symmetries.  For the second sum
one obtains
\begin{eqnarray}
\tilde O_j^{(n_1)}( P,\bar P';\{q,\bar
q\}_1)= \sum_{L_1,L_1'} P(t_j;\{q\}_1;L_1) \hat A_{P,P'}^{(j)}
(\{q,\bar q\}_1;L_1,L')
\nonumber\\
\times
\overline{
P'(t_j;\{q\}_1;L_1')}
B_{L_1,L_1'}^{(n_1)}(t_j,\bar t_j;\{q,\bar q\}_1)
\label{corj2}
\end{eqnarray}
where $\hat A_{P,P'}^{(j)}(\{q,\bar q\}_1;L_1,L')$ possesses the
$SL(2)$ and $\{\tilde G\}$ symmetries.
The $P(t_j;\{q\}_1;L_1)$ function in (\ref{corj1}) and in
(\ref{corj2})  is the one of $P(\{q\}_1;L_1)$ in (\ref{vaccor}),
or one of functions (\ref{etild}), or it is a certain
product of the functions above. For $P$ and
$P'$ in (\ref{cont}) and (\ref{jint}) we use the same symbol as
for the corresponding function on the right side of
(\ref{vaccor}), (\ref{corj1}) or of (\ref{corj2}). As an example,
$O_0^{(n_1)}(\hat S_r,\bar P';\{q,\bar q\}_1)$ is given by
(\ref{vaccor}) for
$P(\{q\}_1;L_1)=\hat S_r^{(n_1)}(\{q\}_1;L_1)$, which is defined
in (\ref{asj}).  The term with $P=\Xi$ in (\ref{jint}) is given
by (\ref{corj1}) for $P(t_j;\{q\}_1;L_1)=
\tilde\Xi_{L_1,L_1'}^{(n_1)}(t,\overline t;\{q,\bar
q\}_1)$ where the right side is defined in (\ref{corhx}). The
term with $P=DX$ is given by (\ref{corj2}) for
$P(t_j;\{q\}_1;L_1)=D(t_j)
X_{L_1,L_1'}^{(n_1)}(t,\overline t;\{q,\bar
q\}_1)$ where $X$ is defined in (\ref{corhx}). And so on.
For the following, below we collect terms due to
holomorphic and anti-holomorphic corrections each are no less
than $\sim\rho_1$ in respect to the leading term. Since $P'$
repeats $P$, only $P$ are listed.

The sum in (\ref{cont}) and the first sum in (\ref{jint})
include $P=1$, $P=\hat S_r$, $P=\hat\Sigma_r$ and
$P=\hat\Sigma_r\Sigma_s$ where $\Sigma_r$ and $S_r$ are defined
by (\ref{asj}). These terms are due to corrections for the
period matrix (\ref{omas}) and corrections for
the correlator (\ref{corr}) when $z$ and $z'$ both are not
nearby $z_0$.  In (\ref{jint}) the discussed terms are due to,
in addition, by the "dots" terms in (\ref{corl}).  Evidently,
$O_0^{(n_1)}(\hat\Sigma_r,\overline P';\{q,\bar q\}_1)$ in
(\ref{cont}) being odd function of $\{\mu,\nu\}$, is nullified
after the integration over the Grassmann variables.  The first
sum in (\ref{jint}) contains also terms with $P=X$ and $P=\Xi$
due to corrections in (\ref{corl}). Also, it includes the
by-linear term with $P=\hat\Sigma_r\Xi$.  Generally, in
(\ref{jint}) we do not assume the integration over
$\vartheta_j$, but for the estimation, it is convenient to
assign the $\sim\sqrt{\rho_1}$ smallness also to $\vartheta_j$.
Finally a proportional to $\vartheta_j$ term must be multiplied
by $\sim 1/\sqrt{\rho_1}$.  From Section 5, the correction for
the pairing (\ref{ngcor}) at $t=t_j$ is $\sim D(t_j) {\cal
X}_{L_1,L_1'}^{(n_1)}(t,\overline t;\vartheta_1;\{q,\bar q\}_1)$
times $\overline{D(t_j){\cal X}_{L_1,L_1'}^{(n_1)}(t,\overline
t;\vartheta_2;\{q,\bar q\}_1)}$, see eq.(\ref{corl}). Hence $P$
for the second sum in (\ref{jint}) runs $P=DX$, $P=D\Xi$,
$P=\hat\Sigma_rD\Xi$ and $P=\Xi D\Xi$. In this case $DX$ and
$D\Xi$ each denote the super-derivative in respect to $t_j$ of
the corresponding function.  All the rest holomorphic
(anti-holomorphic) corrections are less than $\sim\rho_1$.

For $n_1=1$ the singularity of (\ref{cont}) at $\tilde
v_1=v_1-u_1=0$ is canceled locally due to the summation over the
spin structures. Indeed, the terms  with $P=1$ vanish since,
from (\ref{torus}), each a term is proportional to the torus
partition function, the sum over the torus partition functions
being nullified \cite{gsw}.  Terms with $P=\hat S_1$ and
$P=\hat\Sigma_1$ disappear for the same reason since the
genus-1 scalar function (\ref{jr1}) is independent of the spin
structure. As far as  the singularity is canceled separately for
the right movers and for the left ones, spin dependent
corrections proportional to the coefficients in (\ref{asr}) are
not singular. For the integrals over $(\mu_1,\nu_1)$ with
$w_1=v_1-u_1-\nu_1\mu_1$ to be given, the leading singularity is
canceled already for every spin structure since the leading
terms do not depend on $(\mu_1,\nu_1)$, but the cancellation of
the non-leading singularity occurs only for the sum over the
spin structures.

For the $j$-th configuration one can define boundary
(\ref{zbound}) of the fundamental region on the complex $z_j$
plane using the "circles" (\ref{hcirc}). In this case
$\ell_s(t)$ is
the same for all spin structures. Then the spin structure
independent terms are nullified locally due to the summation
over the spin structures. If
one uses "circles" (\ref{hcirc1}) dependent on the
spin structure, the spin structure independent terms are
nullified after the
integration over module variables (but $\tilde
v_1$).
Corrections due to the coefficients
in (\ref{asr}) are too small to originate singular terms.
Further spin structure dependent terms are due to those
corrections for the vacuum correlator (\ref{corl}), which
include the $\tilde E_{L_1}^{(1)}(t;\{q\}_1;\vartheta')$
function (\ref{asymr}), see eq.(\ref{corhx}). For $n_1=1$ the
above function is given by (\ref{assgr}) of Appendix B. From
(\ref{assgr}), the spin structure dependent part of  is
$\sim(\vartheta-\varepsilon(z))$.  At the same time, from
(\ref{jr1}), it follows that
$(\vartheta-\varepsilon(z))D(t)J(t)=0$. So the first sum on the
right side of (\ref{jint}) is non-singular.
From (\ref{assgr}), the singularity of the
second sum disappears for every spin structure after the
integration over the Grassmann variables (with an exception of
any one from them) and after the following integration over
either $u_1$, or over $z_j$. The integral over $u_1$ can be
replaced by the integral over $(z_j-u_1)$ since the integrand
depends on $u_1$ solely through $(u_1-z_j)$.

Instead of $\tilde v_1$, one can use the
$w_1=(v_1-u_1-\nu_1\mu_1)$ variable. To obtain
(\ref{assgr}) in the $(u,w,z)$ variables, the partial derivative
$(\partial_w)_{(u,z)}$ when $(u,z)$ are fixed, is
calculated through $(\partial_z)_{(u,w)}$
with $(u,w)$ are fixed. For this purpose one can
use the
invariance of (\ref{assgr}) under the special $SL(2)$
transformations
\begin{equation} z=\tilde
z+\tilde\vartheta\vartheta_0\,,\quad \vartheta=
\tilde\vartheta-\vartheta_0
\label{trco}
\end{equation}
where $\vartheta_0$ is a parameter common for all the variables.
The above invariance of (\ref{assgr}) follows from the
invariance under (\ref{trco}) of Green function (\ref{zgrin}).
In this case functions (\ref{cfnt}), which determine the
corrections, are found to be
\begin{eqnarray}
\tilde\Sigma^{(1)}(t;\{q\}_1;L_1)= (\vartheta-\varepsilon(z))
\hat W_1-
w^{-1}\biggl[\hat W_b(\mu-\nu)
-\mu\nu\vartheta\partial_z[(z-u)\hat W_1
+\hat W_b]\biggl]\,,
\nonumber\\
\tilde S^{(1)}(t;\{q\}_1;L_1)=\tilde S_{inv}^{(1)}
(t;\{q\}_1;L_1)
-\tilde\Sigma^{(1)}(t;L_1)\vartheta\,,
\nonumber\\
\tilde S_{inv}^{(1)}(t;L_1)=\hat W_b
-w^{-1}\biggl[[(z-u)\hat W_1-\hat W_2
\hat W_b](\mu-\nu)-
\varepsilon(z)\partial_z \hat W_1\biggl](\vartheta-\mu)
\label{exmpl}
\end{eqnarray}
with $\{q\}_1=(u,w,k)$,
$\hat W_1\equiv W_1(z,u,w;L_1)$, $\hat W_2\equiv W_2(z,u,w;L_1)$
and $\hat W_b\equiv W_b(z,u,w)$. So $\hat W_b$ does not depend
on $L_1$.  Both $ \tilde\Sigma^{(1)}(t;\{q\}_1;L_1)$ and $\tilde
S_{inv}^{(1)}(t;\{q\}_1;L_1)$ are invariant under (\ref{trco}).
For the second sum in (\ref{jint}) the singularity at $w_1=0$
disappears for every spin structure after the integration over
$u_1$, $\mu_1$ and $\nu_1$ (if we fix $z_j$ and $\vartheta_j$).
Really (\ref{exmpl}) is required only to verify the vanishing of
the term proportional to
$D(t_j)\tilde\Sigma^{(1)}(t;\{q\}_1;L_1)$.  The rest terms have
fermi statistics and, in addition, they are invariant under
(\ref{trco}) along with step factor (\ref{zbound}).
After the integration over
$\mu_1$ and $\nu_1$ the integrand appears to be the derivative
in $u_1$ of the local function (see Appendix D). Thus the
singularity disappears after the integration over $u_1$.
(Generically, the cancellation of the non-leading singularity in
$w$ forces the cancelation of the singularity in $\tilde v$ and
vice versa, see the next Section).  The
leading singularity of the first sum in (\ref{jint})
disappears for the same reasons while
the non-leading singularities disappear
only after the summation over the spin
structures.  One can define boundary (\ref{zbound}) of the
fundamental region using either the "circles" (\ref{hcirc}), or
(\ref{hcirc1}).  The cancellation of the singular terms
occurs for both cases.

In addition, the singularity of (\ref{jint}) for $n_1=1$
disappears after the integration over $t_j$, the summation over
the spin structures being performed. To see this one transforms
the integration variables by (\ref{tgam}) reducing $\mu_1$ and
$\nu_1$ to zeros. The above change of the variables is correct
since the integral is non-singular. The integral vanishes
locally in the super-Schottky group parameters due to the known
nullification of the 1- and 2-point genus-1 function \cite{gsw}.

\section{Finiteness of the multi-loop superstring amplitudes}

To clarify the calculation of the integrals
for
$n_1>1$, first we consider the $n_1=2$ configuration, the group
limiting points being $U_s=(u_s|\mu_s)$ and $V_s=(v_s|\nu_s)$.
Here $s=1$ and $s=2$ mark the first handle and the 2-nd
one. In the vacuum configuration
(\ref{cont}) we take $z_0=u_2$, while in
(\ref{jint}) we take $z_0=z_j$ assuming $t_j$ to be fixed.  As
before, the size of the configuration is restricted from top by
a cut-off $\Lambda<<\rho$ where $\rho$ is the characteristic
distance from $z_0$ to points associated with the $n_2$
configuration.  As has been discussed in the Introduction,  we
consider the integrals
of functions (\ref{vaccor}),
(\ref{corj1}) or (\ref{corj2}), every function being the sum
over the spin structures of the configuration discussed. The
integrals are taken over the group limiting points of the
configuration keeping $z_0$ and $\tilde v_2$ or
$w_2$ to be fixed.  Calculating the integral over the
variables of the 1-st handle, we first integrate over $u_1$,
$\mu_1$ and $\nu_1$.  Then the singularity at $\tilde v_1\to0$
disappears due to vanishing the integrals of the $n_1=1$
functions. Indeed, at $\tilde v_1\to0$, the integrand is given
by (\ref{cont}) or (\ref{jint}) with $n_1=n_2=1$ where
$Y_{P,\overline P'}^{(n_2=1)}$ is equal to the corresponding
expression among (\ref{vaccor}), (\ref{corj1}) and (\ref{corj2})
calculated for the handle "2". Therefore, the considered $n_1=2$
integrals are convergent. And they are functions of $\tilde v_2$
or $w_2$.  As it was announced in the Introduction, it will be
 shown that the integrals have no the singularity at $\tilde
v_2=0$ or $w_2=0$.  Thus the contribution to the amplitude from
the considered configuration is finite since it is just given by
the additional integration over $\tilde v_2=0$ (or $w_2=0$) of
the integrals considered.  Also we show the vanishing of the 0,-
1-, 2- and 3-point genus-2 amplitudes.

For the $j$-th configuration
we perform $u_2$-boost of the integration variables.  Then the
integral over $u_2$ with fixed $t_j$ is turned to the integral
over $z_j$ with fixed $u_2=0$ and fixed $\vartheta_j$, as well.
So we shall consider the integral with fixed $u_2=0$ and
$\vartheta_j$, the integration over $z_j$ being performed.
Often, as it explained below, we shall transform these
integrals into the integrals with $\vartheta_j$ to be the
integration variable.

By the reasons of the previous paragraph, the
integral with fixed $\tilde v_2$ (or $w_2$)
of the sum over the spin
structures of the first handle
is convergent even without the summation over the spin
structures of the second handle. Moreover, due to the
nullification of the integrals of the $Y_{P,\overline
P'}^{(n_2=1)}$ functions associated with the 2-nd handle, the
integral of the sum over the spin structures of the second
handle is convergent  for the particular spin structure of the
first handle.  Hence the integral of either of the partial spin
structure sums discussed is convergent, as it has been announced
in the Introduction.  By the above reasons, the integral with
fixed both $U_2=(u_2|\mu_2)$ and $V_2=(v_2|\nu_2)$ of the sum
over the spin structures of the 1-st handle is convergent, as
well. If, additionally, the integration over
$t_j=(z_j|\vartheta_j)$ is performed, the integral with fixed
$U_2$ and $V_2$ of the sum over the spin structures of the 2-nd
one is convergent, too.  Indeed, in this case the divergence at
$\tilde v_1=0$ (or $w_1=0$) is canceled due to the nullification
of the 1- and 2-point genus-1 function (see the end of Section
6).  The integrals of the discussed partial spin structure sums
both to be convergent, the integral of the total spin structure
sum admits changes of the integration variables by the spin
structure dependent transformations from the group of its local
symmetries can be performed (see a discussion of this point in
the Introduction).

First we  consider the
integrals of terms in (\ref{vaccor}) and (\ref{corj1}) with
$P=1$. For convenience, these terms in (\ref{jint}) can
be additionally integrated over $\vartheta_j$.  Indeed, solely
the $\sim\vartheta_j$ piece of the integrand contributes to the
integral while the remaining part disappears due to the
integration over the Grassmann module variables.
By
aforesaid, the integral of the  sum over the
spin structures of either of two handles is convergent.
Furthermore, the $P=1$ terms in both (\ref{cont}) and
(\ref{jint}) are invariant under transformations (\ref{trco}) of
the holomorphic variables (the anti-holomorphic ones may be
unchanged). Thus for $w_2=v_2-u_2-\nu_2\mu_2$ to be fixed
and once the integration over the holomorphic Grassmann
variables to be performed, the integrand is transformed to a sum
of derivatives in respect to the boson integration variables,
see (\ref{prp}) in the Appendix D.  Hence the corresponding
integral is nullified apart from the non-singular at $w_2\to0$
terms originated by the $\Lambda$ cut-off.  One could also
change the integration variables but $\mu_2$ by
transformation (\ref{trco}) with $\vartheta_0=\mu_2$ (it
remains the same both $w_2$ and $u_2$). Then $\mu_2$ is removed
from the integrand (with an exception of the $\Lambda$ cut-off
boundary).  In this case the singularity at $w_2=0$ vanishes
after the integration over $\mu_2$. Both calculations give the
same result since the integral is not singular. At the same
time, the singular integral of the single spin structure of
every handle is divergent or finite depending on the variables
used. In particular, the last integral appears to be the finite
once the above change the integration variables to be performed.
For the integral of the sum over
the spin structures of any handle the discussed ambiguity is
absent.

When $\tilde v_2$ is fixed instead of $w_2$
the integral of function (\ref{jint}) with $P=1$ considered is,
for convenience, again
additionally integrated over $\vartheta_j$.
To see for no to be the
singularity at $\tilde v_2=0$,
we represent the discussed integral  by the integral over
$u_1$, $\mu_2$ and $\nu_2$ of the integral
${\cal
A}_{1,P'}^{(2)}(u_1,u_2,\tilde v_2,\mu_2,\nu_2)$
calculated with fixed $u_1$, $u_2$,
$\tilde v_2$, $\mu_2$ and $\nu_2$. By reasons given in the
third paragraph of this Section, this integral is convergent
for the sum over the spin structures of either
of two handles.  The discussed singularity
at $\tilde v_2=0$ might appear solely due
to the region where $u_1$, $u_2$ and $v_2$ both are closely to
each other since the $n_1=1$ integral appearing when
$u_2\to v_2$ and $u_1\neq u_2$, has not
the singularity discussed. Hence we assume both
$u_1$ and $v_2$ to be fixed closely to $u_2$.  In this case we
remove the $\Lambda$ cut-off because, owing to (\ref{bunit}),
the integration $v_1$ variable is bounded inside a small region
nearby $u_2$. Then, due to the symmetry of the integrand under
$SL(2)$, $\{\tilde G\}_2$ and super-Schottky group
transformations,  ${\cal
A}_{1,P'}^{(2)}(u_1,u_2,\tilde v_2,\mu_2,\nu_2)$ is related with
its magnitude ${\cal A}_{1,P'}^{(2)}(u_1,u_2,\tilde v_2,0,0)$ at
$\mu_2=\nu_2=0$ by
(the proof is given in the next paragraph)
\begin{equation}
{\cal
A}_{1,P'}^{(2)}(u_1,u_2,\tilde v_2,\mu_2,\nu_2)=
\biggl(1-\mu_2\nu_2/\tilde v_2
\biggl){\cal
A}_{1,P'}^{(2)}(u_1,u_2,\tilde v_2,0,0)\,.
\label{linvc}
\end{equation}
For the same integral
$\tilde{{\cal
A}}_{1,P'}^{(2)}(u_1,u_2,w_2,\mu_2,\nu_2)$
to be considered as function of $w_2$
(instead of $\tilde v_2$) one obtains that
\begin{equation} {\tilde{\cal
A}_{1,P'}}^{(2)}(u_1,u_2,w_2,\mu_2,\nu_2)=
\biggl(1-\mu_2\nu_2/w_2
-\mu_2\nu_2
\partial_{w_2}
\biggl){\cal
A}_{1,P'}^{(2)}(u_1,u_2,w_2,0,0)\,,
\label{lnvc}
\end{equation}
Eq.(\ref{lnvc}) is obtained by
substituting $\tilde v_2=w_2-\mu_2\nu_2$ into (\ref{linvc}).
Integrating (\ref{lnvc}) over $u_1$, $\mu_2$ and $\nu_2$,
one sees that the absence
of the singularity at $w_2=0$ for the integral on
the left side (proving in the previous paragraph) forces the
vanishing at $w_2=0$ of the integral  of ${\cal
A}_{1,P'}^{(2)}(u_1,u_2,w_2,0,0)$ over $u_1$.
Besides, from (\ref{linvc}) and
(\ref{lnvc}), the singularity at $\tilde v_2=0$  disappears
for the integral of ${\cal
A}_{1,P'}^{(2)}(u_1,u_2,\tilde v_2,\mu_2,\nu_2)$
over $u_1$, $\mu_2$ and $\nu_2$, as
it is required. Relation
(\ref{linvc}) will be proved for the integral of the total spin
structure sum. Hence the discussed cancellation of the
singularity in $\tilde v_2$ is shown only for the integrals
of the sum over all the spin structures of the configuration
considered.

To prove (\ref{linvc}), we
represent the integrand ${\cal O}
(1,\bar P';\{q,\bar q\}_1)$ of the discussed integral
as $[{\cal O}(1,\bar P';\{q,\bar
q\})H(U_2,V_2,U_1)]$ times $H^{-1}(U_2,V_2,U_1)$ where
$H(U_2,V_2,U_1)$ is the factor (\ref{factorg}) for
$t_1^{(0)}=U_2$, $t_2^{(0)}=V_2$ and $t_3^{(0)}=U_1$. As before,
$U_2=(u_2|\mu_2)$, $V_2=(v_2|\nu_2)$ and $U_1=(u_1|\mu_1)$.
Then we  change the holomorphic
integration variables by the $SL(2)$ transformation (\ref{trnsf})
(see Appendix E), which
reduces $\mu_2$ and
$\nu_2$ to zeros, but it does not change both $u_1$, $u_1$ and
$v_2$.
Since the expression inside the square brackets is
$SL(2)$ co-variant, the transformation remains it the same, but
for $\mu_2=\nu_2=0$.  Moreover, we obtain that
$H^{-1}(U_2,V_2,U_1)$ in front of the square brackets receives
only the $[1+\mu_2\nu_2/(u_2-v_2)]$ multiplier.  The $\sim\mu_1$
terms in $H^{-1}(U_2,V_2,U_1)$ does not contribute to the
integral since $[H(U_2,V_2,U_1) {\cal O}_0^{(2)}(1;\bar
P';\{q,\bar q\})]$ at $\mu_2=\nu_2=0$ is even function of
$(\mu_1,\nu_1)$ (after the integration over $\vartheta_j$
the $j$-th configuration). The integration region
boundary (\ref{zbound}) and (\ref{bunit}) is non-invariant
under the transformation that originates additional boundary
integrals.  They, however, cancel each other
since, on equal terms, the integration region can be bounded by
(\ref{zbound}) and (\ref{bunit}) taken as for the former
variables, so for the resulted ones. Indeed, each of the
integration regions is the fundamental region of the local
symmetry group of the integral. Hence the integral is the same
in both cases.  As the result, (\ref{linvc}) appears.  Really
the boundary integrals are canceled to be reduced to each other
by the change of the integration variables by means spin
dependent transformations of the $\{\tilde G\}_2$ group and of
the super-Schottky group (for the $j$-configuration). As it has
been discussed, these transformations can be
surely performed only for the integrals of the whole spin
structure sum. Hence relation (\ref{linvc}) is established only
for the integral of the total sum over the spin structures of
the discussed configuration.

The  ${\cal
A}_{1,1}^{(2)}(u_1,u_2,\tilde v_2,0,0)$ integral assigned to
function (\ref{vaccor}) (for $P=P'=1$) times
$|(u_1-u_2)(u_1-v_2)|^2$, is just the genus-2 vacuum amplitude,
while the corresponding integral of (\ref{corj1}) is the
1-point,
genus-2 one. Indeed, every discussed amplitude
is given by an integral (\ref{ampl})
over
$\mu_1$, $\nu_1$ and $v_1$ and their complex conjugated with
$u_1$,
$u_2$, $v_2$, $\mu_2$ and $\nu_2$ to be fixed.
The integral does not depend on the fixed  variables due to
$SL(2)$, $\tilde G$ and super-Schottky group symmetries.  Taking
$\mu_2=\nu_2=0$, one just obtains ${\cal
A}_{1,1}^{(2)}(u_1,u_2,\tilde v_2,0,0)$ required.  Calculating
its limit at $v_2\to u_2$ under the integral sign, one finds
that ${\cal A}_{1,1}^{(2)}(u_1,u_2,\tilde v_2=0,0,0)$ is
nullified owing to the above discussed properties of the $n_1=1$
functions.  Indeed, at $v_2\to u_2$ it is expressed through the
$n_1=1$ integrals (see the first paragraph of this Section).
The limit under the integral sign is correct since
(see Section 4) at $v_2\to u_2$ both $u_2$ and $v_2$ lay
exterior to the Schottky circles assigned to the handle "1".
Thus the integral vanishes identically since
it is independent
of $\tilde v_2$.  So
the 0- and 1-point amplitudes both are nullified.  As it has
been discussed above, the independence of the integral on the
fixed variables implies a possibility to perform
spin dependent transformations
$\{\tilde G\}_2$ and, for the 1-point amplitude, the
super-Schottky group transformations of the interaction
vertex coordinate. These
transformations can be surely performed only for
the integrals of the whole spin structure sum. Hence the
nullification is shown only for the integral of the
whole sum over the spin structures. To directly verify the
vanishing of the discussed amplitudes $v_2=u_2$ for
arbitrary $(\mu_2,\nu_2)$, it requires the consideration of
corrections $\sim |\tilde v_2|^2$ for each one of movers, which
we did not performed.

The integrals of terms with $P=1$ being non-singular at
$w_2=0$ (or $\tilde v_2=0$), corrections might be
singular when both $P\neq1$ and $P'\neq1$.
In this case only corrections
listed in the previous Sections need to be examined.
For
$\tilde v_2$ to be fixed, the integral differs
from the corresponding integral with fixed $w_2$ by the
additional term $\mu_2\nu_2\partial_{\tilde v_2}{\cal
O}_0^{(2)}(P, \overline P' ;\{q,\bar q\})$. By the dimensional
reasons, the singular part of the integral of ${\cal
O}_0^{(2)}(P, \overline P' ;\{q,\bar q\})$ at $\mu_2=\nu_2=0$
(when $P\neq1$ and $P'\neq1$) is $(\overline
u_2-\overline v_2)^{-1}$ times a non-singular factor. This
singularity is week to give the divergence in the amplitude. So
it is sufficient to verify the absence of the singularity
for only
one of $w_2$ and $\tilde v_2$ to be fixed.

The terms of (\ref{jint}) with $P=\hat\Sigma_r\hat\Sigma$,
$P=\hat\Sigma_r\tilde\Xi$,
$P=\hat\Sigma D\Xi$ and $P=\tilde\Xi D\Xi$
(for definitions, see the text just
below eq.(\ref{corj2}))
can be additionally integrated over $\vartheta_j$ since,
like the terms with $P=1$,
the remaining part disappears due to the
integration over the Grassmann module variables.    The
integrals are
invariant under (\ref{trco}). Indeed,
due to the invariance under (\ref{trco}) of $\tilde
R_L^{(n)}(t,t';\{q\})$, both
$\tilde\Xi_{L_1,L_1'}^{(n_1)}(t_j,\overline
t_j;\{q,\bar q\}_1)$  and
$D(t_j)\Xi_{L_1,L_1'}^{(n_1)}(t_j,\overline
t_j;\{q,\bar q\}_1)$ (see (\ref{corhx})) are invariant under
(\ref{trco}) that, in turn, provides the discussed invariance of
the considered terms.  Thus the singularity at $w_2=0$ is
nullified for the integral of the sum over the spin structures
of any one of the two handles, just as for the $P=1$ terms.
So, by the previous paragraph, the singularity at $\tilde v_2$
disappears, too.  Only the integrals (see
the previous Section) linear in functions
(\ref{asj}) and (\ref{sgras})
need to be more examined.

In doing so, for convenience we again transform the
integrals of functions (\ref{jint}) with fixed $\vartheta_j$
to integrals where $\vartheta_j$ is the integration variable.
For $P=X$, $P=DX$ and $P=\hat S_r$ we can
additionally integrate over $\vartheta_j$ as far
as the $\sim\mu_1\nu_1\mu_2\nu_2$ term of the integrand includes
$\vartheta_j$, as well. With $\vartheta_j$ no to be
fixed, the terms with $P=\hat\Sigma_r$, $P=D\Xi$
and $P=\Xi$ disappear after the integration over Grassmann
variables as far as there is no a piece proportional to all the
Grassmann variables including $\vartheta_j$.  When $\vartheta_j$
is fixed (this is just assumed), we reduce $\mu_2$ to zero by
transformation (\ref{trco}) with $\vartheta_0=\mu_2$.  In this
case we omit the change of the $\Lambda$ cut-off boundary
since it does not originate the singularity at $w_2=0$.  When
$\tilde v_2$ is fixed instead of $w_2$, additional terms
appear due to this transformation, but they can be omitted.
Indeed, by aforesaid, they are non-singular in $\tilde v_2$.
Hence the integrals of terms with  $P=\hat\Sigma_r$, $P=D\Xi$ or
$P=\Xi$ are turned into the integrals over $\vartheta_j$ with
$\mu_2=0$ to be fixed. The integral of the $\vartheta_j$ term in
$\Xi$ (defined by (\ref{corhx})) is nullified being proportional
to a vanishing integral of a term with $P=1$.

For  definiteness, we discuss the integrals with $\tilde
v_2$ to be fixed. As above, $u_2$ is fixed, too (see the second
paragraph of this Section). Really in this
case the third  fixed point being $\infty$, presents.
Indeed, the integrand contains $P$ to be the asymptotics
of the corresponding function. We shall prove
that  every considered integral ${\cal
A}_{P,P'}(u_2,\tilde v_2,\infty)$ is related to an integral
$\tilde{{\cal A}}(u_2,\tilde v_2,u_1^{(0)},0,0; \tilde P,\tilde
P')$ with the fixed $u_1=u_1^{(0)}$, $u_2$, $\tilde v_2$ and
$\mu_2=\nu_2=0$ as it follows
\begin{equation}
{\cal
A}_{P,P'}(u_2,\tilde v_2,\infty)=\frac{1}{|\tilde v_2|^2}
\tilde{{\cal A}}(u_2,\tilde v_2,u_1^{(0)},0,0; \tilde P,\tilde
P')|(u_1^{(0)}-u_2)(u_1^{(0)}-v_2)|^2 +\dots
\label{rel}
\end{equation}
where the "dots" denote terms, which are non-singular
in $\tilde v_2$ at $\tilde v_2=0$. The integrand for
$\tilde{{\cal A}}(u_2,\tilde v_2,u_{1^(0)},0,0;
\tilde P,\tilde P')$ contains
$D(t_0')P(t_0')$ to be
the spinor derivative (\ref{supder}) of the
$P(t_0')$ function
whose asymptotics at
$z_0'\to\infty$ is proportional to the $P$ function in
(\ref{vaccor}), (\ref{corj1}) or (\ref{corj2}).
In the above integral the integration
over $t_0'$ is implied. It will be shown that
$\tilde{{\cal A}}(u_2,\tilde v_2,u_{1^(0)},0,0;
\tilde P,\tilde P')$ is zero at $\tilde v_2=0$
due to vanishing of the $n_1=1$ integrals discussed in Section
6. Hence
the singularity
at $\tilde v_2$ disappears for the considered integral ${\cal
A}_{P,P'}(u_2,\tilde v_2,\infty)$.  Thus
(see the Introduction and the beginning of this
Section) the contribution to the amplitude from the considered
configuration is finite.

Eq.(\ref{rel}) is derived with using a change
of the integration variables by the relevant $SL(2)$
transformation.  In doing so the
integration region (\ref{zbound}) and (\ref{bunit}) is changed.
To reduce it to the former region, the spin structure dependent
transformations of the $\{\tilde G\}_2$ group and of the
super-Schottky group transformations of $t_j$ (for the $j$-th
configuration) are necessary.  As it has been discussed above,
these spin structure dependent transformations can be surely
performed only for the integrals of the total sum over the spin
structures.  Hence eq.(\ref{rel}) is
established only for the integrals of the total sum over the
spin structures of the $n_1=2$ configuration discussed.

To  derive eq.(\ref{rel}) we consider
the integral
${\cal
A}(u_2,\tilde v_2;\tilde P(t_0'),
\tilde P'(t_0'))$, which is obtained by
the replacement $P\to D(t_0')P(t_0')$ and
$P'\to D(t_0')P'(t_0')$ in
${\cal A}_{P,P'}(u_2,\tilde v_2,\infty)$.
Here $D(t_0')P(t_0')$ is the spinor derivative of a
relevant function $P(t_0')$ proportional to $P$
at $z_0'\to\infty$.
In particular,  the integrals of terms with $P=\hat
S_r$ and $P=\hat\Sigma_r$ are calculated from the integral
of term with $P(t_0')=D(t_0')J_r(t_0')$ obtained by the
replacing of $P$ in (\ref{vaccor}) and
(\ref{corj1}) by $D(t_0')J_r^{(2)}(t_0';\{q\};L_1)$.  The
integrals of $P=X$ and $P=\Xi$  are calculated from the integral
of terms with $P(t_0')=D(t_0') {\cal X}(t_0')$. In this case the
corresponding $P$ in (\ref{corj1}) and (\ref{corj2}) are
replaced by $D(t_0')\hat X_{L_1,L_1'}^{(2)}(t_j,\overline
t_j;t_0',\bar t_0';\{q,\bar q\}_1)$ where the vacuum correlator
$\hat X_{L_1,L_1'}^{(2)}(t_j,\overline t_j;t_0',\bar
t_0';\{q,\bar q\}_1)$ is given by (\ref{corr}).
Correspondingly, the integrals of terms with $P=DX$ and $P=D\Xi$
are calculated from the integral containing
$P(t_0')=DD(t_0'){\cal X}(t_0')$. In doing so the
integrals of terms with $P=\hat S_r$, $P=X$ and $P=DX$ are
calculated as the $z_0'\to\infty$ limit of the corresponding
integral multiplied by $(z_0'-u_2)(z_0'-v_2)$, the integration
over $\vartheta_0'$ being performed.  And the integral of terms
with $P=\hat\Sigma_r$, $P=\tilde\Xi$ or $P=D\Xi$ is the
$z_0'\to\infty$ limit of the corresponding integral with
$\mu_2=\vartheta_0'=0$
multiplied by $(z_0'-v_2)$.
We want to relate the above
integrals to the
integrals with $u_1$ to be fixed instead of $z_0'$.
AS the first step,
we  calculate the
considered integrals in terms of the
integrals with fixed $\mu_2=\nu_2=0$ and with the same
$u_2$, $v_1$ and $z_0'$ (in the last integrals the
integration over $\vartheta_0'$ is performed).
Then we use the relevant $L(2)$
transformation $g(z)$ to fix $u_1$ instead of $z_0'$.
To do the first step above,
we represent ${\cal
A}(u_2,\tilde v_2;\tilde P(t_0'),
\tilde P'(t_0'))$
by the integral over
either $(\mu_2,\nu_2)$, or
$\nu_2$ (with $\mu_2=0$) of the
${\cal
A}(u_2,\tilde v_2,\mu_2,\nu_2;\tilde P(t_0'),
\tilde P'(t_0'))$ integral with
$(\mu_2,\nu_2)$  to be fixed.
By the third paragraph of this Section, all the
integrals of the sum over the spin structures either of
the handles are convergent. Once we calculate the
$(\mu_2,\nu_2)$ dependence of the integrals, the desired ${\cal
A}(u_2,\tilde v_2;\tilde P(t_0'), \tilde P'(t_0'))$ integrals
appear to be given though the ${\cal A}(u_2,\tilde
v_2,\mu_2,\nu_2;\tilde P(t_0'), \tilde P'(t_0'))$ integral with
$\mu_2=\nu_2=0$, as it is required.  To calculate the
$(\mu_2,\nu_2)$ dependence of the ${\cal A}(u_2,\tilde
v_2,\mu_2,\nu_2;\tilde P(t_0'), \tilde P'(t_0'))$ integral, we
change the integration variables by transformation (\ref{trnsf})
(see Appendix E), which does not change $u_2$, $v_1$ and $z_0'$.
In doing so we proceed like the deriving of
(\ref{lnvc}).

The considered transformation (\ref{trnsf})
to be performed for
the term including $P(t_0')=D(t_0')\hat
X_{L_1,L_1'}^{(2)}(t_j,\overline t_j;t_0',\bar t_0';\{q,\bar
q\}_1)$, the integral receives the add since the vacuum
correlator (\ref{corr}) is, generally, not invariant under
$SL(2)$ transformations. Really, the above correlator receives
two additional terms, every term being dependent on only one of
the points. For arbitrary Grassmann parameters these terms are
rather tremendous. To avoid the direct calculation of the
additional terms, we replaces the integral by the integral of
the difference
\begin{eqnarray}
\Delta_{L_1,L_1'} (t_j,\overline
t_j;t_0',\bar t_0';\{q,\bar q\}_1)= D(t_0'){\cal
X}_{L_1,L_1'}^{(2)}(t_j,\overline t_j;t_0';\{q,\bar q\}_1)
\nonumber\\
-\biggl[D(t_0'){\cal X}_{L_1,L_1'}^{(2)}(t_j,\overline t_j;
t_0';\{q,\bar q\}_1)\biggl]_{t_j=t_0',\bar t_j=\bar t_0'}
\label{difx}
\end{eqnarray}
where the correlator at the same point is defined as
usually, see Section 2. Indeed, being at $\tilde z\to\infty$
smaller than $\sim\rho_1$, the last term of the difference does
not originate the singularity at $\tilde v_2=0$ in the integral.
The add to (\ref{difx}) under the $SL(2)$ change is due to only
the singular term due to is absent for the correlator at the
same points.  So the add to (\ref{difx}) is $D(t_0')\ln Q(t_0')$
where $Q(t_0')$ is given by (\ref{supder}) for the
transformation (\ref{trnsf}) considered.  Being independent of
the module variables, this addition  term originates the $P=1$
integral, which vanishes.  Finally, the correlator at the same
point $t_0'$ can be omitted since it is not contribute to the
singular term.  Hence all the specifics due to the
non-invariance of the vacuum correlator can be neglected.

As before, the $\Lambda$ cut-off boundary does
not contributes to the singularity due to the vanishing of the
integrals of the $n_1=1$ functions. Thus the singular part of
the integral associated with $P=\hat S_r$, $P=X$ or $P=DX$ is,
like the $P=1$ case, related by (\ref{lnvc}) with the
corresponding integral at $\mu_2=\nu_2=0$. To calculate
${\cal
A}(u_2,\tilde v_2\tilde P(t_0'),
\tilde P'(t_0'))$ desired, one integrates the above relation
over $(\mu_2,\nu_2)$. Thus ${\cal
A}(u_2,\tilde v_2\tilde P(t_0'),
\tilde P'(t_0'))$  appears to be
${\cal A}(u_2,\tilde v_2,0,0;\tilde P(t_0'),
\tilde P'(t_0'))$ times
$1/|v_2|^2$.
For the integrals with
$\vartheta_0'=\mu_2=0$ we reduce $\nu_2$ to zero and
$\vartheta_0'$ to $\vartheta'$. In this case the integration
over $\nu_2$ is replaced by the integration over $\vartheta'$
and,  simultaneously, the integral receives  factor
$(z_0'-u_2)/(u_2-v_2)$. In any case the singular at
$v_2=0$ part of the desired ${\cal A}_{P,P'}(u_2,\tilde
v_2,\infty)$ integral in (\ref{rel}) is the
limit at $z_0'\to\infty$ of the
integral over
$\vartheta'$  of
${\cal A}(u_2,\tilde v_2,0,0;\tilde P(t_0'),
\tilde P'(t_0'))$ times
$|(z_0'-u_2)(z_0'-v_2)|^2/|v_2|^2$.
The $(z_0'-u_2)(z_0'-v_2)$ factor is just the factor
(\ref{factorg}) for fixed $z_0'$, $u_2$, $v_2$
and $\mu_2=\nu_2=0$.  Furthermore, we can remove the cut-off
and, simultaneously, to restrict the
integration region by the $B_0^{(2)}(t_0',\bar t_0';\{q,\bar
q\}_1)$ step factor (\ref{zbound}). Indeed, like the $P=1$ case,
the integral may be non-vanishing at $\tilde v_2=0$ solely due
to the region where $u_1$, $v_1$ and $v_2$ both go to $u_2$. The
resulted integral with the above constraint of the integration
region is convergent.  Truly, when either
$\tilde v_1\to0$, or  $u_1$ and $v_1$ both go to the infinity,
the singular part of the integrand disappears since it is
proportional to the vanishing integral of $n_1=1$ function
assigned to the handle "2". When $z_j\to z_0'$, the integral is
again convergent due to the vanishing of the $P=1$ integrals.
We fix
$u_1=u_1^{(0)}$ (nearby $u_2$)
by a relevant $L(2)$ transformation $g(z)$ of the integration
variables. Then $z_0'$ turns into $z'$, and the
integration over $u_1$ is replaced by the integration over $z'$.
Once the $L(2)$ transformation being performed, the
$(z_0'-u_2)(z_0'-v_2)$ factor in the integral is replaced by
$(u_1^{(0)}-u_2)(u_1^{(0)}-v_2)$, and eq.(\ref{rel}) appears.

Except of the
integral of the term with either $P'(t_0')$, or
$P(t_0')$ to be $D(t_0'){\cal X}(t_0')$, the integrand for the
right side integral in (\ref{rel}) is obtained by the $P\to
P(t_0')$ replacement in the integrand for the left side one.
Taking the $\tilde v_2\to0$ limit, one can see that the
right side integral vanishes at
$\tilde v_2=0$ due to the nullification of the integrals of the
$n_1=1$ function associated with $(u_2,v_2)$ limiting points.
The
integral is invariant under $SL(2)$ and $\tilde G$
transformations as well the super-Schottky transformations of
the interaction vertex coordinate. Hence it vanishes
identically in $\tilde v_2$.

The integral  of the term with
$P(t_0')=D(t_0'){\cal X}(t_0')$ is non-invariant under
the considered
transformation $g(z)=(az+b)/(cz+d)$. Truly, it receives the
additional term $-D(t')\hat
X_{L_1,L_1'}^{(2)}(t(\infty),\overline t(\infty);t',\bar
t';\{q,\bar q\}_1)$, which appears due to the corresponding
addition to the vacuum correlator, see Appendix E. In this case
$t'=(z'|\vartheta')$ and
$t(\infty)=(z(\infty)|\vartheta(\infty))$ with $z(\infty)=-d/c$
and $\vartheta(\infty)=0$. The discussed $g(z)$ transformation
has $u_2$ and $v_2$ to be the limiting points, and the
multiplier $k$ to be
$k=(z_0'-u_2)(z'-v_2)/(z_0'-v_2)(z'-u_2)$.  So $z(\infty)=
z'+(z'-u_2)(z'-v_2)/(z_0'-z')$.  The integration region
is changed under the $g(z)$ transformation. To calculate the
right side integral in (\ref{rel}) at $\tilde v_2=0$ by taking
the $\tilde
v_2=0$ limit under the integral sign, the integration region
needs to be is reduced to the (\ref{zbound}) and (\ref{bunit}).
In doing so $t(\infty)$ is changed.  Nevertheless, since the
addition term discussed is invariant under the super-Schottky
transformations of $t(\infty)$, one can replace it by its
super-Schottky group image laying interior to
region (\ref{zbound}). In this case the desired $\tilde v_2=0$
limit under the integral sign can be performed. Moreover, the
discussed additional term is not singular at $\tilde v_2\to0$.
Thus the integral is nullified at $\tilde v_2\to0$ due the
nullification of the $n_1=1$ function, just as the integrals
discussed above.  Since considered integral is not
$SL(2)$-invariant, it does not vanish identically, unlike the
integrals considered before. In any case, from (\ref{rel}) the
 desired integrals of (\ref{cont}) and of (\ref{jint})
have no the singularity at $\tilde v_2\to0$. So the
contribution to the amplitude of the discussed $n_1=2$
configuration is finite.

The nullification of the 2- and 3-point amplitudes is verified
as for the 0- and 1-point ones discussed
above. Every amplitude is given by the corresponding integral
(\ref{ampl}) of the 2- or 3-point function. The integral is
taken over $\mu_1$, $\nu_1$ and $v_1$ and their complex
conjugated, both $\mu_e$, $\nu_2$, $u_1$, $u_2$ and $v_2$
being fixed.  The integral of
the sum over the spin structures of either of two handles is
convergent for the given spin structure of the remaining handle.
In this case the integral of the spin structure sum of the 2-nd
handle is convergent due properties 1-, 2- and 3- point genus-1
functions.  The 3-point genus-1 function are
examined just as the $n_1=1$ functions in the previous
Section. Since the desired amplitude are independent of the
fixed parameters (the total sum over the spin structures is
implied), they can be calculated at $\mu_2=\nu_2=0$. It allows
to avoid a large number of the corrections
$\sim |\tilde v_2|^2$ in each one of movers. The amplitude is
nullified at $v_2\to u_2$ due to vanishing the integrals of
$n_1=1$ functions associated with the handle "2".
Being independent of the fixed variables, the considered
amplitude is zero for any $v_2$.

For general $n_1$, the cancellation of the divergences can be
derived by the mathematical induction.  Assuming
the $n_1'$ integrals to be convergent for all $n_1'<n_1$,
one verifies the cancellation of the singularity
at $\tilde v_{n_1}=0$ or $w_{n_1}=0$ for the integrals over
variables of the remaining $(n_1-1)$ handles.  In doing so the
consideration like  given for the $n_1=2$ case, is
performed.  In particular, it is verified   that the integrals
of the spin structures of the $(n_1-1)$ handles
are convergent for every spin structure of the $n_1$-th handle.
And the integrals of the sum over the spin
structures of the $n_1$-th handle are convergent for every spin
structure of the remaining $(n_1-1)$ handles.  As for $n_1=2$,
the integrals with $P(t_0')$
instead of $P$ are considered. Like the $n_1=2$ case, the total
cancellation of the singularity is verified for the whole sum
over the spin structures. Step-by-step, the
nullification of the 0-, 1, 2- and 3-point amplitudes is
verified.

Due to $L(2)$ symmetry, the cancellation of the divergences for
the finite $z_0$ forces the same for $z_0\to\infty$.
This case could also be considered like the finite $z_0$ case.
When the $\{N_0\}$ set in (\ref{ampl}) for
the $m>3$ point amplitude is formed by the limiting group
points, further singular configurations appear to be with  either
$(m-1)$, or $m$ vertices go to the same point. In this case the
leading approximated integrand is proportional either to the 1-
function or, respectively, to the 0-point one.  By aforesaid,
the above 0- and1-point integrals are
nullified along with the integrals due to the leading
corrections.  Hence the considered configurations originate no
divergences in the amplitude.

So, the divergences do not appear when one integrate,
step-by-step, the sum over the spin structures of the given handle.
over its limiting points.
For the
considered handle, the integration over the difference $\tilde
v$ or over the super-difference $w$ between the limiting points
is performed in the last turn.  In this case the $SL(2)$
symmetry is preserved  along with the $\{\tilde G\}$ symmetry
and with the symmetry under super-Schottky group changes of every
particular interaction vertex coordinate. These spin
structure dependent transformations can be performed
due to the convergence of the integrals
over
$n_1>1$ configurations
of the above discussed partial sums over the spin
structures.

The total group of the local symmetries of the amplitude
contains, in addition, the modular group.
Naively, the modular symmetry is provided due in this case to
terms of the sum over the spin structures in amplitude
(\ref{ampl}) are correctly transformed  into each other
\cite{dannph}. So the invariance under
the $\sqrt k_s\to-\sqrt k_s$ change ($k_s$ is the super-Schottky
group multiplier) is evident since the integration
variables are not touched. For non-zero Grassmann moduli, all
the other modular transformations (including the addition
$\pm2\pi$ to each of the remaining arguments in (\ref{argum}))
are, however, accompanied by the spin structure dependent change
\cite{dannph} of the integration variables.
The integral of the single spin structure
over the $n_1$ configurations being divergent,
both a possibility to perform the modular
transformations and the modular invariance
need an argumentation.
Like
$\{\tilde G\}$ and  super-Schottky group
transformations.
the modular transformations can be performed
due to the convergence of the
integrals of the above discussed partial sums over the spin
structures.  And the modular symmetry is preserved.
The argumentation is outlined below.

The transition functions for modular transformations depend on
the super-spin structure by terms proportional to Grassmann
super-Schottky group parameters \cite{dannph}.
The discussed terms are calculated \cite{dannph} from the set
of integral equations, the kernels being given through ghost
Green functions for zero module parameters.  The integration is
performed along contours where every contour rounds the Schottky
circles of the given handle together with the cut between the
circles. For the $n_1$ configuration the leading approximated
transition functions coincide with the transition functions of
the relevant modular transformation of the genus-$n_1$.
And the remaining $n_2=n-n_1$ configuration is changed by the
relevant transition functions of the modular genus-$n_2$
transformation.  The corrections for the transition functions
are due to corrections for the ghost Green functions, which are
calculated either using the expressions \cite{dannph}, or using
representation of the Green functions in terms of the genus-1
ones at zero Grassmann module parameters, see Appendix C of the
present paper. In the last case the method of Section 4 can be
employed. For the $n_2$ configuration, the spin structure
dependent corrections are no larger than
$\sim\rho_1^2$ for the Grassmann transition function and no
larger than $\sim\rho_1^3$ for the boson transition one,
$\rho_1\to0$ (for definitions, see the previous Section).
So, they are negligible.

For the $n_1=1$ case ($\tilde v\to0$)
the leading approximated transition functions are independent of
the spin structure.  The  spin structure dependent corrections
for the Grassmann transition function are not larger than
$\sim|\tilde v|^3$, and they are not more than $\sim|\tilde
v|^4$ for the boson transition one.  Before the modular
transformation to be performed, we can previously cut with below
the integration region $\tilde v\to0$ by some cut-off
$\tilde\rho$ and then to perform the required change of the
variables.  At $\tilde\rho\to0$ the additional terms due to spin
structure dependent corrections are nullified since the integral
of these terms is found to be convergent.  So the resulted
integral coincides with the former one.  For $n_1=2$ we, as
before, consider the integral of the given spin structure of the
first handle over its variables, $U_2$ and $V_2$ being fixed.
We transform the variables of the 1-st handle for the sum over
the spin structures of the second one.  Then we transform the
variables of the 2-nd handle for the integral of the sum over
the spin structures of the first one. Hence
for the
integral of the total sum over the spin structures of the
configuration the modular transformation can be performed, the
modular symmetry being preserved.  For $n_1>2$ the mathematical
induction is employed So all the local symmetries of the
amplitude are preserved.

\subsection*{Acknowledgments}

The research described in this publication was made
possible in part by
Award No. RP1-2108 of the U.S. Civilian Research and
Development Foundation
for the Independent States of the Former Soviet Union
(CRDF),
and in part by Grants No. 00-02-16691
and No. 00-15-96610  from
the Russian Fundamental Research
Foundation.

\def\thesection{Appendix \Alph{section}}
\def\theequation{\Alph{section}.\arabic{equation}}
\setcounter{equation}{0}

\appendix
\section{Integration region and the unitarity}
As an example, we consider the genus-2 forward scattering
amplitude. The vacuum expectation of the vertex product is
proportional to $\exp S$ where, being calculated for zero
Grassmann parameters, $S$  is given through the field vacuum
correlators (\ref{corr}) of the boson string \cite{divfl,dan89}.
The Schottky multipliers to go to zeros, the leading
approximated holomorphic Green function $R(z,z')$ is $\ln(z-z')$
while the scalar function $J_r^{(2)}(z;\{q\})$ is
$\ln[(z-u_r)/(z-v_r)]$. The period matrix is given by
(\ref{kzero}). Then for the tachyon-tachyon  forward scattering
amplitude in the boson string theory
\begin{eqnarray}
S=-\frac{s}{4}\left[\ln\left|\frac{(z_1-z_2)(z_3-z)}
{(z_1-z_3)(z_2-z_3)}\right|-\ln\left|
\frac{(z_1-u_j)(z_3-v_j)}
{(z_1-v_j)(z_3-u_j)}\right|
\frac{\hat\omega_{jl}}{\det\hat\omega}
\ln\left|\frac{(z_2-u_l)(z-v_l)}
{(z_2-v_l)(z-u_l)}\right|\right]
\nonumber\\
-2\ln\left|\frac{(z_1-u_j)(z-u_j)(z_2-v_j)
(z_3-v_j)}{(z_1-v_j)
(z-v_j)(z_2-u_j)(z_3-u_j)}\right|
\frac{\hat\omega_{jl}}{\det\hat\omega}
\ln\left|\frac{(z_1-u_l)(z-u_l)(z_2-v_l)
(z_3-v_l)}{(z_1-v_l)
(z-v_l)(z_2-u_l)(z_3-u_l)}\right|
\nonumber\\
-4\ln|(z_1-z_2)(z_3-z)|
\label{sapp}
\end{eqnarray}
where $s=-(p_1+p_2)^2=-(p_3+p_4)^2$ and $\hat\omega_{jl}$ is
given by
\begin{equation}
\hat\omega_{11}=\ln|k_2|\,,\quad\hat\omega_{22}=
\ln|k_1|\,,\quad
\hat\omega_{12}=\hat\omega_{21}=-\ln\left|\frac{(u_1-u_2)
(v_1-v_2)}{(u_1-v_2)(v_1-u_2)}\right|\,.
\label{hatom}
\end{equation}
For the massless boson scattering amplitude only the
proportional to $s$ term presents on the right side of
(\ref{sapp}).  First we discuss the case where both $v_1$ and
$v_2$ does not go to $z_3$.  The configurations $|k_1|\geq|k_2|$
and $|k_1|\leq|k_2|$ give the same contribution to the
amplitude. So we  consider $|k_1|\geq|k_2|$.  We define new
variables by
\begin{eqnarray}
\ln|k_1|=x,\quad\ln|k_2|=x\alpha,\quad
|u_2-u_1|=|k_1|^{\beta}|v_1-z_3|,
\nonumber\\
u_1-z_3=
y_1(u_2-u_1),\quad z-u_1=y(u_2-u_1)
\label{new}
\end{eqnarray}
where $\alpha\geq1$. From (\ref{ndgbound}), it follows that
$0\leq\beta\leq1/2$.  Generically, $y\sim y_1\sim1$. In the
boson string theory \cite{divfl,dan89} at $k_1\to0$ the
holomorphic partition function is $\sim|k_1k_2|^{-4}$ while in
the superstring theory (see Section 5 of the present paper) it
is $\sim|k_1k_2|^{-2}$. Terms of the expansion of the integrand
(\ref{ampl}) over the powers of the small variables corresponds
to different thresholds while the expansion over
powers of $1/x$ corresponds to the expansion over powers of the
center mass space momentum of the intermediate state.  Near the
given threshold the leading approximated amplitude
$A_4^{(2)}(s)$ discussed is found to be
\begin{equation}
A_4^{(2)}(s)=\frac{(4\pi)^{D}}{8} \int\limits_1^\infty
d\alpha\int\limits_0^{1/2}d\beta
\frac{\hat
A_1(\alpha,\beta)\hat A_2(\alpha,\beta)}
{[\alpha-\beta^2]^{D/2}}
\int\limits_{-\infty}^{-\kappa}\frac{e^{x\tilde S}}{x^{D-2}}dx
\label{sngamp}
\end{equation}
where the variables are defined by (\ref{new}). In this case
$D=10$ for the superstring and $D=26$ for the boson string. The
cutoff $\kappa>>1$ bounds the region of small $|k_1|$.
Furthermore, $\hat A_1(\alpha,\beta)$ is an
integral over real and imaginary parts of both  $y$ and $y_1$
while $\hat A_2(\alpha,\beta)$ is an integral over real and
imagery parts of $v_1$ and of $v_2$. By using (\ref{sapp}) one
find that
\begin{equation}
\tilde
S=\biggl(\beta-\frac{\alpha\beta^2+\beta^2
-2\beta^3}{\alpha-\beta^2}\biggl)
\biggl[-\frac{s}{4} +\tilde
p(\alpha,\beta)\biggl(\beta-\frac{\alpha\beta^2+\beta^2
-2\beta^3}{\alpha-\beta^2}\biggl)^{-1}\biggl]
\label{tildes}
\end{equation}
where $\tilde p(\alpha,\beta)$ is a linear function of its
arguments with the coefficients depending on the threshold
discussed. In the superstring theory all the coefficients are
non-negative. At $\tilde S\geq0$ the integral (\ref{sngamp}) is
divergent. Thus the cut begin with that $s$, which is the
minimal value of $s$ where $\tilde S$ is nullified in the
integration region. To calculate the discontinuity, we go to
$\tilde x=-x\tilde S$ integrated from  $\kappa\tilde S$ till
$\infty$.   When $\tilde S$ rounds the $\tilde S=0$ point, the
initial point $\kappa\tilde S$ of the integration contour gets
about the pole at $\tilde x=0$ to be either above the $\tilde
x=0$  point, or below it, depending on the sign of the
$Im\,\tilde S$. Thus the discontinuity $[A_4^{(2)}(s)]_{disc}$
of $\tilde A_4^{(2)}(s)$ is given by the integral over $\tilde
x$ along the closed contour surrounding the $\tilde x=0$ point
to be
\begin{equation}
[A_4^{(2)}(s)]_{disc}=\frac{(4\pi)^{D}}{8} \int\limits_1^\infty
d\alpha\int\limits_0^{1/2}d\beta
\frac{\hat
A_1(\alpha,\beta)\hat A_2(\alpha,\beta)}
{[\alpha-\beta^2]^{D/2}}\theta(-\tilde S)\frac{2\pi
\tilde S^{(D-3)}}{(D-3)!}
\label{disamp}
\end{equation}
Every threshold $s=s_i$ determine the minimum of the second term
in the square brackets on the right side of (\ref{tildes}) at
corresponding $\alpha=alpha_i$ and $\beta=\beta_i$, which, among
other, can be on the boundary of the region. At $(s-s_i)\to0$
only small $(\alpha-\alpha_i)$ and $\beta-\beta_i$ contribute to
(\ref{disamp}).  Being smooth functions of $\alpha$ and $\beta$,
in the leading approximation $\hat A_1(\alpha,\beta)$ and $\hat
A_2(\alpha,\beta)$ both are replaced by $\hat
A_1(\alpha_i,\beta_i)$ and $\hat A_2(\alpha_i,\beta_i)$. It is
naturally to expect that  every one from $\hat
A_1(\alpha_i,\beta_i)$ and $\hat A_2(\alpha_i,\beta_i)$ present
the threshold values of the corresponding $2\to3$ amplitude as
this is for three tachyon cut of the tachyon-tachyon amplitude
in the boson string theory. In this case $\tilde
p(\alpha,\beta)=m_{th}^2(1+\alpha-\beta)/4$ where $m_{th}^2=-8$
is the square of the tachyon mass.  As far as $m_{th}^2<0$, the
calculation includes rather subtle matters, which are not
discussed here. Instead we calculate the discontinuity for a
$m_{th}^2>0$ continuing the obtained result to $m_{th}^2=-8$.
There is only one threshold to be at $s=9m_{th}^2$. In this case
$\beta\approx1/2$ and $\alpha\approx1$.  So (\ref{tildes}) is
approximated by
\begin{equation}
\tilde S\approx
\frac{1}{3}\biggl(-\frac{s-9m_{th}^2}{6}
+8m_{th}^2[(\alpha-1)^2-2(\alpha-1)(\frac{1}{2}-\beta)+
2(\frac{1}{2}-\beta)^2]\biggl)
\label{tilds}
\end{equation}
Other factors in (\ref{disamp}) are taken at $\alpha=1$ and
$\beta=1/2$.  One can check that $\hat A_1(1,1/2)=\hat
A_2(1,1/2) =A_5^{(0)}$ is really the $2\rightarrow3$ tree
tachyon interaction amplitude \cite{gsw} at $s=9m_{th}^2$.  In
this case $\hat A_1(1,1/2)$ given by the integral where three
vertex coordinates are fixed to be 0, 1 and $\infty$. In $\hat
A_2(1,1/2)$ the fixed coordinates are $z_1$, $z_2$ and $z_3$.
The integral is none other than $[A_5^{(0)}]^2$ times the phase
volume.

For the configuration $v_2\to z_3$, $u_1\to z_3$ and $u_2\to z_3$
we discuss, as an example, the case $k_1\geq k_2$.
We define new integration variables as it follows
\begin{eqnarray}
\ln|k_1|=x,\quad\ln|k_2|=x\alpha,\quad
|u_2-u_1|=|k_1|^{\beta+\delta}|v_1-z_3|,\quad
|v_2-u_1|=|k_1|^\delta|v_1-z_3|,
\nonumber\\
|u_1-z_3|=
|k_1|^\eta|v_1-z_3|,\quad z-u_1=|k_1|^\eta y(v_1-z_3)
\label{newvar}
\end{eqnarray}
where $1\leq\alpha$ and $0\leq\delta\leq\eta$. It follows from
(\ref{ndgbound}) that $0\leq\beta\leq1/2$.  In addition,
$0\leq\eta\leq1/2$, as far as $z_3$ lies out of the $C_{u_1}$
circle.  Then, by using eq.(\ref{sapp}), for the tachyon-tachyon
forward scattering amplitude the expression of $\tilde S$ in
(\ref{disamp}) is found to be
\begin{eqnarray}
\tilde
S=\biggl(\eta-\frac{\alpha\eta^2-(\eta-\delta)^2
-2\beta\eta(\eta-\delta)}{\alpha-\beta^2}\biggl)
\biggl[-\frac{s}{4}+\frac{m_{th}^2}{4}(1+\alpha-\beta)
\nonumber\\
\times
\biggl(\eta-\frac{\alpha\eta^2-(\eta-\delta)^2
-2\beta\eta(\eta-\delta)}{\alpha-\beta^2}\biggl)^{-1}\biggl]
\label{tlds}
\end{eqnarray}
In this case the false threshold appears
at $s=6m_{th}^2$, which corresponds to the minimum of
the last term on the right side of (\ref{tlds}) to be at
$\alpha=1$, $\beta=\eta=1/2$ and $\delta=\eta(1-\beta)=1/4$.
The discussed configuration is removed from the integral
as it is proposed in Section 4.

\section{Explicit Scalar superfield Green functions}
\setcounter{equation}{0}

Explicitly $R_L^{(n)}(t,t';\{q\})$ can be given through the
genus-1 Green functions $R_{l_s}^{(1)}(t,t';s)$ calculated for
the Schottky parameters $\hat m_s=(k_s,u_s,v_s)$ along with the
Grassmann ones $\mu_s$ and $\nu_s$, the spin structure being
$l_s=(l_{1s},l_{2s})$. Due to (\ref{gamab}), the above genus-1
function is written through $z_s$ and $\vartheta_s$ in
(\ref{tgam}) using the boson Green function $R_b^{(1)}(z,z';\hat
m_s)$ and the fermion Green one $R_f^{(1)}(z,z';\hat
m_s;l_{1s},l_{2s})$ as it follows \cite{danphr}
\begin{eqnarray}
R_{l_s}^{(1)}(t,t';s)=R_b^{(1)}(z_s,z_s';\hat m_s)
-\vartheta_s\vartheta_s'
R_f^{(1)}(z_s,z_s';\hat m_s;l_{1s},l_{2s})
\nonumber\\
+\tilde\varepsilon_s'\vartheta_s'\Upsilon_s(\infty,z_s')
+\tilde\varepsilon_s'\vartheta_s\Upsilon_s(z_s,\infty)\,,
\nonumber\\
\Upsilon_s(z,z')=(z-z')R_f^{(1)}(z,z';\hat m_s;l_{1s},l_{2s})
\label{zgrin}
\end{eqnarray}
The proportional to $\Upsilon_s$ terms are added to provide
decreasing  $K_{l_s}^{(1)}(t,t';s)$ at $z\to\infty$ or
$z'\to\infty$. The above $K_{l_s}^{(1)}(t,t';s)$ is related with
$R_{l_s}^{(1)}(t,t';s)$ by (\ref{kr}). The Poincar{\'e} series
for  boson and fermion Green functions are given in the end of
this Appendix.  For the odd spin structure $l_m=(1/2,1/2)$ we
use \cite{danphr} the Green function  with the property that
\begin{equation}
R_{l_m}^{(1)}(t_m^b,t';m)=R_{l_m}^{(1)}(t,t';m)+
J_m^{(1)}(t')-
\varphi_m(t)\varphi_m(t')
\label{trans1}
\end{equation}
where $J_m^{(1)}(t)$ the genus-1 scalar function and
$\varphi_m(t)$ is the spinor zero mode given by
\begin{equation}
\varphi_m(t)=\frac{\vartheta_m(u_m-v_m)^{1/2}}
{[(z_m-u_m)(z_m-v_m)]^{1/2}}+
\varepsilon_m'(u_m-v_m)^{1/2}
\label{phi}
\end{equation}
Here $(z_m|\theta_m)$ variables are determined by (\ref{tgam}).
The last term in (\ref{phi}) provides vanishing the spinor zero
mode at $z\to\infty$. The genus-1 scalar function $J_s^{(1)}(t)$
is
\begin{equation}
J_s^{(1)}(t)=\ln\frac{z_s-u_s}{z_s-v_s}
\label{jr1}
\end{equation}
where $z_r$ is given by (\ref{tgam}).

For the genus-$n$ superspin structure without the odd genus-1
spin structures, the desired Green function is directly given by
(\ref{rpart}) $\tilde R_{L_r}^{(n_r)}(t,t';\{q\}_r)$ are
replaced by $\tilde R_{l_s}^{(1)}(t,t';s)$. The kernel of the
operator $\hat K=\hat K_{sr}$ for $s\neq r$ is $\tilde
K_{l_s}^{(1)}(t,t';s)dt'$, which is the non-singular part of
$K_{l_s}^{(1)}(t,t';s)$ related with
$\tilde R_{l_s}^{(1)}(t,t';s)$ by (\ref{kr}). The non-singular
part of the Green function is defined by (\ref{lim}).
The integration over $t'$ is performed along $C_r$-contour
surrounding the limiting points $u_r$ and $v_r$ and, for the
Ramond handle, the cut between them.

Below we denote $R_L^{[n]}(t,t';\{q\})$ the Green function given
it terms of the $R_{l_s}^{(1)}(t,t';s)$ genus-1 functions by
(\ref{rpart}). If odd spin structure handles present, the
change of $R_{l_s}^{(1)}(t,t';s)$ under $t\to t_s^{(b)}$
is different from (\ref{rtrans}) due to the last term in
(\ref{trans1}).  We show that in this case the Green function
satisfying (\ref{rtrans}), is given by
\begin{equation}
R_L^{(n)}(t,t';\{q\})= R_L^{[n]}(t,t';\{q\}) -\frac{1}{2}
\sum_{m,m'}\Phi_m(t;L;\{q\})
\hat V_{mm'}^{-1}\Phi_{m'}(t';L;\{q\})
\label{rprta}
\end{equation}
where the last term appears, if genus-1 odd spin structures
present.  The $\hat V_{mm'}$ matrix elements are defined only
for those $(m,m')$ that label the odd genus-1 spin structures.
Both $\hat V_{mm'}$ and $\Phi_m(t;L;\{q\})$ are calculated in
terms of the genus-1 zero spinor modes $\varphi_m(t)$ defined by
(\ref{phi}).  In so doing
\begin{equation} \Phi_m(t;L;\{q\})=
\varphi_m(t)+\sum_p\int_{C_m}[(1-\hat K)^{-1}\hat
K]_{pm}(t,t')dt'\varphi_m(t')
\label{bphi}
\end{equation}
where $\delta_{mm'}$ is the Kronecker symbol
while the $\hat V_{mm'}$ matrix is found
to be\footnote{This matrix is slightly different from the
corresponding matrix in \cite{danphr}. Eq.(\ref{rprta})
can be also obtained from eq.(59) of \cite{danphr}
using the second of eqs.(52) in \cite{danphr}.}
\begin{eqnarray}
\hat V_{mm'}=-\frac{1}{2}\sum_{p\neq m}\int_{C_p}D(t)\varphi_m(t)dt
\int_{C_{m'}}[(1-\hat K)^{-1}\hat
K]_{pm'}(t,t')dt'\varphi_{m'}(t') \nonumber\\
-\frac{1}{2}(1-\delta_{mm'})\int_{C_{m'}}D(t)\varphi_m(t)dt
\varphi_{m'}(t)\,,
\label{vmatr}
\end{eqnarray}
To check that (\ref{rprta}) satisfies eqs.(\ref{rtrans}), the
first term on the right side of (\ref{rprta}) is presented by
(\ref{rpart}), and $\Phi_m(t;L;\{q\})$ in (\ref{rpart}) is
presented in the like way as
\begin{eqnarray}
\Phi_m(t;L;\{q\})=
\sum_{p\neq r}\int_{C_p}K_{l_r}^{(1)}(t,t_1;r)dt_1\int_{C_m}
[(1-\hat K)^{-1}\hat
K]_{pm}(t_1,t_2)dt_2\varphi_m(t_2)
\nonumber\\
+\delta_{mr}\varphi_m(t)+(1-\delta_{mr})
\int_{C_m}K_{l_r}^{(1)}(t,t_1;r)dt_1\varphi_m(t_1)
\label{bigphi}
\end{eqnarray}
Indeed, calculating the contribution to (\ref{bigphi}) of the
pole term in $K_{l_r}^{(1)}(t,t_1;r)$, one obtains (\ref{bphi})
using (\ref{opr}).  One can see from  (\ref{bigphi}) the
relations (\ref{rtrans}) to be true, the scalar function
$J_r^{(n)}(t;\{q\};L)$ in (\ref{rtrans}) being
\begin{equation}
J_r^{(n)}(t;\{q\};L)=\tilde J_r^{(n)}(t;\{q\};L)
-\frac{1}{2}\sum_{m,m'}\Phi_m^{(r)}(L;\{q\})
\hat V_{mm'}^{-1}\Phi_{m'}(t;L;\{q\})
\label{jr}
\end{equation}
where $\tilde J_r^{(n)}(t;\{q\};L)$ is given by (\ref{tjr})
through the genus-1 functions, and
\begin{eqnarray}
\Phi_{m'}^{(r)}(L;\{q\})=
\Phi_{m'}(t_r^b;L;\{q\})-\Phi_{m'}(t;L;\{q\})=
(1-\delta_{mr})
\int_{C_m}D(t_1)J_r^{(1)}(t_1)dt_1\varphi_m(t_1)
\nonumber\\
+\sum_{p\neq r}\int_{C_p}D(t_1)J_r^{(1)}(t_1)dt_1\int_{C_m}
[(1-\hat K)^{-1}\hat
K]_{pm}(t_1,t_2)dt_2\varphi_m(t_2)\,.
\label{phir}
\end{eqnarray}
The genus-1 function $J_r^{(1)}(t)$ is given by (\ref{jr1}).
To calculate the period matrix,
one chooses a fixed parameter $t_0$, as it is discussed in
Section 3. Then one can check that
\begin{eqnarray}
J_r^{(n)}(t;\{q\};L)-J_r^{(n)}(t_0;\{q\};L)=
[\tilde J_r^{(n)}(t;\{q\};L)-\tilde J_r^{(n)}(t_0;\{q\};L)]
\nonumber\\
-\frac{1}{2}\sum_{m,m'}\Phi_m^r(L\{q\})
\hat V_{mm'}^{-1}\{\Phi_{m'}(t;L;\{q\})-
\Phi_{m'}(t_0;L;\{q\})\}
\label{jara}
\end{eqnarray}
where the term in square brackets is calculated  by (\ref{jar})
for $j_r=r$ through the genus-1 functions.  To prove
(\ref{jara}), one calculates the contribution to (\ref{jara}) of
the singular term in the Green function by the method given
in Section 3. In addition, eq.(\ref{bphi}) for
$\Phi_{m'}(t;L;\{q\})$ is used.  From (\ref{jara}), the period
matrix elements are found to be
\begin{eqnarray}
2\pi i\omega_{rs}^{(n)}(\{q\};L)=
2\pi i\tilde\omega_{rs}^{(n)}(\{q\};L)
-\frac{1}{2}\sum_{m,m'}\Phi_m^{(r)}(L;\{q\})
\hat V_{mm'}^{-1}\Phi_{m'}^{(s)}(L;\{q\})
\nonumber\\
+(1-\delta_{rs})\int_{C_s}D(t_1)J_r^{(1)}(t)dtJ_s^{(1)}(t)\,.
\label{omjra}
\end{eqnarray}
where $\tilde\omega_{rs}^{(n)}(\{q\};L)$ is presented by
(\ref{omjr}) at $j_r=r$ and $j_s=s$ in terms of the genus-1
functions.  The boson Green function $R_b^{(1)}(z,z';\hat m)$ in
(\ref{zgrin}) is given by (the symbol "$s$" is omitted)
\begin{equation}
\partial_{z'}R_b^{(1)}(z,z';\hat m)=
-\sum_n\frac{1}{(z-g_n(z'))(c_nz'+d_n)^2}
\label{bsc}
\end{equation}
where the sum is performed  over the group products of the
Schottky transformation $g(z)$. In so doing $g_0(z)=z$, and
that negative values of $n$ are associated with the inverse
transformations. The fermion Green functions are given by
\begin{eqnarray}
R_f^{(1)}(z,z';\hat m;l_1=1/2,l_2)=
\sum_{n=0}^\infty(-1)^{(2l_2+1)n}
\left[\frac{1}{1-k^n\frac{(z-v)(z'-u)}{(z-u)(z'-v)}}-
\frac{1}{1-k^n\frac{(z-u)(z'-v)}{(z-v)(z'-u)}}\right]
\nonumber\\ \times
\frac{(u-v)}
{2\sqrt{(z-u)(z-v)(z'-u)(z'-v)}}\,,
\nonumber\\
R_f^{(1)}(z,z';\hat m;l_1=0,l_2)=
\sum_n\frac{(-1)^{(2l_2+1)n}}{(z-g_n(z'))(c_nz'+d_n)}.
\label{fsc}
\end{eqnarray}
where like (\ref{bsc}), the summation is performed  over the
group products of $g(z)$. At $z'\to\infty$
\begin{equation}
R_f^{(1)}(z,z';\hat m;l_1,l_2)\to \frac{1}{z-z'}+
\frac{\hat W_1(z;\hat m;l_1,l_2)}{z'-u}+
\frac{\hat W_2(z;\hat m;l_1,l_2)}{(z'-u)^2}\,.
\label{asfgf}
\end{equation}
with  corresponding $\hat W_1(z;\hat m;l_1,l_2)$ and
$\hat W_2(z;\hat m;l_1,l_2)$. At $z\to\infty$
\begin{equation}
\hat W_2(z';\hat m;l_1,l_2)\to-(z-u)
\hat W_1(z';\hat m;l_1,l_2)\,,
\qquad
\hat W_1(z;\hat m;l_1,l_2)\to\frac{\hat
a(k;l_1,l_{2})(u-v)^2}{(z'-u)^2}
\label{assig}
\end{equation}
where $\hat a(k;l_1,l_2)$ depends on the multiplier and on the
spin structure. Up to the unessential constant term,
the non-singular part (\ref{lim}) of the genus-1 Green function
(\ref{zgrin})  is given at $z'\to\infty$  by (see
eq.(\ref{tgam}) for definitions)
\begin{eqnarray}
\tilde R_{l_s}^{(1)}(t,t';s)(z'-u)=(\vartheta-\varepsilon(z))
(1+\varepsilon\varepsilon')[
\hat W_1(z;\hat m;l_1,l_2)(\vartheta'-\mu)
+\hat W_2(z;\hat m;l_1,l_2)\varepsilon']
\nonumber\\
+[\hat W_b(z;\hat m)-\vartheta\varepsilon(z)\partial_z
\hat W_b(z;\hat m)]
(1-\varepsilon'\vartheta')
\label{assgr}
\end{eqnarray}
where $\hat W_b(z;\hat m)$ determines the asymptotics of
the boson Green function (\ref{bsc}).

\section{Integration measures}
\setcounter{equation}{0}

In this Appendix we reduce to the convenient for application
form eq.(127) from \cite{danphr} expressing
$\tilde Z^{(n)}(\{q\},L)$ through genus-1 functions. First, we
transform the scalar superfield contribution on the right side
of eq.(112) of \cite{danphr} as it follows (in notations of
\cite{danphr})
\begin{eqnarray}
trace\ln(I-\hat K^{(1)}+\hat\varphi\hat f)
=trace\ln(I-\hat
K^{(1)})
-trace\ln[I+\hat f_1(I-\hat K^{(1)})^{-1}\varphi]
\nonumber\\
=trace\ln(I-\hat K^{(1)}) -\ln\det \hat V \,.
\label{dfln}
\end{eqnarray}
We have used that for any operators $A_1$, $A_2$ and for
$A$ with $\det A\neq0$,
\begin{eqnarray}
trace\ln[A+A_1A_2]=trace\ln A+trace\ln[1+A^{-1}A_1A_2]
\nonumber\\
=trace\ln A+(-1)^Ptrace\ln[1+A_2A^{-1}A_1]
\label{trace}
\end{eqnarray}
where $P=1$ when both $A_1$ and $A_2$ are the Fermi operators,
otherwise $P=0$. In addition, eq.(50) and eq.(52) of
\cite{danphr} are used. The $K^{(1)}$ operator in
(\ref{dfln}) is the same as Appendix B.  In addition
the ghost contribution we express now in terms of function $\hat
G^{(1)}$ given below instead of $G_\sigma^{(1)}(z,z')$ in
eq.(127) from \cite{danphr}.  The desired factor in
(\ref{zhol}) is given by
\begin{equation}
\ln\tilde
Z^{(n)}(\{q\},L)=-5trace\ln(I-\hat K)+5\ln\det \hat V
+trace\ln(I-\hat G)-\ln\det \hat U
\label{tzh}
\end{equation}
where the $\hat K$ operator is the same as in (\ref{rprt})
and the $\hat V$ matrix is defined by
(\ref{vmatr}). The matrix operator
$\hat G$ is defined in terms of a genus-1
ghost correlator $G_{l_s}^{(1)}(t,t';s)$ defined below in the
same manner as $\hat K$ is given in terms of
$K_{l_s}^{(1)}(t,t';s)$.  So $\hat G=\{\hat G_{sr}\}$ where
$\hat G_{sr}$ is an integral operator vanishing at $s=r$.  For
$s\neq r$, the kernel of $\hat G_{sr}$ is $\tilde
G_{l_s}^{(1)}(t,t';s)dt'$ defined by (\ref{greg}) for the
genus-1 case.  Like $\hat V$, the elements $\hat
U_{mn}$ of  $\hat U$ are defined only for $(m,m')$ assigned to
the odd genus-1 spin structures. They are given it terms of 3/2
zero modes $\chi_m^{(1)}(t)$ and in terms of -1/2 genus-1 zero
modes $\phi_m^{(1)}(t)$ as it follows
\begin{eqnarray}
\hat
U_{mm'}=-\frac{1}{2}\sum_{p\neq m}\int_{C_p}\chi_m^{(1)}(t)dt
\int_{C_{m'}}[(1-\hat
G)^{-1}\hat G]_{pm'}(t,t')dt'\phi_{m'}^{(1)}(t')
\nonumber\\
-\frac{1}{2}(1-\delta_{mm'})\int_{C_{m'}}\chi_m^{(1)}(t)dt
\phi_{m'}^{(1)}(t)\,.
\label{umatr}
\end{eqnarray}
The above genus-1 zero modes are given by
\begin{equation}
\chi_m^{(1)}(t)=-\frac{(u_m-v_m)^2}
{[(z_m-u_m)(z_m-v_m)Q_m^2(t)]^{3/2}}\,,\quad
\phi_m^{(1)}(t)=
\frac{\vartheta_mQ_m^2(t)\sqrt{(z_m-u_m)(z_m-v_m)]}}
{(u_m-v_m)}
\label{uf}
\end{equation}
where $Q_m$ is defined in (\ref{supder}) and the
$(z_m|\theta_m)$ variables are defined by (\ref{tgam}) at $s=m$.
The genus-1 function $G_{l_s}^{(1)}(t,t';s)$ is given through
the boson Green function  $G_b^{(1)}(z,z';\hat m_s)$
and the fermion Green one $G_f^{(1)}(z,z';\hat
m_s;l_{1s},l_{2s})$ as \cite{danphr}
\begin{eqnarray}
G_{l_s}^{(1)}(t,t';s)=Q_s^2(t)
[G_b^{(1)}(z_s,z_s';\hat m_s)
\theta_s'+
\theta G_f^{(1)}(z_s,z_s';\hat m_s;l_{1s},l_{2s})
\nonumber\\
-\tilde\varepsilon_s'\Upsilon_s^{(gh)}(\infty,z_s')]Q_s^{-3}(t')\,,
\nonumber\\
\Upsilon_s^{(gh)}(z,z')=
(z-z')G_f^{(1)}(z,z';\hat m_s;l_{1s},l_{2s})
\label{zghgri}
\end{eqnarray}
where $z_s$ and $\vartheta_s$ are defined by (\ref{tgam}) while
$\hat m_s=(k_s,u_s,v_s)$. The
proportional to $\Upsilon_s^{(gh)}$  terms
provide decreasing $G_{l_s}^{(1)}(t,t';s)$ at
$z\to\infty$ or at $z'\to\infty$.
The boson part of the ghost Green function in (\ref{zghgri}) is
(see eq.(68) in \cite{danphr})
\begin{equation}
G_b^{(1)}(z,z';\hat m)=
-\sum_n\frac{1}{(z-g_n(z'))(c_nz'+d_n)^4}
\label{bgsc}
\end{equation}
where the summation is performed  over the
group products of $g(z)$.
For even spin structures the fermion part in (\ref{zghgri})
expressed in terms of (\ref{fsc}) as
\begin{eqnarray}
G_f^{(1)}(z,z';\hat m;l_1,l_2)=
\frac{(z-u)(z-v)}{(z'-u)(z'-v)}R_f^{(1)}(z,z';\hat m;l_1,l_2)
\nonumber\\
-\frac{(z-v)\Sigma_1(z';\hat m;0,l_2)+
\Sigma_2(z';\hat m;0,l_2)}{(z'-u)(z'-v)}
\label{fgsc}
\end{eqnarray}
where the last term is calculated in terms of
(\ref{asfgf}). In this case
\begin{equation}
\Sigma_1(z;\hat m;l_1,l_2)=1+
\hat\Sigma_1(z;\hat m;l_1,l_2)\,,\qquad
\Sigma_2(z;\hat m;l_1,l_2)=z-u+
\hat\Sigma_2(z;\hat m;l_{1},l_{2})\,.
\label{asrel}
\end{equation}
One can check that the function (\ref{fgsc}) goes to zero
at $z\to\infty$. Furthermore,  at $z'\to\infty$
\begin{equation}
G_f^{(1)}(z,z';\hat m;l_1,l_2)\to \frac{1}{z-z'}+
\frac{\Sigma_{gh}(z;\hat m;l_1,l_2)}{(z'-u)^3}
\label{azpeg}
\end{equation}
where the numerator in the last term is a function os $z$. If
both $z\to\infty$ and $z'\to\infty$, then
\begin{equation}
G_f^{(1)}(z,z';\hat m;l_1,l_2)\to \frac{1}{z-z'}-
\frac{a_{gh}(k;l_1,l_2)(u-v)^4}{(z-u)(z'-u)^3}
\left[\frac{3}{z'-u}-
\frac{1}{z-u}\right]
\label{azzeg}
\end{equation}
where $a_{gh}(k;l_1,l_2)$ depends on the multiplier and on the
spin structure. Moreover,
\begin{eqnarray}
(cz+d)G_f^{(1)}(g(z),z';\hat m;l_1,l_2)-
G_f^{(1)}(z,z';\hat m;l_1,l_2)+\frac{1-\sqrt k}{\sqrt k}p_\mu(z)
\chi_\mu(z';\hat m;l_1,l_2)
\nonumber\\
-(1-\sqrt k)
\chi_\nu(z';\hat m;l_1,l_2)
\label{evfcon}
\end{eqnarray}
where $p_\mu(z)$ and $p_\nu(z)$ are given by
\begin{equation}
p_\mu(z)=\frac{2(z-v)}{u-v}\,,\quad
p_\nu(z)=-\frac{2(z-u)}{u-v}
\label{pol}
\end{equation}
while the depending on $z'$ functions are defined to be
\begin{eqnarray}
\chi_\mu(z;\hat m;l_1,l_2))=-
\frac{(u-v)\Sigma_1(z;\hat m;l_1,l_2)+
\Sigma_2(z;\hat m;l_1,l_2)}{2(z-u)(z-v)}\,,
\nonumber\\
\chi_\nu(z;\hat m;l_1,l_2))=-
\frac{\Sigma_2(z;\hat m;l_1,l_2)}{2(z-u)(z-v)}\,.
\label{chistev}
\end{eqnarray}
Eq.(\ref{evfcon}) is non other than eq.(63) of \cite{danphr} in
the genus-1 case.  For the odd spin structure, due to the
$(-1/2)$ mode, there is no the Green function obeying
(\ref{evfcon}).  In this case we define the ghost Green function
by
\begin{eqnarray}
G_f^{(1)}(z,z';\hat m;1/2,1/2)=G_{(\sigma=1)}^{(1)}(z,z')
-2\left(\sqrt{\frac{z-u}{z-v}}-1\right)
\chi_\nu(z';\hat m;1/2,1/2)\,,
\nonumber\\
\chi_\nu(z;\hat m;1/2,1/2)=-\frac{1}{2\sqrt{(z-u)(z-v)}}\sum_n
\frac{1}{(c_nz'+d_n)^2}
\label{ghostodd}
\end{eqnarray}
where $G_{(\sigma=1)}^{(1)}(z,z')$ is defined
by eq.(69) in \cite{danphr}. At $z'\to\infty$
\begin{equation}
G_f^{(1)}(z,z';\hat m;1/2,1/2)\to\frac{1}{z-z'}+
\frac{\Sigma_{gh}(z;\hat m;1/2,1/2)}{(z'-u)^2}\,.
\label{azpog}
\end{equation}
Moreover, at $z\to\infty$
\begin{equation}
\Sigma_{gh}(z;\hat m;1/2,1/2)\to\frac{(u-v)^2}{8(z-u)}
\label{azzpog}
\end{equation}
Under the Schottky transformation $g(z)$ the above function
(\ref{ghostodd}) is changed as
\begin{eqnarray}
(cz+d)G_{(f)}^{(1)}(g(z),z';\hat m;1/2,1/2)-
G_f^{(1)}(z,z';\hat m;1/2,1/2)=
\nonumber\\
\left[\frac{1-\sqrt k}{\sqrt k}p_\mu(z)
-(1-\sqrt k)p_\nu(z)\right]
\chi_\nu(z';\hat m;1/2,1/2)
\nonumber\\
-\frac{\sqrt{
(z-u)(z-v)}(u-v)}{[(z'-v)(z'-u)]^{\frac{3}{2}}}\,.
\label{trghodd}
\end{eqnarray}
Eq.(\ref{trghodd}) follows from (\ref{ghostodd}) along with
eq.(82) in \cite{danphr}. We show that in this case the left
side of (\ref{trghodd}) is obtained in a form of series, which
are given in terms of (\ref{trghodd}) as it follows
\begin{eqnarray}
-\frac{(u-v)^2}{[(z-v)(z-u)]^{\frac{3}{2}}}=
c\sqrt{\frac{z-v}{z-u}}\sum_{n=-\infty}^{\infty}
\frac{1}{k^{(n+1)/2}Q_n(z)Q_{n+1}^2(z)}+
\frac{1}{k^{n/2}Q_n^2(z)Q_{n+1}(z)}\,,
\nonumber\\
\chi_\nu(z;\hat m;1/2,1/2)=
\frac{c}{2}\sqrt{\frac{z-v}{z-u}}\sum_{n=-\infty}^{\infty}
\frac{1}{k^{(n-1)/2}Q_n(z)Q_{n+1}^2(z)}+
\frac{1}{k^{n/2}Q_n^2(z)Q_{n+1}(z)}\,.
\label{munuod}
\end{eqnarray}
Here $Q_n(z)=c_nz+d_n$ for the group product $g^n$. Indeed, the
right side in the first line of (\ref{munuod}) is $\sim
1/z^{3/2}$ at $z\to\infty$, and so it is proportional to
3/2-zero mode on the left side. To verify the coefficient, the
both parts are multiplied by $\sqrt{(z-u)(z-v)}$. Then the
equation is integrated along the corresponding Schottky circle.
In the second line, the leading at $z\to\infty$ term on the left
side is equal to corresponding term on the right side.  So the
right side of the discussed relation may differ from its left
side only by the term proportional to $\chi_\mu(z;\hat
m;1/2,1/2)$. To verify the equation, it is again multiplied
by $\sqrt{(z-u)(z-v)}$ and then it is integrated along the
Schottky circle.  For even spin structures the ghost function
(\ref{fgsc}) is related to $G_{(\sigma=1)}^{(1)}(z,z')$ in
\cite{danphr} by
\begin{eqnarray}
G_{(\sigma=1)}^{(1)}(z,z')=
G_f^{(1)}(z,z';\hat m;l_1,l_2)
-[p_\mu(z)\chi_\mu(z';\hat m;l_1,l_2)+
p_\nu(z)\chi_\nu(z';\hat m;l_1,l_2)]
\nonumber\\
+\frac{1}{2}\sqrt{\frac{z-u}{z-v}}
[p_\mu(z)(3\chi_\mu(z';\hat m;l_1,l_2)
-\chi_\nu(z';\hat m;l_1,l_2))
\nonumber\\
+p_\nu(z)(\chi_\mu(z';\hat m;l_1,l_2)+
\chi_\nu(z';l_1,l_2))]
\label{ghostev}
\end{eqnarray}
with $p_\mu(z)$ and $p_\nu(z)$ being defined by (\ref{pol}).

To derive the ghost terms in (\ref{tzh}) one represents $\hat
S_\sigma^{(1)}$ in eq.(127) of \cite{danphr} through $\hat
G^{(1)}$.  For this purpose one uses eq.(81) of \cite{danphr}
along with eqs.  (\ref{ghostodd}) and (\ref{ghostev}) of the
present paper.  The result expression being arranged by
(\ref{trace}), the desired terms in (\ref{tzh}) are obtained.

To verify eq.(\ref{ipf}), we represent
$G_L^{(n)}(t,t';\{q\})$ in (\ref{ipf})
as it follows
\begin{eqnarray}
G_L^{(n)}(t,t';\{q\})=
\sum_{s=1}^n\tilde
G_{l_s}^{(1)}(t,t';s)
+\sum_{r,s}\int_{C_s}[(1-\hat G)^{-1}\hat
G]_{rs}(t,t_1)dt_1\tilde G_{l_s}^{(1)}(t_1,t';s)
\nonumber\\
+\frac{\vartheta-\vartheta'}{z-z'}
-\sum_{m,m'}\phi_m^{(n)}(t;L;\{q\})
\hat U_{mm'}^{-1}\chi_{m'}^{(n)}(t';L;\{q\})
\label{gprt}
\end{eqnarray}
where the sum in the last term is performed over genus-1 odd spin
handles. The $\hat U_{mm'}$ matrix elements are defined by
(\ref{umatr}) and the functions in the last term on the right
side are calculated in terms of the genus-1 zero modes
(\ref{uf}) by
\begin{eqnarray}
\chi_m^{(n)}(t;L;\{q\})=\chi_m^{(1)}(t)+
\sum_{p,p'}\int_{C_p}\chi_m^{(1)}(t')dt'\int_{C_p'}[(1-\hat
G)^{-1}]_{pp'}(t',t_1)dt_1\tilde G(t_1,t)\,,
\nonumber\\
\phi_m^{(n)}(t;L;\{q\})=\tilde\phi_m^{(1)}(t)+
\sum_p\int_{C_m}[(1-\hat G)^{-1}\hat
G]_{pm}(t,t')dt'\phi_m^{(1)}(t')
\label{bphig}
\end{eqnarray}
where $\tilde\phi_m^{(1)}(t)$ is as it follows
\begin{equation}
\tilde\phi_m^{(1)}(t)=\phi_m^{(1)}(t)
-\frac{1}{2}\vartheta_mQ_m^2(t)[2z_m-u-v]-\frac{\varepsilon'}{8}
\label{phireg}\,.
\end{equation}
Due to (\ref{tgam}) and (\ref{uf}),
$\tilde\phi_m^{(1)}(t)$ vanishes at
$z\to\infty$.
To prove (\ref{gprt}) one uses the same trick as in
(\ref{rprta}). Then one can verify that
$\phi_m^{(n)}(t;L;\{q\})$ is the superconformal 3/2 tensor under
the transformations of the super-Schottky group, and that
(\ref{gprt}) obeys the conditions (63) of the paper
\cite{danphr}.  Thus (\ref{gprt}) is the correct expression for
$G_L^{(n)}(t,t';\{q\})$. In (\ref{umatr}) and (\ref{bphig}) the
$\phi_m^{(1)}(t)$ zero mode can be replaced by
$\tilde\phi_m^{(1)}(t)$. Indeed the difference of the above
quantities has not singularities inside the $C_r$ contour, the
integral of it along $C_r$ being equal to zero.  So one
substitutes in eq.(\ref{ipf}) the Green functions given by
(\ref{rprta}) and by (\ref{gprt}) through the genus-1 functions.
Further one uses (\ref{trace}) considering the $(m,m')$ sum in
(\ref{rprta}) and  in (\ref{gprt}) as the separable operator
$A_1A_2$. In so doing (\ref{tzh}) appears. So (\ref{ipf}) is
proved.

Applying (\ref{trace}) to the non-holomorphic factor in
(\ref{hol}), one can obtains that
\begin{equation}
5trace\ln \hat V-5\ln\det
\Omega_{L,L'}^{(n)}(\{q,\overline q
\})= -5\ln\det\tilde
\Omega_{L,L'}^{(n)}(\{q,\overline q
\})
+5trace\ln\tilde V
\label{trvm}
\end{equation}
where $V$ is the same as in (\ref{tzh}) and the summation over
$(r,s)$ is implied. The $\tilde
\Omega_{L,L'}^{[n]}(\{q,\overline q \})$ matrix is calculated
for the period one $\tilde\omega_{rs}^{(n)}(\{q\};L)$ in
(\ref{omjra}). The $\tilde V_{mm'}$ of $\tilde V$ is defined by
\begin{equation}
\tilde V_{mm'}=
\hat V_{mm'}+\Phi_m^{(r)}(L;\{q\})
[\tilde\Omega_{L,L'}^{(n)}(\{q,\overline q
\})]_{rs}^{-1}
\Phi_{m'}^{(s)}(L;\{q\})
\label{tildv}
\end{equation}
where $\Phi_m^{(r)}(L;\{q\})$ is defined by (\ref{phir}). From
(\ref{vmatr}) one can see that $\hat V^T=-\hat V$ and therefore,
$\tilde V^T=-\tilde V$. So $\hat V_{mm}=\tilde V_{mm}=0$. Then
for the degenerated configuration of Section 5  one obtains
using (\ref{vmatr}) (\ref{phir}), that $det\tilde
V\sim\rho_1/\rho$ when both $L_1$ and $L_2$ are odd super-spin
structures.  From (\ref{umatr}) $\det\hat U$ in (\ref{tzh}) is
not nullified when $1<n_1<n-1$, all the rest factors being
finite.  In this case the integration measure is
$\sim(\rho_1/\rho)^5$.  When both $L_1$ and $L_2$ are odd and
either $n_1=1$, or $n_2=n-n_1=1$, then $\det\hat U$ is
nullified due to a presence of the ghost (-1/2) zero genus-1
mode.  Then  $\det\hat U\sim(\rho_1/\rho)^3$
when $n_1=1$, and $\det\hat U\sim(\rho_1/\rho)^2$ for $n_2=1$
and $n>2$. In this case the integration measure is
$\sim (\rho_1/\rho)^2$ and, respectively, $\sim
(\rho_1/\rho)^3$. Hence the integration measure is nullified
when $L_1$ and $L_2$  are odd. Due to the second term on the
right side of (\ref{rprta}), the Green function  is
$\sim1/\sqrt{\rho_1}$ when one of its argument lies near $z_0$
(see Section 5) another argument being at a finite distance from
$z_0$. So the integrand decreases also for the configuration
where one from the vertex coordinates goes to $z_0$. So $L_1$
being odd, the degenerated configurations discussed are not able
to originate divergences, as it has already been noted in the
end of Section 5. In the $1<n_1<n-1$ case the
$\sim(\rho_1/\rho)^5$ smallness in the integration measure is
compensated when a number of the vertices is $\geq10$ and 5
vertices lay near $z_0$.  This case is relevant for obtaining
the contribution to (\ref{ampl}) of odd super-spin structures
\cite{dan96}.

\section{ Property of the function
invariant under  the super-boosts}

\setcounter{equation}{0}

We consider the function
$\psi(\{(x_r|\xi_r)\},\{(w_s|\iota_s)\})$ invariant
under change (\ref{trco}) of its $p$ arguments $(x_r|\xi_r)$
where $r=1,\dots,p$. Here $\iota_s=\nu_s-\mu_s$ while
$w_s=v_s-u_s-\nu_s\mu_s$. Due to the invariance under the
boosts, it depends on the differences $\{(x_r-x_s)\}$. We
presents the above function as it follows
\begin{eqnarray}
\psi(\{(x_r|\xi_r)\},\{(w_s|\iota_s)\})
=\xi_i\left(\prod_{j=2}^p
(\xi_j-\xi_1)\right)
\psi_0(\{x_r\},\{(w_s|\iota_s)\})
\nonumber\\
+\sum_{i=2}^p
\left(\prod_{j=2,j\neq i}^p
(\xi_j-\xi_1)\right)
\psi_i(\{\tilde x_r\},\{(w_s|\iota_s)\})
+\dots
\label{func}
\end{eqnarray}
where $\tilde x_r=x_r-x_1$ while the dots encode lower powers of
$\{\xi_r\}$. Really $\{(x_r|\xi_r)\}$ are identified with
$(u_r|\mu_r)$ and with the vertex coordinates of interest.
Applying (\ref{trco}) to (\ref{func}), one obtains the desired
relation
\begin{equation}
\psi_0(\{(x_r)\},\{(w_s|\iota_s)\})
=\sum_{j=2}^p(-1)^j\partial_{\tilde
x_j}\psi_j(\{(\tilde x_r)\}),\{(w_s|\iota_s)\}\,.
\label{prp}
\end{equation}

\section{SL(2) transformations}
\setcounter{equation}{0}

The general super-conformal transformation is given by
\begin{equation}
z={\it f}(\hat z)+{\it f}'(\hat z)\hat\vartheta\xi(\hat z)\,,
\quad
\vartheta=\sqrt{{\it f}'(\hat z)}[(1+\frac{1}{2}\xi\xi')
\hat\vartheta+\xi(\hat z)]
\label{trn}
\end{equation}
where ${\it f}'(z)=\partial_z {\it f}(z)$ while
${\it f}(z)$ is the transition function and $\xi(z)$ is the
Grassmann partner. For the transformation, which preserving
$z_1$, $z_2$ and $z_3$, reduces $\vartheta_1$
and $\vartheta_2$  to zeros,
\begin{equation}
{\it f}(\hat z)=\hat z-\frac{(\hat z-z_1)(\hat z-z_2)}
{(z_3-z_1)(z_3-z_2)}\hat\vartheta_3\xi_0(z_3)\,,
\quad
\xi(\hat z)=\frac{\vartheta_1(\hat z-z_2)}{(z_1-z_2)\sqrt{{\it
f}'(z_1)}} -\frac{\vartheta_2(\hat z-z_1)}{(z_1-z_2)\sqrt{{\it
f}'(z_2)}}
\label{trnsf}
\end{equation}
where
$\xi_0(z)=[\vartheta_1(z-z_2)-\vartheta_2(z-z_1)]/(z_1-z_2)$.
Evidently, ${\it f}'(z_1){\it f}'(z_2)=1$.
If $\vartheta_1=\vartheta_3=0$, then
${\it f}(\hat z)=\hat z$ and
$\xi(\hat z)=-\vartheta_2(\hat z-z_1)/(z_1-z_2)$.

Under the $L(2)$ transformation given by (\ref{trn}) with
${\it f}(\hat z)=g(\hat z)$ and
$\xi(\hat z)\equiv0$, the spinor derivative of the vacuum
correlator (\ref{corr}) is changed as
\begin{eqnarray}
D(t')\hat X_{L,L'}(t,\overline t;t',\overline t';\{q\})=
[g'(\hat z)]^{-1/2}D(\hat t')\biggl[
\hat X_{L,L'}(\hat t,\overline{\hat t};t',\overline{
\hat t'};\{q\})
\nonumber\\
-\hat X_{L,L'}(\hat t(\infty),\overline{\hat
t(\infty)};t',\overline{ \hat t'};\{q\})\biggl]
\label{trcorr}
\end{eqnarray}
where $\hat t(\infty)=(\hat z=g^{(-1)}(\infty)|
\hat\vartheta=0)$ while
$g^{(-1)}$
is the transformation inverse to $g$. The addition
is determined by the requirement that the right side of
(\ref{trcorr}) vanishes at $z\to\infty$, just as its left part.

\end{document}